\documentclass[final,3p,times]{elsarticle}
\usepackage[utf8]{inputenc}
\usepackage{amssymb}
\usepackage{amsmath}
\usepackage{mathtools}
\usepackage{tabulary, booktabs, siunitx}
\usepackage{caption}
\usepackage{subcaption}
\usepackage{hyperref}
\usepackage{geometry}

\hypersetup{colorlinks=true, linkcolor=blue}

\begin{document}

\begin{frontmatter}

\title{Discerning media bias within a network of political allies and opponents: the idealized example of a biased coin}

\journal{Physica A}

\author[add1]{Nicholas Low Kah Yean\corref{mycorrespondingauthor}}
\cortext[mycorrespondingauthor]{Corresponding author}
\ead{nlow1@student.unimelb.edu.au}

\author[add1,add2]{Andrew Melatos} 
\ead{amelatos@unimelb.edu.au}

\date{\today}
\address[add1]{School of Physics, University of Melbourne, Parkville, VIC 3010, Australia}
\address[add2]{Australian Research Council Centre of Excellence for Gravitational Wave Discovery (OzGrav), University of Melbourne, Parkville, VIC 3010, Australia}

\begin{abstract}
    Perceptions of political bias in the media are formed directly, through the independent consumption of the published outputs of a media organization, and indirectly, through observing the collective responses of political allies and opponents to the same published outputs. A network of Bayesian learners is constructed to model this system, in which the bias perceived by each agent obeys a probability density function, which is updated according to Bayes's theorem given data about the published outputs and the beliefs of the agent's political allies and opponents. The Bayesian framework allows for uncertain beliefs, multimodal probability distribution functions, and antagonistic interactions with opponents, not just cooperation with allies. Numerical simulations are performed to test the idealized example of inferring the bias of a coin. It is found that some agents converge on the wrong conclusion faster than others converge on the right conclusion under a surprisingly broad range of conditions, when antagonistic interactions are present which ``lock out'' some agents from the truth, e.g. in Barabási–Albert networks. It is also found that structurally unbalanced networks routinely experience turbulent nonconvergence, where some agents fail to achieve a steady-state belief, e.g. when they are allies of two agents who are opponents themselves. The subtle phenomenon of long-term intermittency is also explored.
\end{abstract}


\end{frontmatter}


\newcommand{\mean}{\langle \theta \rangle}
\newcommand{\meanSub}[1]{\langle \theta \rangle_{#1}}
\newcommand{\stddev}{\sqrt{\langle ( \theta - \mean ) ^2 \rangle}}
\newcommand{\stddevSub}[1]{\sqrt{\langle ( \theta - \mean ) ^2 \rangle}_{#1}}
\newcommand{\barabasi}{Barabási–Albert }
\newcommand{\erdos}{Erdős–Rényi }

\section{Introduction}

Media bias is defined as a ``portrayal of reality that is significantly and systematically (not randomly) distorted'' (verbatim quote) \cite{groeling2013media}. It must be ``volitional'', ``putatively influential'', ``be reasonably, plausibly threatening to conventional values or institutions'' and ``sustained'', according to the classical definition by Williams \cite{Williams1975}. Substantial resources have been invested recently to refine the theoretical definition of media bias and devise algorithms to detect it in practice \cite{eberl2017, kulshrestha2017quantifying, peng2018, fan2019, Hamborg2019} (see \cite{groeling2013media} for a review). Media bias can manipulate and skew the opinions of consumers of media products and trigger a change in voting behavior \cite{druckman2005impact, eberl2017}. The evolution of human opinion due to media bias is complex, dynamic, and embedded in a network, because the opinions of a consumer are formed partly through communication with other consumers. 

Studies in opinion dynamics aim to understand the diffusion and evolution of human opinions via interacting agents in a network \cite{xia2011multidiscipline}. Network models have been proposed to investigate the influence of the media on public opinion \cite{martins2010mass, sirbu2013opinion, quattrociocchi2014opinion, pineda2015, brooks2020model}. Particularly, the convergence of consumers' opinions onto one or more published media opinions has been the primary focus of this research and has yielded some counterintuitive results. For example, a media organization that is too assertive, say by broadcasting its message too frequently, fails to convince the optimal number of consumers to adopt its message \cite{martins2010mass, sirbu2013opinion, pineda2015, brooks2020model}. Furthermore, the media tend to slow down \cite{brooks2020model} and impede \cite{quattrociocchi2014opinion} consensus among consumers, under certain circumstances. See \cite{sirbu2016survey} for a review on studies of media influence using opinion dynamics models.

In most of the foregoing work, the opinion of a networked individual evolves deterministically; the variable defining their opinion takes a unique, fully determined value at every instant. In reality, opinions seldom form without uncertainty. In this paper, we address the role of uncertainty explicitly by modeling the opinion of a person as a random variable and calculating its probability distribution. That is, we allow the central tendency and dispersion of a person's opinion to evolve stochastically in the face of uncertainty within a Bayesian framework. Comparatively few papers have considered uncertainty in this way (see \cite{jadbabaie2012non, fang2019social,fang2020opinionBayes} for some of these papers). Furthermore, we allow for the reality of antagonistic interactions, where consumers develop perceptions of media bias in defiant reaction to the public comments of other consumers with the opposite ideological stance. Unlike the reinforcement of opinions by interaction with political allies, which has been studied extensively \cite{degroot1974reaching, deffuant2000mixing, hegselmann2002opinion}, the role of antagonistic interactions with political opponents is understood less clearly \cite{shi2016evolution}.

In this paper, we map the real-world problem of inferring media bias onto the idealized, emblematic problem of inferring the bias in a coin. Agents in a network observe multiple tosses of the coin and progressively refine their estimate of the bias from the results of the tosses and the opinions of their networked allies and opponents, just as media consumers read multiple articles in (say) a newspaper and progressively refine their estimate of the newspaper's political bias. A new Bayesian update rule is proposed, building on previous literature, which encodes the antagonistic interactions between political opponents by discouraging opinion overlap. The generalized model has two advantages. (i) It acknowledges the reality that opinions are often multi-faceted, e.g. a person can have equal confidence in two contradictory views, or unequal confidence in multiple states. Bayes's theorem accommodates such multi-modal probability distributions, if the data support them. (ii) It captures the subtle frustrated dynamics, which occur in a network containing both allies and opponents, which often prevent evolution towards a steady state. This occurs, for example, when an individual strives to align their opinion with two political allies, who are themselves opposed.

The paper is structured as follows. In Section \ref{sec:model_theory}, the opinion dynamics model, which describes the opinion of an agent as a probability distribution and captures antagonistic interactions, is introduced. In Section \ref{sec:dynamics_algorithm}, a discrete-time automaton that numerically simulates the model is discussed. In Section \ref{sec:smaller_networks}, numerical simulations are performed in networks with two or three agents as a validation exercise, to compare against previous published work and establish baseline behaviors, which can then be used to interpret the dynamics of larger networks. In Section \ref{sec:larger_networks}, numerical simulations are performed on larger, scale-free networks. In Section \ref{sec:balance}, we interpret the asymptotic learning of larger, scale-free networks within the framework of structural balance theory. In Section \ref{sec:other_networks}, we investigate the influence of a network's topology on the dynamics of opinion formation. In Section \ref{sec:discussion}, we provide an introductory discussion of the social implications of the model.

\section{Inferring the bias of a coin: Bayes's rule in a network}
\label{sec:model_theory}

In the idealized model of media bias analyzed in this paper, a network of $n$ people strive to infer the hidden bias $\theta_0$ of a coin, where $0\leq \theta_0 \leq 1$ is the probability that the coin lands heads up after a single toss. Their beliefs concerning $\theta_0$ evolve in two ways. First, they observe a series of coin tosses and update their beliefs about $\theta_0$ independently via Bayes' theorem, as described in Section \ref{sec:external_signal}. Second, peer pressure molds their beliefs based on the beliefs of their allies and opponents in the network, as described in Sections \ref{sec:network} and \ref{sec:internal_signals}. The structure of the idealized model is compared with other models in the literature in Section \ref{sec:other_models}.

\subsection{External signal: observations of multiple coin tosses}
\label{sec:external_signal}

The outcome of each coin toss is a public external signal. It is witnessed by everybody in the network simultaneously and it is unfiltered; that is, everybody agrees on the outcome itself, even though it has different implications for individual beliefs in general. The coin tosses, for instance, could represent an editorial published daily in a newspaper. Using Bayes' rule, the beliefs of the $i$-th person in the network are updated based on the coin toss $S(t) \in \{\text{heads}, \text{tails}\}$ at time $t$:

\begin{equation}
\label{eq:bayes}
    x'_i(t+1/2,\theta) = \frac{P[S(t)|\theta]}{P[S(t)]}  x_i(t,\theta).
\end{equation}

\noindent In (\ref{eq:bayes}), $P[S(t)|\theta]$ is the likelihood, $x_i(t,\theta)$ is the prior probability distribution at time $t$, $x'_i(t+1/2,\theta)$ is the posterior probability distribution and $P[S(t)]=\sum_\theta P[S(t)|\theta] x_i(t,\theta)$ is the normalizing constant. The likelihood encodes the probability of observing a heads, given $\theta$. Because there are only two possible outcomes in a coin toss, the likelihood takes the following explicit form:

\begin{equation}
\label{eq:actual_bayes}
    P[S(t)|\theta] =
    \begin{dcases}
        \theta&, \quad \text{if } S(t)  \text{ is heads} \\
        1-\theta&, \quad \text{if } S(t)  \text{ is tails.}
    \end{dcases}
\end{equation}

Note that the first argument of $x_i'$ in (\ref{eq:bayes}) is written as $t+1/2$ rather than $t+1$, in order to indicate that (\ref{eq:bayes}) is the first half of a two-step update rule, the second half of which involves network interactions and is described in Section \ref{sec:network}. The full time-step from $t$ to $t+1$ is complete, when both halves of the update rule are executed. Iterating (\ref{eq:bayes}) and (\ref{eq:actual_bayes}) without network interactions would generate the standard binomial distribution.

\subsection{Network of allies and opponents}
\label{sec:network}

The political relationships between $n$ people in a network can be described by the adjacency matrix $A_{ij}$, an $n\times n$ matrix where $A_{ij} > 0$ indicates that persons $i$ and $j$ are allies, $A_{ij} < 0$ indicates that they are opponents and $A_{ij} = 0$ indicates that they do not know each other. Note that $A_{ij} = 0$ does not necessarily mean that persons $i$ and $j$ are disconnected --- they could be connected indirectly via a mutual ally or opponent. In the smaller networks considered in Section \ref{sec:smaller_networks}, we have $A_{ij} \neq 0$ for all $i$ and $j$. In contrast, we do have $A_{ij}=0$ for some $i$ and $j$ in the larger networks studied in Section \ref{sec:larger_networks}, but every agent is connected to someone in the network by construction. Two assumptions are made about $A_{ij}$:

\begin{enumerate}
    \item The adjacency matrix is symmetric, with $A_{ij}=A_{ji}$. Physically, this occurs because political relationships are mutual; if Alice leans the same way as Bob, then perforce Bob leans the same way as Alice. 
    \item $A_{ij}$ can only take three possible values: $-1$, 1 or 0. These three options are sufficient to study the main network topologies, especially those with internal tensions (e.g. $A_{12}=1$, $A_{13}=1$, and $A_{23}=-1$). Intermediate weights with $|A_{ij}|<1$ (e.g. $A_{ij}=1/2$) will be studied in future work. Intermediate weights are covariant with the functional form of the update rule, which is not unique (see Section \ref{sec:internal_signals}).
\end{enumerate}

The network topologies that are considered in this paper are scale-free networks that follow a power-law degree distribution, where the degree of an agent is the number of allies and opponents of the agent. Many real-world networks, such as web links on the Internet and citations in academic papers \cite{clauset2009power}, are thought to be approximately scale-free \cite{barabasi2003scale}. In this paper, the \barabasi model \cite{barabasi1999emergence} is used to generate random scale-free networks. The \barabasi model has a tunable parameter that varies the ratio of links with $A_{ij}=0$ to $A_{ij}\neq 0$ and guarantees that the generated network does not contain isolated agents or communities of agents. We also examine \erdos and square lattice networks briefly for completeness.

\subsection{Internal signals: influence of allegiances}
\label{sec:internal_signals}

The beliefs of each person in the network are shaped by the beliefs of their allies and opponents through a form of ``peer pressure''. There are many valid ways to capture these internal influences within a Bayesian framework. For example, one can form a weighted combination of the beliefs of others and treat the result as a prior, which is then updated to give a posterior using the external data and Bayes's theorem as in Section \ref{sec:external_signal}. The latter approach will be investigated in forthcoming work. In this paper, we stick to a standard approach, which is favored in the literature, and update the posterior probability according to the rule

\begin{equation}
\label{eq:update}
    x_i(t+1, \theta) \propto \max \left[0, \; x'_i(t + 1/2,\theta) + \mu \Delta x'_i(t + 1/2, \theta) \right].
\end{equation}

\noindent In (\ref{eq:update}), the proportionality constant is set via the condition $\sum_\theta x_i(t+1,\theta) = 1$. Here $0.0<\mu\leq 0.5$ (termed the learning rate) quantifies the susceptibility of an agent to its neighbors' beliefs and sets the time scale of the model. The displacement

\begin{equation}
\label{eq:update_diff}
    \Delta x'_i(t+1/2, \theta) = 
        a_i^{-1} \sum_{j \neq i} A_{ij} \left[ x'_j(t+1/2,\theta) - x'_i(t+1/2,\theta) \right]
\end{equation}

\noindent with

\begin{equation}
\label{eq:a_i_def}
    a_i = \sum_{j \neq i} \left| A_{ij} \right|,
\end{equation}

\noindent quantifies the average difference in belief between agent $i$ and its allies and opponents. Note that we are only interested in connected networks, i.e. there are no isolated agents or communities of agents, which guarantees $a_i > 0$ and that zero division in (\ref{eq:update_diff}) do not occur. As we have $A_{ij} = 1$ for allies and $A_{ij}=-1$ for opponents, equation (\ref{eq:update_diff}) moves the opinion of agent $i$ towards allies and away from opponents. Left untouched, the antagonistic interactions can yield $x_i(t+1,\theta) < 0$ via (\ref{eq:update_diff}) and (\ref{eq:a_i_def}), which is meaningless for a probability density. The maximisation operation in (\ref{eq:update}) is introduced to avoid this outcome. Note that the first argument of $x'_i$ and $\Delta x'_i$ in (\ref{eq:update_diff}) is written as $t+1/2$ to emphasize that (\ref{eq:update_diff}) is the second half of the two-step update rule. Applying equation (\ref{eq:bayes}) to every agent simultaneously, then applying equation (\ref{eq:update}) to every agent simultaneously completes the full time-step from $t$ to $t+1$.

The condition $0.0< \mu \leq 0.5$ prevents the beliefs from overshooting. The same condition is imposed in the popular Deffuant-Weisbuch model \cite{deffuant2000mixing}. Consider a trivial network with two agents and $A_{12}=1$. Suppose we start with $x'_1(t+1/2,\theta) > x'_2(t+1/2,\theta)$ for some $\theta$. For $\mu > 0.5$, we obtain $x_1(t+1,\theta) < x_2(t+1,\theta)$ after the interaction. Physically, $\mu>0.5$ implies that, if agent 1 is more confident in $\theta$ than agent 2 before they interact, then agent 1 becomes less confident than agent 2 after interacting, even though the two are allies, which is unrealistic. We also average the difference of beliefs, rather than the beliefs themselves, in order to properly account for antagonistic interactions when setting $A_{ij}=-1$. Suppose instead, just to make the point, we average the beliefs in equation (\ref{eq:update_diff}) in a trivial network with $n=2$ and $A_{12}=-1$. Then, equation (\ref{eq:update_diff}) always yields $\Delta x'_i(t+1/2, \theta) \leq 0$ and causes the beliefs of agents 1 and 2 to tend towards zero, which is unrealistic. 

\subsection{Relation to other models}
\label{sec:other_models}

The model in Sections \ref{sec:external_signal}--\ref{sec:internal_signals} draws inspiration from related models in the literature. Notably, equations (\ref{eq:update}) and (\ref{eq:update_diff}) are combinations of the popular DeGroot \cite{degroot1974reaching} and Deffuant-Weisbuch \cite{deffuant2000mixing} models. In the Deffuant-Weisbuch model, the opinions of the agents (which are not probabilistic) are updated in a pairwise manner. At each time step, two random agents $i$ and $j$ are picked and their scalar opinions $x_{i,j}(t)$ are updated as follows:

\begin{align}
    x_i (t+1) &= x_i(t) + \mu A_{ij} \left[ x_j(t) - x_i(t) \right] \label{eq:DW1},\\
    x_j (t+1) &= x_j(t) + \mu A_{ij} \left[ x_i(t) - x_j(t) \right] \label{eq:DW2}.
\end{align}

\noindent Equation (\ref{eq:update})  resembles equations (\ref{eq:DW1}) and (\ref{eq:DW2}) because the adjustment term is proportional to the difference in opinions. In order to extend equations (\ref{eq:DW1}) and (\ref{eq:DW2}) to simultaneously recognize the influence of every agent in the network, we look to the DeGroot model, where one averages the opinions of an agent's allies. Analogously, we average the difference in opinions across every ally and opponent in this paper, according to equation (\ref{eq:update_diff}). Additionally, the bounds on the learning rate, $0.0<\mu\leq0.5$, are taken directly from the Deffuant-Weisbuch model.

The update rule in the Deffuant-Weisbuch model also updates the adjacency matrix $A_{ij}$. The agents in these two models only interact with other agents if the difference in their opinions falls within a tunable threshold, $\delta$. Specifically, one has

\begin{equation}
\label{eq:DW_adj}
    A_{ij}(t) = 
    \begin{dcases}
        1, \quad \text{if } |x_j(t) - x_i(t)| < \delta\\
        0, \quad \text{otherwise}
    \end{dcases}
\end{equation}

\noindent This differs from Sections \ref{sec:external_signal}--\ref{sec:internal_signals}, where $A_{ij}$ is static, i.e. $A_{ij}$ does not vary with $t$. Note that the dynamics of $A_{ij}$ described in equation (\ref{eq:DW_adj}) are endogenous, i.e. $A_{ij}$ varies due to the internal dynamics of the network. They are not exogenous, or caused by external influences. We will consider the evolution of $A_{ij}$ in the context of the media bias application in future work.

There are relatively few opinion dynamics models that describe the opinion of an agent as a probability distribution. The few that do resemble aspects of the model in Section \ref{sec:external_signal}--\ref{sec:internal_signals} \cite{jadbabaie2012non, fang2019social, fang2020opinionBayes}. In Fang et al.'s model \cite{fang2019social}, which extends \cite{jadbabaie2012non} by considering the possibility that the same external signal may not be interpreted equally by all agents, an external signal is included exactly through equation (\ref{eq:bayes}). The mixing of the beliefs of an agent with its neighbors is a weighted average, similar to equations (\ref{eq:update}) and (\ref{eq:update_diff}). However,  there are some notable differences. The differences are specified in detail in \ref{apendix:fang}.

Extensions to the Deffuant-Weisbuch model \cite{carletti2006propaganda, quattrociocchi2011opinions, pineda2015} have been proposed to study specifically the influence of media organizations on the opinions of networked agents. In these extensions, the media have been modeled as a special agent whose opinion never changes. This contrasts the role contemplated in Section \ref{sec:external_signal}, where the media are modeled as a random, fluctuating external signal which embodies a hidden bias $\theta_0$. The DeGroot and Deffuant-Weisbuch models originally do not incorporate antagonistic interactions, but there exist extensions that do \cite{shi2016evolution, martins2010mass, chen2019opinion, he2019discrete}. The extensions focus on the convergence of opinions when antagonistic interactions are present, but other outcomes (including turbulent nonconvergence, also reported in \cite{shi2016evolution}) are possible, as we show in Section \ref{sec:triad_unbalanced}.

\section{Probabilistic discrete-time automaton}
\label{sec:dynamics_algorithm}

In this section we convert the model in Sections \ref{sec:external_signal}--\ref{sec:internal_signals} into an automaton, which is iterated $T$ times in response to $T$ consecutive occurrences of the external signal, i.e. $T$ coin tosses. The automaton is probabilistic, in the sense that the external signal is a random variable, and the beliefs of every agent are described by a probability distribution. However, the update rules at every time-step are deterministic in response to the external signal and peer pressure from other agents of the network. 

The automaton operates as follows:

\begin{enumerate}
    \item Generate or specify the network topology $A_{ij}$, the initial beliefs $x_1(t=0,\theta),...,x_n(t=0,\theta)$, the learning parameter $\mu$ and the true bias of the coin $\theta_0$.
    
    \item Simulate an independent coin toss and update the opinions of each agent according to Bayes' theorem, given in equations (\ref{eq:bayes}) and (\ref{eq:actual_bayes}).
    
    \item Mix the opinions of each agent with their respective neighbors using equations (\ref{eq:update}) and (\ref{eq:update_diff}).
    
    \item Repeat steps 2 and 3 until $t=T$.
\end{enumerate}

The long-run behavior of the automaton is a central concern of this paper and the media bias application more generally. Several scenarios exist. First, the system may settle to an equilibrium, in which the beliefs of every agent do not change, i.e. $x_i(t,\theta)$ is independent of $t$ for $t\rightarrow \infty$. If the belief of an agent does not change, the agent is said to achieve asymptotic learning. If every agent achieves asymptotic learning, the system is said to achieve asymptotic learning. In practice, we say that agent $i$ achieves asymptotic learning, once we obtain $\max_\theta | x_i(t+\tau,\theta)-x_i(t,\theta) | < 0.01 \max_\theta x_i(t, \theta)$ for all $1\leq \tau \leq 99$ and $0\leq \theta \leq 1$.\footnote{We prefer the error condition, $\max_\theta | x_i(t+\tau,\theta)-x_i(t,\theta) | <  0.01 \max_\theta  x_i(t, \theta)$, over an alternative such as $\max_\theta | x_i(t+\tau,\theta)-x_i(t,\theta) | < 0.01 x_i(t, \theta) $ because $|x_i(t,\theta)|$ tends to zero rapidly for some $\theta$, and the right-hand side of the condition becomes prone to numerical error.} The earliest time when the system achieves asymptotic learning is labelled $t_{\rm a}$. Another scenario is when the network does not settle during the (finite) duration of the automaton --- and indeed shows no signs of doing so, if the automaton runs longer. This ``turbulent''\footnote{We avoid terming this outcome `chaotic', as `chaotic' is reserved mathematically for dynamical systems with sensitive dependence on initial conditions. Testing systematically for the latter property falls outside the scope of this paper.} outcome is said to be achieved if at least one agent fails to achieve asymptotic learning within $t=T$ timesteps.

\begin{figure}[h!t]
    \centering
    \includegraphics[width=0.5\textwidth]{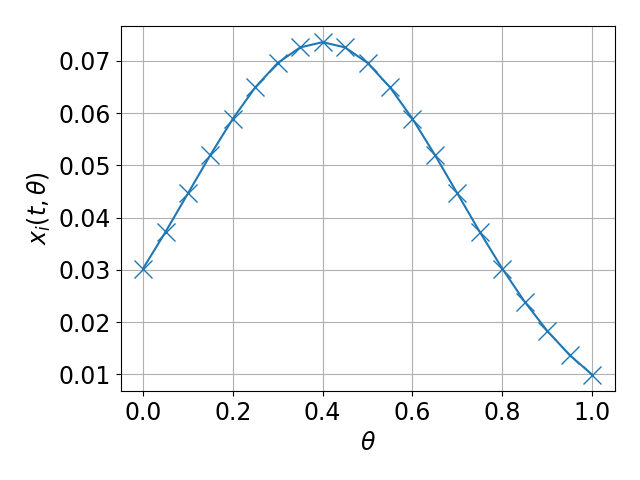}
    \caption{A discretized prior generated by sampling a Gaussian with mean 0.4 and standard deviation of 0.3 before truncation to the domain $0 \leq \theta \leq 1$. The Gaussian is sampled at $\theta = \{0.00, 0.05, ..., 1.00\}$, as indicated by the crosses. The mean of the discretized prior, $\mean = \sum_\theta \theta x_i(t,\theta)$, is larger than 0.4 due to sampling only in the range $0 \leq \theta \leq 1$. This truncated, discretized Gaussian is used as agent 1's prior in Figures \ref{fig:pair_ally} and \ref{fig:pair_opponents}.}
    \label{fig:gaussian}
\end{figure}

The continuous variable $\theta$ is discretized into a grid of 21 regularly-spaced values, i.e. $\theta = \{0.00, 0.05, ..., 1.00\}$. The grid is chosen partly to manage computational cost and partly because there is a limit to the granularity of human opinions in reality. Unless stated otherwise, for simplicity, we assume $\theta_0=0.6$, $\mu=0.25$ and $T=10^4$ time steps. Each initial prior, $x_i(t=0,\theta)$, is independently generated by sampling a Gaussian with mean and standard deviation in the ranges $[0.0, 1.0]$ and $[0.1, 0.4]$ respectively. We do not start with uniform priors for all agents because the beliefs of interacting agents must differ in order for the agents to change each other's beliefs, as seen in equations (\ref{eq:update}) and (\ref{eq:update_diff}). Figure \ref{fig:gaussian} illustrates how the priors are generated from a Gaussian template through discretization and truncation to the domain $0 \leq \theta \leq 1$.

\section{Validation: building blocks with $n=2$, $3$}
\label{sec:smaller_networks}

Before studying networks of realistic sizes, we validate the automaton by running it on networks with $n=2$ and $n=3$. We consider a pair of allies in Section \ref{sec:pair_ally}, a pair of opponents in Section \ref{sec:pair_opponents}, and an unbalanced triad in Section \ref{sec:triad_unbalanced}. The exercise has three aims. The first aim is to compare the results against analogous models in the literature, where the beliefs are not probabilistic \cite{martins2010mass, shi2016evolution}, in order to verify that the automaton behaves reasonably. The second aim is to develop intuition through simple cases about what changes, when deterministic beliefs are supplanted by probabilistic beliefs. The third aim is to identify key behaviors in small networks, because small subnetworks are the building blocks for larger networks like those studied in Sections \ref{sec:larger_networks} and \ref{sec:balance}.

\subsection{Pair of allies: consensus}
\label{sec:pair_ally}

We begin by looking at a network with $n=2$ and $A_{12}=1$, which represents a pair of political allies. The initial beliefs of agents 1 and 2 are sampled from Gaussian distributions with means of 0.4 and 0.8 respectively, and standard deviations of 0.3.

\begin{figure}[h!t]
    \centering
    \begin{subfigure}[b]{0.49\linewidth}
        \includegraphics[width=\linewidth]{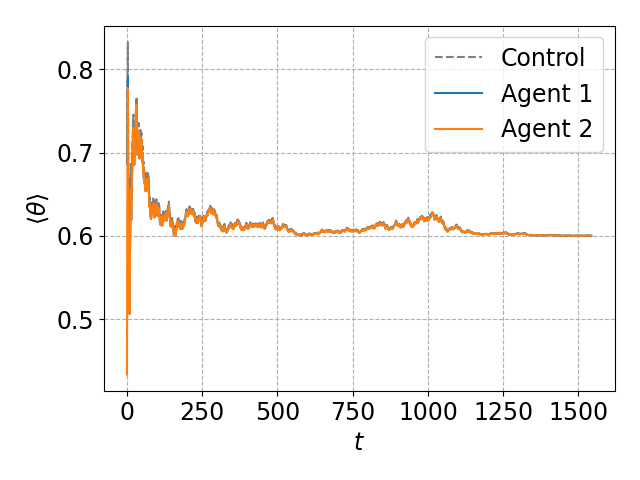}
        \caption{}
        \label{subfig:pair_ally_mean}
    \end{subfigure}
    \begin{subfigure}[b]{0.49\linewidth}
        \includegraphics[width=\linewidth]{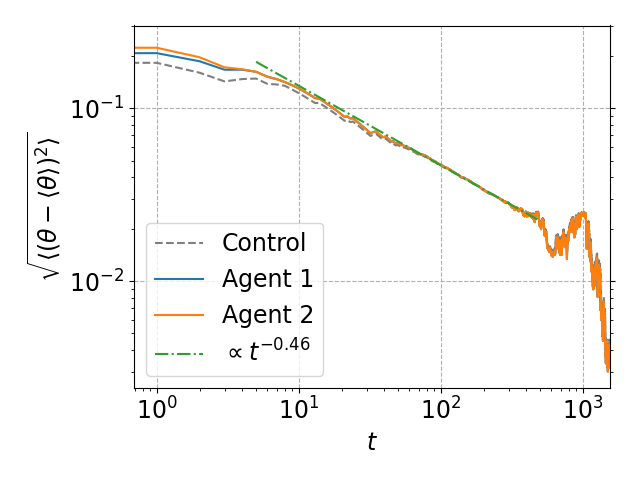}
        \caption{}
        \label{subfig:pair_ally_std}
    \end{subfigure}
    \caption{Evolution of the (a) mean and (b) standard deviation of the beliefs of a pair of allies (solid curves). Only $ t \leq t_{\rm a}=1543$ time steps are shown. As a control, we also evolve the beliefs of agent 1 independently, such that it ignores the internal signal of its ally (dashed curve). For reference, a power law with exponent $\approx -0.46$ is plotted in (b) (dashed-dotted curve). The means and standard deviations of the control and both agents are sufficiently similar to be indistinguishable by eye. Parameters: $\theta_0=0.6$, $\mu=0.25$, $A_{12}=1$, $T=10^4$.}
    \label{fig:pair_ally}
\end{figure}

Figure \ref{fig:pair_ally} shows the evolution of the mean and standard deviation of the beliefs of agents 1 and 2 (denoted $\langle \theta \rangle_{1,2}$ and $\stddevSub{1,2}$ respectively) in a particular simulation. As a control, we also show the evolution of $\langle \theta \rangle_{1}$ and $\stddevSub{1}$, when agent 1 ignores the beliefs of agent 2. At $t=t_{\rm a}=1543$, the beliefs of both agents settle on $\theta=\theta_0$, with $x_1(t=t_{\rm a}, \theta=\theta_0)=x_2(t=t_{\rm a}, \theta=\theta_0) = 1$.\footnote{In networks with allies only, $x_{1,2}(t,\theta)$ does not equal unity exactly for any $\theta$, unless we start with $x_1(t=0,\theta=\theta_1)=x_2(t=0,\theta=\theta_1)=1$ for some $0<\theta_1<1$, but machine precision rounds $x_{1,2}(t,\theta=\theta_0)$ to unity.} At every other $\theta$, we have $x_{1,2}(t=t_{\rm a}, \theta \neq \theta_0) \ll 1$. We say that the beliefs of agent $i$ and $j$ converge if they satisfy

\begin{equation}
    \label{eq:belief_convergence}
    \max_\theta |x_i(t,\theta)-x_j(t,\theta)| < \epsilon \max_\theta [x_i(t,\theta), x_j(t,\theta)].
\end{equation}

\noindent In (\ref{eq:belief_convergence}), $\epsilon$ is a user-selected tolerance. Agents 1 and 2 converge onto each other after $t=9$ time steps with $\epsilon=10^{-3}$. Because the strength of the internal signals is proportional to the difference in $x_i(t, \theta)$ as seen in equation (\ref{eq:update}), the interaction between agents 1 and 2 has negligible influence after $\sim 10$ time steps. By running $10^4$ independent simulations with randomized coin tosses and priors as described in Section \ref{sec:dynamics_algorithm}, we find that the pair of allies always infer the correct coin bias, assuming that at least one of them starts with $x_i(t=0, \theta=\theta_0) \neq 0$. The consensus of allies is a common result in many previous works, such as in \cite{degroot1974reaching, deffuant2000mixing, hegselmann2002opinion}. The convergence of allies in assessing media bias in the real world is also expected \cite{carletti2006propaganda, martins2010mass, pineda2015, fang2020opinionBayes}. Thus, Figure \ref{fig:pair_ally} shows that the model described in Sections \ref{sec:external_signal}--\ref{sec:internal_signals} produces sensible results with a pair of allies.

The evolution of the mean beliefs in Figure \ref{subfig:pair_ally_mean} is identical to the evolution of the beliefs of allies in deterministic models with mass media \cite{carletti2006propaganda, martins2010mass, pineda2015} --- the opinions of allies grow closer as they interact. However, the probabilistic treatment in this paper provides additional information about the confidence of the agents in their beliefs, via the standard deviation plotted in Figure \ref{subfig:pair_ally_std}. As the change in beliefs of two interacting allies is proportional to the difference in their beliefs, as seen in equation (\ref{eq:update}), internal signals do not affect $\stddevSub{1,2}$, once the beliefs of agents 1 and 2 converge. At $t>9$, the decrease of $\stddevSub{1,2}$ is entirely driven by the external coin tosses. Hence, the bump in $\stddevSub{1,2}$ at $t \approx 1000$ can be attributed to the coin returning more heads than expected by chance. Intuitively, once the beliefs of the two allies converge, they no longer learn any new information off each other. As they observe more coin tosses, they become increasingly confident in $\theta=\theta_0$. A least squares fit to $\stddevSub{1,2}$ for $50<t<500$ returns the approximate scaling $t^{-0.46}$ for the first 500 time steps.

\subsection{Pair of opponents: reaching the wrong conclusion first}
\label{sec:pair_opponents}

We now investigate the behavior of a pair of opponents, with $n=2$ and $A_{12}=-1$. We repeat the simulation described in Section \ref{sec:pair_ally}, with the same priors and an identical sequence of coin tosses as before.

\begin{figure}[h!t]
    \centering
    \begin{subfigure}[b]{0.49\linewidth}
        \includegraphics[width=\linewidth]{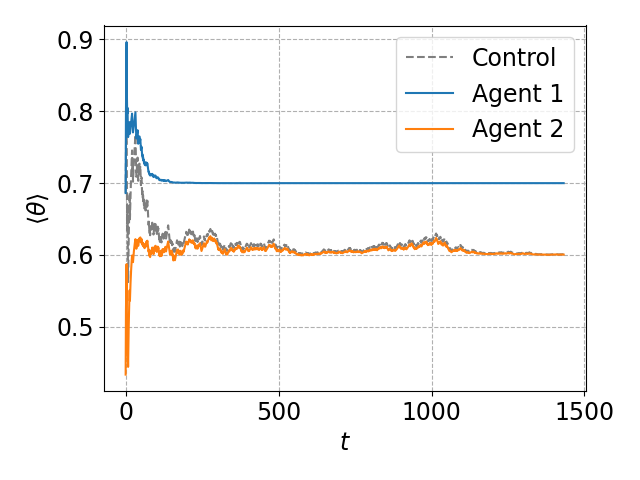}
        \caption{}
    \end{subfigure}
    \begin{subfigure}[b]{0.49\linewidth}
        \includegraphics[width=\linewidth]{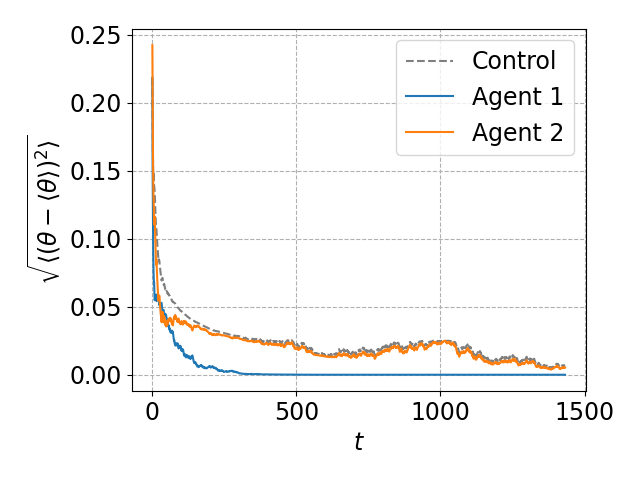}
        \caption{}
    \end{subfigure}
    \caption{Evolution of the (a) mean and (b) standard deviation of the beliefs of a pair of opponents (solid curves). Only $t \leq t_{\rm a}=1432$ is shown. Agents 1 and 2 achieve asymptotic learning at $t_{{\rm a}, 1}=256$ and $t_{{\rm a}, 2}=1432$ respectively. The control (dashed curve) is defined as in Figure \ref{fig:pair_ally}. Parameters: $\theta_0=0.6$, $\mu=0.25$, $A_{12}=-1, T=10^4$.}
    \label{fig:pair_opponents}
\end{figure}

Figure \ref{fig:pair_opponents} shows the evolution of the mean and standard deviation of the beliefs of the two opponents. In this particular simulation, the two opponents disagree, which is a common result in previous works \cite{shi2016evolution, martins2010mass, chen2019opinion, he2019discrete}. One agent correctly infers $\theta$ (with $\langle \theta \rangle \rightarrow 0.6 = \theta_0$), while the other infers $\theta$ incorrectly (with $\langle \theta \rangle \rightarrow 0.7 \neq \theta_0$). An analogous outcome is observed in a deterministic model \cite{martins2010mass}, where opponents may cluster around, but not exactly on, the opinion promoted by the media. The cluster spreads over an increasing range of $\theta$, as the percentage of antagonistic links in the network increases. Note that this spread is not quantified in \cite{martins2010mass}, so quantitative comparisons cannot be made. Intuitively, agents trust the external signals of the media, but they distrust the beliefs of their opponents. The media ``pulls'' the opinions of the agents together, but antagonistic relations ``push'' the opinions apart, creating a cluster around the true media bias.  The shape of the beliefs of both agents in Figure \ref{fig:pair_opponents}, when asymptotic learning is reached, is similar to that of the two allies in Section \ref{sec:pair_ally}; the beliefs peak at the agents' respective $\langle \theta \rangle $, with $x_{i}(t=t_{\rm a}, \theta = \langle \theta \rangle _{i}) = 1$ and $x_{i}(t=t_{\rm a}, \theta \neq \langle \theta \rangle _{i}) \ll 1$ for $i=1,2$.

Figure \ref{fig:pair_opponents} reveals a peculiar behavior: agent 1 gains confidence in the wrong conclusion (at $t_{{\rm a}, 1}=256$) quicker than agent 2 does in the right conclusion (at $t_{{\rm a}, 2}=1432$), as $\stddevSub{1}$ drops quicker than $\stddevSub{2}$. Neither standard deviation can be fitted accurately with a power law of the form $\stddevSub{1,2} \propto t^\beta$, unlike in Figure 1b. After achieving asymptotic learning, $\stddevSub{1}$ plateaus at zero. The evolution of the control reveals that agent 1 would have inferred the correct coin bias if it ignored the beliefs of its opponent. Agent 2 matches the control for $t \gtrsim 500$; their beliefs converge at $t=556$ time steps with $\epsilon=0.05$ as defined in equation (\ref{eq:belief_convergence}). Hence, we can say that the evolution of $\stddevSub{2}$ at $t \gtrsim 500$ is dominated by the external coin tosses. 

This behavior, where the wrong conclusion is reached first, can be explained as follows. Suppose by chance the coin initially returns more heads than a $\theta_0=0.6$ coin is expected to. The beliefs of the two opponents are driven towards $\theta>\theta_0$. Because agent 1 starts with an initial belief centered at a higher $\theta$, it is now more confident than agent 2 concerning $\theta>\theta_0$, viz. $x_1(t,\theta) > x_2(t,\theta)$ for $\theta>\theta_0$, but less confident for $\theta \leq \theta_0$. In Figure \ref{fig:pair_opponents}, the initial run of heads is sufficient to cause $x_1(t,\theta) > x_2(t,\theta)$ for $\theta > 0.65$ and $x_1(t,\theta) < x_2(t,\theta)$ otherwise. At this time ($t=7$ for this simulation), antagonistic interactions in equations (\ref{eq:update}) and (\ref{eq:update_diff}) cause agent 1 to be absolutely sure that $\theta_0$ cannot be $\theta \leq 0.65$, i.e. $x_1(t, \theta) = 0$ for $\theta \leq 0.65$. Likewise, we have $x_2(t, \theta) = 0$ for $\theta > 0.65$. The beliefs of agent 1 and 2 no longer have any overlap, i.e. $x_1(t, \theta) x_2(t, \theta) = 0$ for all $t \geq 7$ and $\theta$, so any further internal signals have no effect. 

The beliefs of both agents now evolve purely due to the coin tosses, which generates a binomial distribution centered at $\theta_0$ as the likelihood of the external signal. However, neither agent sees the entire likelihood due to part of their beliefs being zeroed out by their opponent. Agent 1 only sees the $\theta \geq 0.70$ tail of the likelihood and agent 2 sees the whole likelihood except the $\theta \geq 0.70$ tail. Because the likelihood is exponentially suppressed as $\theta$ deviates from $\theta_0$, and the beliefs of the agents are always normalized, $x_1(t,\theta>0.70)$ gets suppressed quicker than $x_2(t, \theta \neq \theta_0)$. This causes agent 1 to achieve asymptotic learning toward the wrong $\theta_0$ quicker than its opponent. Said differently, the initial run of heads causes agent 2 to ``lock'' agent 1 out of $\theta<0.70$. As the coin tosses makes it increasingly clear that the true bias satisfies $\theta_0 < 0.70$, agent 1 quickly settles on its best option, $\theta=0.70$, as any other choice becomes increasingly unlikely.

\begin{figure}[h!t]
    \centering
    \begin{subfigure}[b]{0.32\linewidth}
        \includegraphics[width=\linewidth]{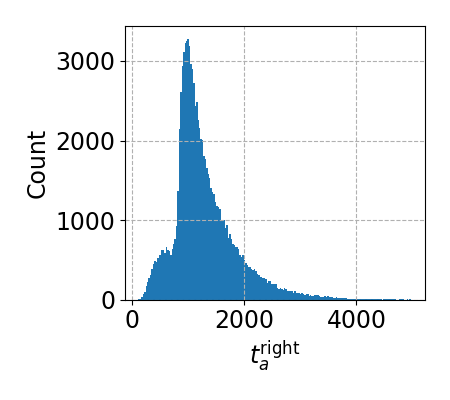}
        \caption{}
    \end{subfigure}
    \begin{subfigure}[b]{0.32\linewidth}
        \includegraphics[width=\linewidth]{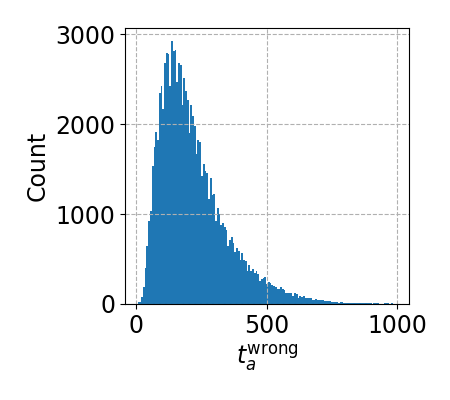}
        \caption{}
    \end{subfigure}
    \begin{subfigure}[b]{0.32\linewidth}
        \includegraphics[width=\linewidth]{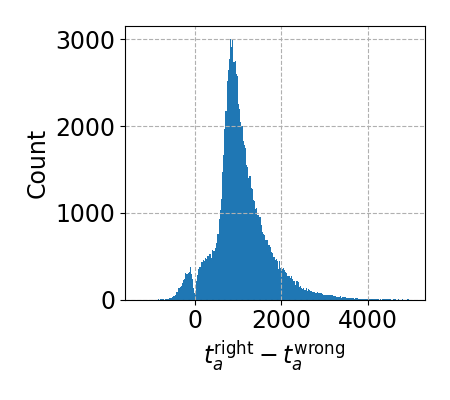}
        \caption{}
    \end{subfigure}
    \caption{Reaching the wrong conclusion first: histogram of (a) $t_a^\text{right}$, (b) $t_a^\text{wrong}$ and (c) $t_a^\text{right}-t_a^\text{wrong}$ for $10^5$ independent simulations with a pair of opponents. 787 simulations are excluded as both agents reached the wrong conclusion. Parameters: $\theta_0=0.6$, $\mu=0.25$, $A_{12}=-1, T=10^4$.}
    \label{fig:pair_wrong_hist}
\end{figure}

We now ask: how often is the wrong conclusion reached first? As before, consider a network with $n=2$ and $A_{12}=-1$. Let $t_{\rm a}^{\rm right}$ and $t_{\rm a}^{\rm wrong}$ denote the times when asymptotic learning is achieved (in the sense defined in Section \ref{sec:dynamics_algorithm}) by the agents that settle on the right $(\theta=\theta_0)$ and wrong $(\theta \neq \theta_0)$ coin bias respectively. Upon randomizing the priors and executing $T=10^4$ tosses per simulation, an ensemble of $10^5$ independent simulations is performed. Figure \ref{fig:pair_wrong_hist} and Table \ref{tab:pair_wrong_conclusion} shows the histogram and summary statistics respectively of $t_{\rm a}^{\rm right}$, $t_{\rm a}^{\rm wrong}$ and $t_a^\text{right}-t_a^\text{wrong}$ for these $10^5$ simulations. In 787 of these simulations, both agents infer the wrong $\theta_0$ and are excluded from summary statistics in Table \ref{tab:pair_wrong_conclusion}. In each of these 787 realizations, the coin tosses favor $\theta \neq \theta_0$ by chance during the run. Additionally, there are no instances of both agents agreeing with each other.

\begin{table}[h!t]
\centering
\begin{tabular}{@{}lrrr@{}}
\toprule
Property of $t_{\rm a}$ & $t_{\rm a}^{\rm right}$ & $t_{\rm a}^{\rm wrong}$ & $t_{\rm a}^{\rm right}-t_{\rm a}^{\rm wrong}$    \\ \midrule
Positive &  99213 & 99213  & 95370 \\
Negative & 0 & 0    & 3842  \\
Zero  & 0 & 0   & 1    \\
Mean (time steps) & 1300 & 222   & 1078   \\
Standard deviation (time steps) & 594 & 135 & 638  \\
First quartile (time steps) & 945 & 127  & 745  \\
Median  (time steps) & 1160 & 190   & 981  \\
Third quartile (time steps) & 1541 & 284  & 1351  \\ \midrule
Total & 99213 & 99213 & 99213 \\ \bottomrule
\end{tabular}
\caption{Summary statistics of $t_{\rm a}^{\rm right}$, $t_{\rm a}^{\rm wrong}$ and $t_a^\text{right}-t_a^\text{wrong}$ for $10^5$ independent simulations whose histograms appear in Figure \ref{fig:pair_wrong_hist}. 787 data points are excluded as both agents inferred the wrong $\theta_0$.}
\label{tab:pair_wrong_conclusion}
\end{table}

The distributions of $t_{\rm a}^{\rm right}$ and $t_{\rm a}^{\rm wrong}$ differ. This is clear from Figure \ref{fig:pair_wrong_hist} and Table \ref{tab:pair_wrong_conclusion}. The modes of $t_{\rm a}^{\rm right}$ and $t_{\rm a}^{\rm wrong}$ are at 1000 and 150 time steps respectively. There is a bump in the histogram of $t_{\rm a}^{\rm right}$ to the left of the modal value, whereas the histogram of $t_{\rm a}^{\rm wrong}$ does not have the bump. Agents settle on the wrong conclusion quicker than the right conclusion $\approx 96\%$ of the time. On average, the wrong conclusion is reached 1078 time steps quicker, with a standard error of $ \approx \pm 2$ time steps. The histogram of $t_a^\text{right}-t_a^\text{wrong}$ is skewed right and can be split into several regimes. The histogram rises at $t_a^\text{right}-t_a^\text{wrong} < -100$. In the region $|t_a^\text{right}-t_a^\text{wrong}| < 100$, there is a ``notch'' in the histogram, where it drops to a minimum at $t_a^\text{right} = t_a^\text{wrong}$, followed by a ``bump'' at $t_a^\text{right}-t_a^\text{wrong} = 300$. The histogram then peaks at $t_a^\text{right}-t_a^\text{wrong} = 860$ and decays to zero as $t_a^\text{right}-t_a^\text{wrong}$ increases.

\begin{figure}[h!tbp]
    \centering
    \begin{subfigure}[b]{0.45\linewidth}
        \includegraphics[width=\linewidth]{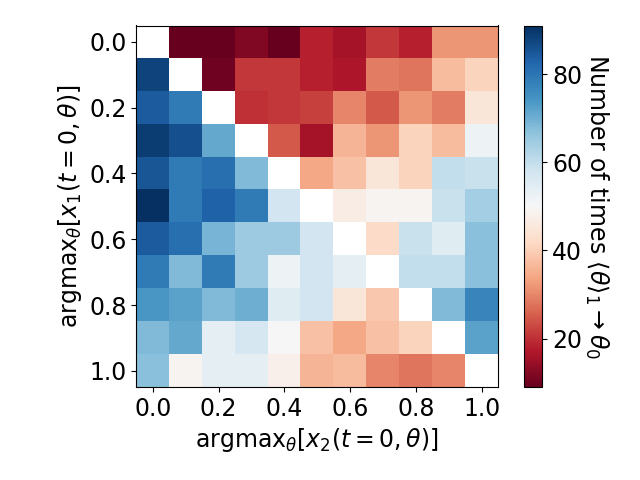}
        \caption{}
        \label{subfig:opponents_stats_prior}
    \end{subfigure}
    \begin{subfigure}[b]{0.45\linewidth}
        \includegraphics[width=\linewidth]{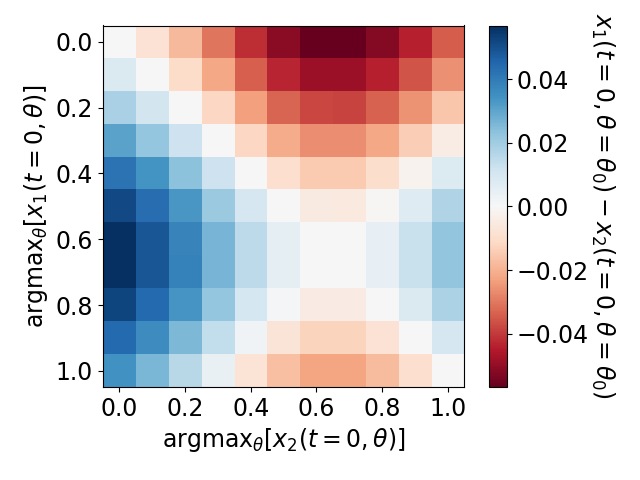}
        \caption{}
        \label{subfig:opponents_stats_prior_diff}
    \end{subfigure}
    \caption{The effects of the priors on what agent infers the correct bias. (a) A heat map of the number of times out of 100 that agent 1 infers the correct coin bias as a function of agent 1 and 2's priors. A red color in the heat map shows that agent 1 is less likely to infer the correct bias compared to agent 2, while the opposite is true for a blue color. (b) A heat map of the difference in priors at $\theta=\theta_0$. A red color shows that agent 1 is initially less confident than agent 2 at $\theta = \theta_0$, while the opposite is true for a blue color. The priors are generated by varying the mean of the pre-truncated Gaussian while keeping its standard deviation fixed. We exclude scenarios where the priors of both agents are exactly equal because the internal signals have no effect, which would lead to both agents always inferring the correct bias.}
    \label{fig:opponents_stats_prior}
\end{figure}

Another related question we may ask is: how much do the initial conditions influence what agent infers the correct bias? To answer this question, we vary the priors of agents 1 and 2 by varying the mean of the Gaussian distribution that they sample. Specifically, we fix the standard deviation of the Gaussian at 0.3 and vary the mean of the Gaussian in the range $[0.0, 1.0]$ for both agents. For each pair of priors, we perform 100 simulations with $A_{12}=-1$ and randomized coin tosses, and track how many times agent 1 infers the correct coin bias with $\mean \rightarrow \theta_0$. Figure \ref{subfig:opponents_stats_prior} shows a heat map of the number of times out of 100 that agent 1 infers the correct bias as a function of agent 1 and 2's priors. In each of these simulations, one agent always infers the correct bias while the other infers the wrong bias. For reference, we show a heat map of what agent is initially more confident in $\theta = \theta_0$ as a function of the agents' priors in Figure \ref{subfig:opponents_stats_prior_diff}. We note that there is a symmetry in these heat maps --- there is nothing special about the labels of ``1'' and ``2'' of the agents, so we expect the heat map shown in Figure \ref{subfig:opponents_stats_prior_diff} to be equal to the negative of its transpose.

Figures \ref{subfig:opponents_stats_prior} and \ref{subfig:opponents_stats_prior_diff} show an interesting connection between the agents' priors and who infers the correct bias: the agent who is initially more confident in $\theta_0$ is more likely to infer the correct bias, which is an intuitive and expected behavior. However, Figure \ref{fig:opponents_stats_prior} also shows that the choice of priors does not completely determine who infers the correct bias, if the priors are constrained to sample from Gaussians of a fixed standard deviation. No matter what the means of the Gaussians are, the coin tosses can cause either agent to infer the correct coin bias. Because the coin tosses always imply $\theta=\theta_0$ in the long run, the specific sequence of coin tosses in the short run must be important in determining who infers the correct bias. This has an interesting social implication: the bias of a media organization, as perceived by a pair of opponents, is more reliant on earlier published media products compared to those published later.

We now investigate the role played by the learning parameter, $\mu$, for a pair of opponents. Because $\mu$ quantifies how susceptible an agent is to the beliefs of its neighbor, one expects the converging influence of the coin tosses to overpower the diverging influence of the antagonism, when $\mu$ is sufficiently small. We vary $\mu$ in the range $1\times 10^{-4} \leq \mu \leq 0.5$ and perform independent simulations with randomized priors and coin tosses until we obtain $10^3$ simulations where the opponents disagree significantly for each value of $\mu$. The opponents are said to disagree significantly when we have $|\langle \theta \rangle_1 - \langle \theta \rangle_2| \geq 0.05$ at $t=t_{\rm a}$; the arbitrary tolerance $0.05$ corresponds to the grid resolution in $\theta$. We track $D$, the percentage of the simulations for each $\mu$ where the agents disagree significantly. $D$ gives a rough sense of how influential internal signals are relative to external signals. If the opponents ultimately agree, it is because the ``herding'' influence of the media overpowers the agents' antagonism. We also track $t_{\rm a}^{\rm right}-t_{\rm a}^{\rm wrong}$ when the opponents disagree significantly to check if $\mu$ has any effect on the tendency to reach the wrong conclusion.

\begin{figure}[h!t]
    \centering
    \begin{subfigure}[b]{0.49\linewidth}
        \includegraphics[width=\linewidth]{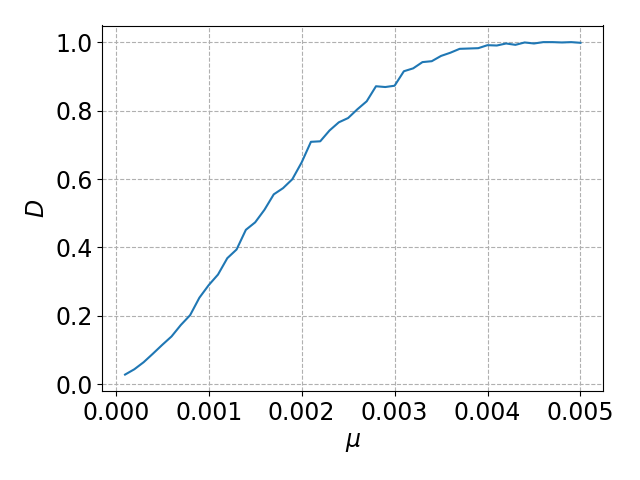}
        \caption{}
        \label{fig:pair_mu_disagree}
    \end{subfigure}
    \begin{subfigure}[b]{0.49\linewidth}
        \includegraphics[width=\linewidth]{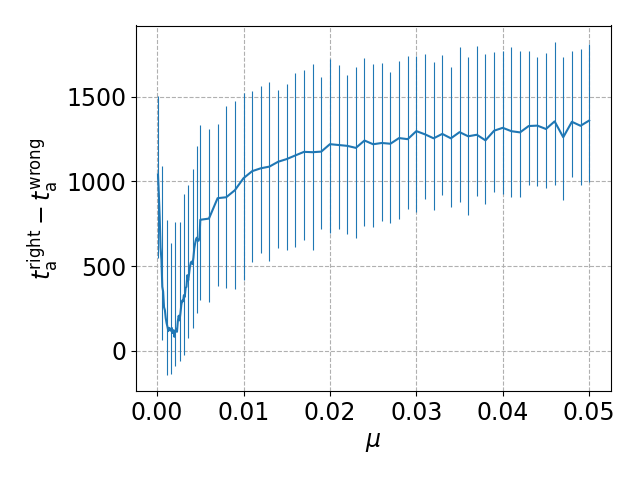}
        \caption{}
        \label{fig:pair_mu_t}
    \end{subfigure}
    \caption{The effects of the learning rate $\mu$ on a pair of opponents. (a) $D$, the percentage of simulations where the two opponents disagree on the coin bias, as a function of $\mu$. Only $10^{-4} < \mu < 5 \times 10^{-3}$ is shown, as $D$ for $\mu>5 \times 10^{-3}$ is a horizontal line at unity. (b) The medians of $t_{\rm a}^{\rm right}-t_{\rm a}^{\rm wrong}$ for $10^3$ simulations when opponents disagree. The error bars extend to the first and third quartiles.}
    \label{fig:pair_mu}
\end{figure}

Figure \ref{fig:pair_mu_disagree} shows a plot of $D$, the percentage of the simulations where the opponents disagree significantly, against $\mu$, while Figure \ref{fig:pair_mu_t} plots $t_{\rm a}^{\rm right}-t_{\rm a}^{\rm wrong}$ against $\mu$ when the opponents disagree significantly. $D$ plateaus at unity for $\mu \geq 0.004$, where the influence of the internal signal overwhelms the external signal. For $\mu < 0.004$, $D$ decreases as $\mu$ tends towards zero and the network effects weaken. We also observe a steady decrease of the medians $t_{\rm a}^{\rm right}-t_{\rm a}^{\rm wrong}$ as $\mu$ decreases in the range $2 \times 10 ^{-3} < \mu < 0.5$, where a minimum is reached at $\mu = 2 \times 10 ^{-3}$. At $\mu < 2 \times 10 ^{-3}$, $t_{\rm a}^{\rm right}-t_{\rm a}^{\rm wrong}$ rises towards low $\mu$. When the initial beliefs do not overlap much, one of the opponents can quickly ``lock'' the other out of $\theta_0$. For $\mu=10^{-4}$, this can happen over $\sim 10$ time steps, which is a similar order of magnitude as the $\mu=0.25$ case in Figure \ref{fig:pair_wrong_hist}. Once an opponent is locked out of $\theta_0$, they quickly achieve asymptotic learning toward the wrong conclusion, as discussed before. In contrast, for $\mu=2 \times 10 ^{-3}$, which has the lowest $t_{\rm a}^{\rm right}-t_{\rm a}^{\rm wrong}$, the opponent gets locked out of $\theta_0$ over $\sim 100$ time steps.

\subsection{Unbalanced triad: internal contradictions}
\label{sec:triad_unbalanced}

\begin{figure}[h!t]
    \centering
    \begin{subfigure}[b]{0.24\linewidth}
        \includegraphics[width=\linewidth]{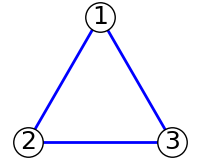}
        \caption*{$G_0$}
    \end{subfigure}
    \begin{subfigure}[b]{0.24\linewidth}
        \includegraphics[width=\linewidth]{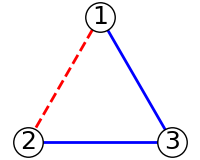}
        \caption*{$G_1$}
    \end{subfigure}
    \begin{subfigure}[b]{0.24\linewidth}
        \includegraphics[width=\linewidth]{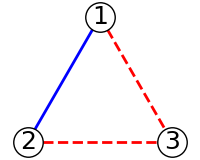}
        \caption*{$G_2$}
    \end{subfigure}
    \begin{subfigure}[b]{0.24\linewidth}
        \includegraphics[width=\linewidth]{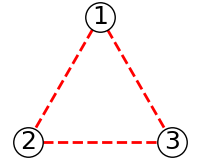}
        \caption*{$G_3$}
    \end{subfigure}
    \caption{The four possible triads of allies and opponents. The blue solid lines represent allies, while the red dashed lines represent opponents. $G_1$ is the triad with internal tensions.}
    \label{fig:types_of_triads}
\end{figure}

We now consider networks with $n=3$ agents. There are four unique triads that can be constructed with the ally and opponent relations depicted in Figure \ref{fig:types_of_triads}. We specifically consider the triad $G_1$, with $A_{23}=A_{31}=1$ and $A_{12}=-1$, because this network contains internal tensions --- agent 3 allies with two other agents, who are opposed to each other. This is reminiscent of frustrated configurations in the study of phase transitions (e.g. a preference for anti-parallel nearest-neighbor spins on a triangular Ising lattice\footnote{The Ising model has been adapted into opinion dynamics models \cite{sirbu2016survey} when opinions are binary, such as voting in two-party systems.} \cite{baxter1982exactly}), which exhibit glassy behavior and take a long time to choose between multiple equilibria.

\begin{figure}[h!t]
    \centering
    \begin{subfigure}[b]{0.49\linewidth}
        \includegraphics[width=\linewidth]{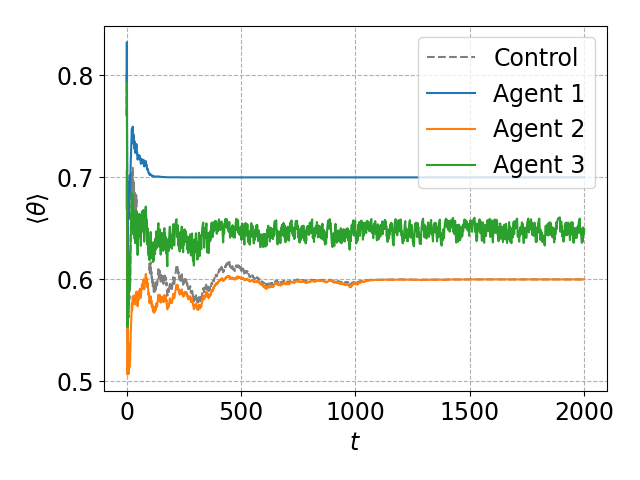}
        \caption{}
    \end{subfigure}
    \begin{subfigure}[b]{0.49\linewidth}
        \includegraphics[width=\linewidth]{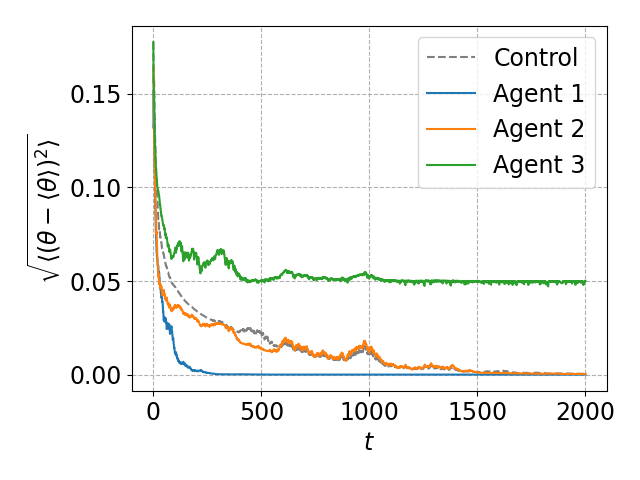}
        \caption{}
    \end{subfigure}
    \begin{subfigure}[b]{0.49\linewidth}
        \includegraphics[width=\linewidth]{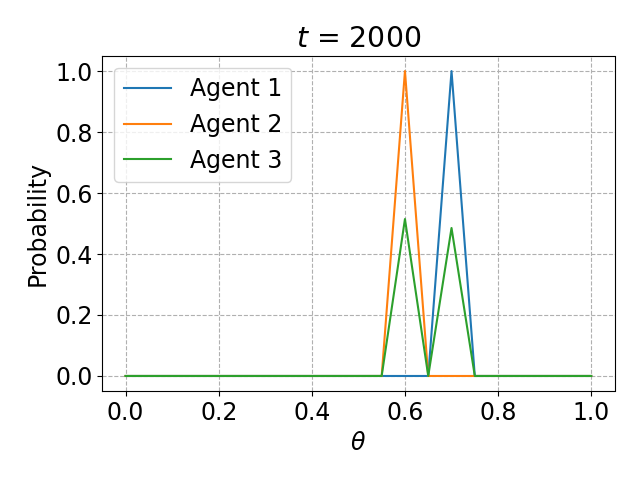}
        \caption{}
    \end{subfigure}
    \caption{Triad with internal tension $(A_{23}=A_{31}=1, A_{12}=-1)$. (a) Mean bias $\langle \theta \rangle$ versus time for agents 1 (blue solid curve), 2 (orange solid curve) and 3 (green solid curve). The control (grey dashed curve) shows $\mean$ for agent 3, if it ignores the beliefs of its allies. Agents 1 and 2 achieve asymptotic learning at $t_{{\rm a}, 1}=240$ and $t_{{\rm a}, 2}=1117$ time steps respectively, while agent 3 does not achieve asymptotic learning. Agent 2 converges on the true bias whereas agents 1 and 3 does not. Only $t \leq 2000$ time steps are shown for readability. (b) Standard deviation $\stddev$ versus time, with the same color scheme as panel (a). (c) Probability distribution functions $x_i(t=2000,\theta)$ versus $\theta$ for each agent. The probability distribution function of agent 3 has two peaks. Parameters: $\theta_0=0.6$, $\mu=0.25$, $T=10^4$.}
    \label{fig:triad_unbalanced}
\end{figure}

Figure \ref{fig:triad_unbalanced} shows how $\mean$ and $\stddev$ evolve for one specific realization, with randomized initial beliefs and coin tosses. Agent 3 is used as a control here, as defined in Section \ref{sec:pair_ally}. Turbulent nonconvergence is observed, as agent 3 does not settle on a particular $\theta$ for $0\leq t\leq T$. The control shows that agent 3 would have settled on $\theta=\theta_0$ without the influence of its allies. Agent 1 infers the wrong bias (with $\meanSub{1} \rightarrow 0.7 \neq \theta_0$), while agent 2 infers the correct bias (with $\meanSub{2} \rightarrow 0.6 = \theta_0$). The belief of agent 3 at $t> \max (t_{{\rm a}, 1}, t_{{\rm a}, 2})$ is multimodal --- it peaks at both $\theta=0.6$ and $\theta=0.7$. However, the relative confidence in $\theta=0.6, 0.7$ oscillates. The switching period of the oscillation, defined to be the number of consecutive time steps that the condition $x_3(t,\theta=0.6) > x_3(t,\theta=0.7)$ or $x_3(t,\theta=0.6) < x_3(t,\theta=0.7)$ holds, is small compared to $t_{{\rm a}, 1}=240$ and $t_{{\rm a}, 2}=1117$, with 75\% of switches being less than 5 time steps. 

The standard deviation of agent 3 achieves a steady state at $\stddevSub{3} \approx 0.05$, which corresponds to $|\meanSub{1} - \meanSub{2}| / 2$, half the separation in $\theta$ of the two peaks in $x_3(t,\theta)$. At $t=300$, the standard deviations of the beliefs of agent 2 and the control become similar enough to be indistinguishable by eye. At $t=500$, $\stddevSub{2}$ dips below the curve for the control. A run of heads (evidenced by the bump in $\meanSub{\rm control}$), pushes the beliefs towards $\theta > 0.6$. Since agent 2 is ``locked out'' of $\theta \geq 0.7$ by its opponent, it does not ``see'' the $\theta \geq 0.7$ tail of a binomial likelihood, while the control ``sees'' the whole binomial likelihood. This suppresses $x_2(t,\theta<0.6)$ compared to $x_{\rm control}(t,\theta<0.6)$. Similarly, at $t=1000$, $\stddevSub{\rm control}$ drops below $\stddevSub{2}$ during a run of tails.

Behaviors similar to turbulent nonconvergence have been reported in some models with deterministic beliefs. In Shi et al.'s model, \cite{shi2016evolution}, opinions are restricted to the range $[-1,+1]$. With a $G_1$ triad (Figure \ref{fig:types_of_triads}), the opinion of agent 3 oscillates between $\pm 1$ indefinitely. However, there is no information on how certain agent 3 is about its beliefs and how the uncertainty evolves. In Figure \ref{fig:triad_unbalanced}, agent 3 also has oscillatory beliefs, but it is confident in both $\theta=0.6, 0.7$ and not in any other $\theta$.

Why does agent 3 in Figure \ref{fig:triad_unbalanced} fail to convince both of its allies to overcome their antagonism and settle on the same bias? Suppose, initially, agent 1 is more confident than agent 2 in $\theta_1$, and agent 2 is more confident in $\theta_2$ than agent 1. Antagonistic interactions causes agent 1 to grow more confident in $\theta_1$, while agent 2 becomes more confident in $\theta_2$. Agent 3, being allies with both agent 1 and 2, attempts to align its belief in both options by making its belief a superposition of the beliefs of agents 1 and 2. This in turn causes agent 3 to be less confident in $\theta_1$ compared to agent 1, and less confident in $\theta_2$ compared to agent 2, because the beliefs are always normalized via equation (\ref{eq:update}). Due to agent 3's relatively lower confidence in $\theta_1$ and $\theta_2$ compared to its allies, the antagonism between agents 1 and 2 outweighs their solidarity with agent 3. Intuitively, agent's 3 indecisiveness in ``picking a side'' reduces its persuasive force when convincing its allies to compromise with each other.

In order to study how often turbulent nonconvergence occurs like in Figure \ref{fig:triad_unbalanced}, and whether it occurs for other triads, an ensemble of $10^5$ independent simulations is performed on each of the four possible triads, $G_0$, $G_1$, $G_2$ and $G_3$ illustrated in Figure \ref{fig:types_of_triads}. Turbulent nonconvergence only occurs in the triad with internal contradictions ($G_1$), and it does so in every one of the $10^5$ realizations. Analogous yet distinct behavior is reported in Shi et al.'s model \cite{shi2016evolution}. Shi et al.'s model differs from the model described in Sections \ref{sec:external_signal}--\ref{sec:internal_signals} in three ways: (i) there is no external signal, (ii) opinions are deterministic, and (iii) opinions are updated asynchronously (two agents per time step). In Shi et al.'s model, turbulent nonconvergence only occurs in ``unbalanced'' networks, where the balance of a network is described by structural balance theory (originally proposed by Heider \cite{heider1946attitudes}, later extended by Cartwright and Harary \cite{cartwright1956structural} and Davis \cite{davis1967clustering}), where certain combinations of ally-opponent relations are either socially stable or unstable. According to Davis \cite{davis1967clustering}, only $G_1$ in Figure \ref{fig:types_of_triads} is unbalanced, consistent with the results above. Structural balance theory is not restricted to triads \cite{cartwright1956structural}. In Section \ref{sec:balance}, larger networks are studied in the context of structural balance.

The wrong conclusion is reached first in an unbalanced triad, consistent with the behavior of a pair of opponents studied in Section \ref{sec:pair_opponents}. The summary statistics and distribution of $t_a^\text{right}-t_a^\text{wrong}$ for the unbalanced triad are similar to the pair of opponents and are not repeated here for the sake of brevity. For the unbalanced triad, the histogram of $t_a^\text{right}-t_a^\text{wrong}$ is also skewed right like Figure \ref{fig:pair_wrong_hist}. The bump at the left-hand side of the histogram is twice as tall as in Figure \ref{fig:pair_wrong_hist}.

\section{Scale-free networks with $n>3$}
\label{sec:larger_networks}

We now turn to larger networks with more realistic sizes. There are many valid ways to generate larger networks. In this section, we illustrate the behavior of the automaton in Section \ref{sec:dynamics_algorithm} in the context of scale-free networks as one representative example; other types of networks will be considered in future work. In scale-free networks, the degree distribution of the network follows a power law (where the degree of an agent equals the total number of its allies and opponents). It is thought that many real-world networks are approximately scale-free \cite{barabasi2003scale}. Additionally, there is some empirical evidence that the positive and negative degrees (defined in terms of links with $A_{ij} > 0$ and $A_{ij} < 0$ respectively) of real-world networks are individually approximately scale-free \cite{tang2016survey, kumar2016citations, maniu2011signed}. We generate these networks using the \barabasi model \cite{barabasi1999emergence}, as implemented in the \texttt{barabasi\_albert\_graph} function in the \texttt{Python} package \texttt{networkx} \cite{networkx}. This function accepts two integer tunable parameters, the number of agents $n$ and the attachment factor $m<n$, and yields a network with $m(n-m)$ connections. By construction, \barabasi networks are never disconnected. We emphasize that we focus on \barabasi networks merely as one concrete example. In Section \ref{sec:other_networks}, we briefly examine \erdos and square lattice networks. Many other kinds of networks are equally suitable for the media bias application and will be studied in future work.

\begin{figure}[h!t]
    \centering
    \includegraphics[width=0.6\linewidth]{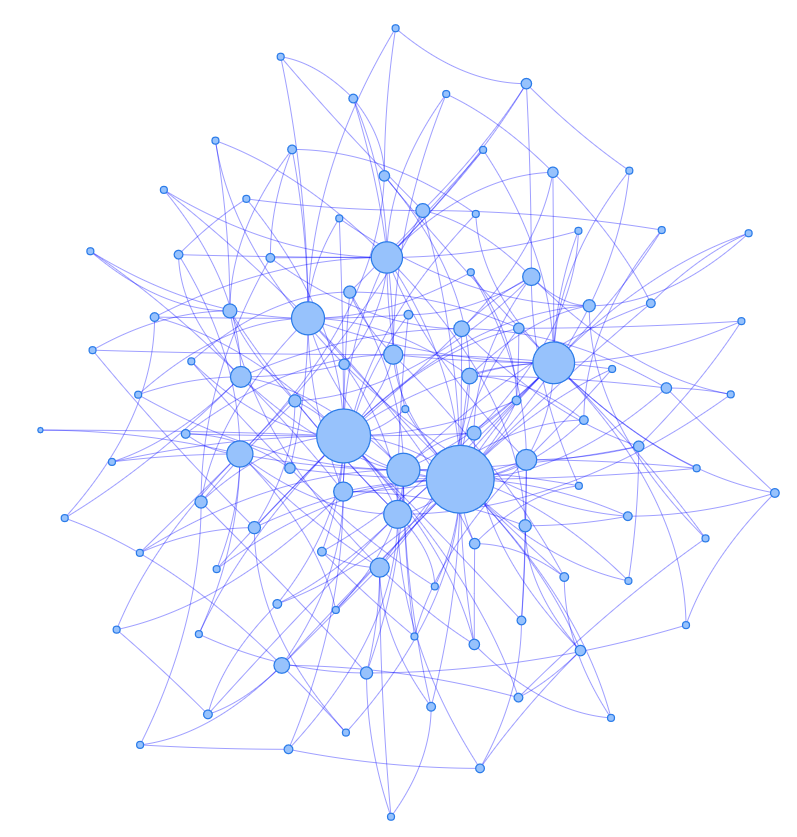}
    \caption{A particular \barabasi network generated with $n=100$ agents and attachment parameter $m=3$. The circles represent the agents and the lines between them represent their connections. The size of a circle is proportional to the agent's degree. This network is visualized using the \texttt{Python} package \texttt{pyvis} \cite{pyvis}.}
    \label{fig:BA_graph}
\end{figure}

The \texttt{networkx} package implements the \barabasi model as follows. The first $m$ agents do not know each other, i.e. $A_{ij}=0$ for all $1 \leq i,j \leq m$. The $(m+1)$-th agent is connected to all previous $m$ agents, i.e. $A_{i,(m+1)}=1$ for all $1 \leq i \leq m$. More agents are iteratively added, one at a time, by randomly connecting to $m$ pre-existing agents. The probability that the added agent connects to a pre-existing agent $i$ is $k_i / \sum_j k_j$, where $k_i$ is the degree of agent $i$ and the sum is taken over all pre-existing agents. Said differently, the newly added agent prefers to connect to agents with more connections. The algorithm terminates, when there are $n$ agents in the network. Note that the \barabasi model generates $A_{ij}$ with entries 0 or $+1$. To generate a random network that also has $A_{ij}=-1$ for some $i,j$, further modifications are required, which are discussed in each subsection. In this section, we focus on \barabasi networks with $n=100$ and $m=3$. Figure \ref{fig:BA_graph} represents the connections in a particular \barabasi network visually.

As a validation exercise, we start by generalizing the allies-only study in Section \ref{sec:pair_ally} to $n=100$ agents. We do not detect any behavior in the $n=100$ network that differs qualitatively from the analogous $n=2$ network; asymptotic learning and consensus are achieved in both cases. We summarize the validation exercise for completeness in \ref{sec:ba_allies}.

\subsection{Opponents only: reaching the wrong conclusion first}
\label{sec:ba_opponents}

In a network with two opponents, there is a strong tendency for one agent to reach a wrong conclusion, and to do so before their opponent reaches the right conclusion, as shown in Section \ref{sec:pair_opponents}. It is important to test whether the tendency persists in larger opponents-only networks. Here we consider a network with $n=100$ and $A_{ij}=-1$ for all nonzero entries, i.e. all agents are mutually opposed. We use the same link structure, initial beliefs, and sequence of coin tosses as in Figure \ref{fig:BA_graph} and \ref{sec:ba_allies}. The only difference with the test in \ref{sec:ba_allies} is the sign of $A_{ij}$.

\begin{figure}[!ht]
    \centering
    \begin{subfigure}[b]{0.49\linewidth}
        \centering
        \includegraphics[width=\linewidth]{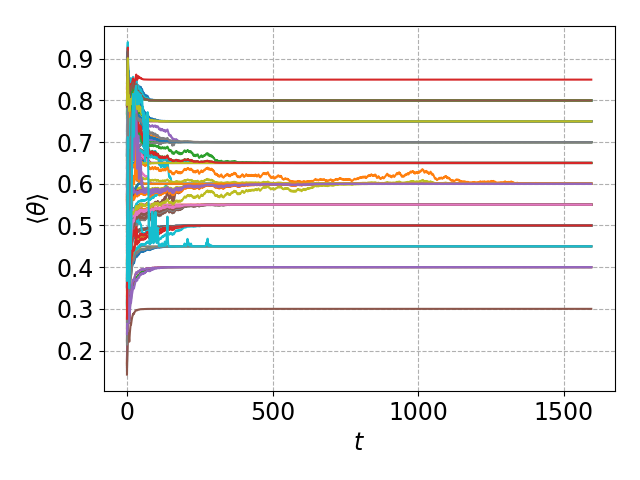}
        \caption{}
    \end{subfigure}
    \begin{subfigure}[b]{0.49\linewidth}
        \centering
        \includegraphics[width=\linewidth]{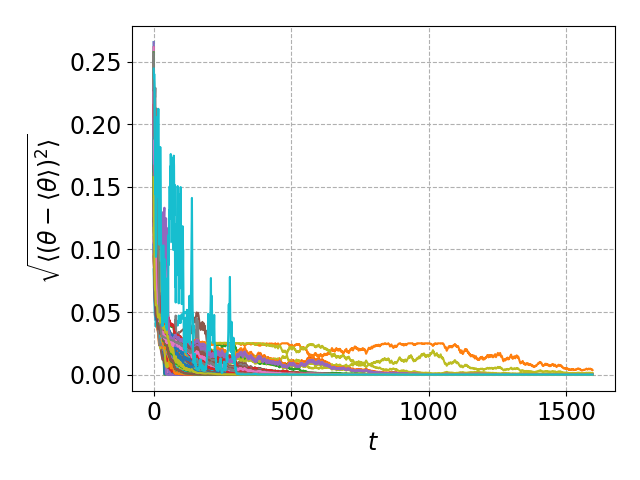}
        \caption{}
        \label{subfig:BA_opponents_stddev}
    \end{subfigure}
    \caption{Opinion formation in the \barabasi network in Figure \ref{fig:BA_graph} with opponents only, i.e. $A_{ij} \leq 0$ for all $i$ and $j$, to be compared with the opponents-only study in Section \ref{sec:pair_opponents} and Figure \ref{fig:pair_opponents}. Evolution for $t \leq t_{\rm a}=1595$ of the (a) mean $\mean$ and (b) standard deviation $\stddev$ of the beliefs of each agent, plotted as differently colored curves for each agent. Some but not all curves are indistinguishable by eye. Parameters: $n=100$, $m=3$, $\theta_0=0.6$, $\mu=0.25$, $A_{ij} \in \{0, -1\}$ for all $1 \leq i,j \leq n$.}
    \label{fig:BA_opponents}
\end{figure}

The evolution of $\mean$ and $\stddev$ in a particular realization is shown in Figure \ref{fig:BA_opponents}. We observe that the beliefs of the agents settle in the range $0.30 \leq \mean \leq 0.85$. The beliefs of each agent at $t=t_{\rm a}$ peak at unity at their corresponding $\langle \theta \rangle$, i.e. $x_i(t=t_{\rm a}, \theta=\langle \theta \rangle_i) \approx 1$ for all $1 \leq i \leq n$. 17 agents infer the correct true bias (with $\langle \theta \rangle \rightarrow \theta_0$), while the other 83 agents do not (with $|\langle \theta \rangle - \theta_0| \geq 0.05$). There is only one instance, where two agents agree despite being opponents, with $\meanSub{i}=\meanSub{j}=0.55$ and $A_{ij}=-1$. 

Figure \ref{subfig:BA_opponents_stddev} shows some interesting features in the evolution of $\stddev$. At $t=1000$, a bump in $\stddev$ is visible for the two agents who have the highest $\stddev$, which coincides with a run of heads. At $t \approx 100$, we see an agent with $\stddev \approx 0.15$, which is at least three times higher than the $\stddev$ of any other agent. This agent has their belief zeroed out at $0.50 \leq \theta \leq 0.75$ by its opponents, but not at any other $\theta$, creating a multimodal belief that peaks at both $\theta=0.45$ and $\theta=0.80$. The initial run of heads at $t=100$ (as seen in the bump in $\mean$ in Figure \ref{fig:BA_allies}) initially favors $\theta=0.80$ over $\theta=0.45$, but as more coin tosses are observed, $\theta=0.80$ is suppressed compared to $\theta=0.45$, as $P[S(t)|\theta]$ is greater at $\theta=0.45$ than $\theta=0.80$.

Some comparisons can be drawn with a deterministic model proposed by Martins et al. \cite{martins2010mass} that incorporates antagonistic interactions and the influence of the media. In Martins et al.'s model, the media is modeled as a special agent whose opinion, $\theta_{\rm media}$, never changes. The agents' beliefs are updated as in a Deffuant-Weisbuch model with antagonistic interactions, as described by equations (\ref{eq:DW1})--(\ref{eq:DW_adj}), except that $A_{ij}$ can take a value of $-1$. The media is always an ``ally'' to every agent and the beliefs can take any real-valued number in the range $[0,+1]$ inclusive. When the network consists of opponents only, the beliefs of some opponents cluster around the media's opinion, some fluctuate throughout the whole opinion space and some settle on the boundary of the opinion space (0 or $+1$). However, the agents that cluster around the media opinion are not steady in their beliefs; they fluctuate around the media opinion. This contrasts the behavior seen in Figure \ref{fig:BA_opponents}, where (i) the opinions settle stably at the corresponding $\mean$, and (ii) none of the agents settle at the extremes $\theta=0,1$. The clustering is discussed qualitatively in \cite{martins2010mass}, so a quantitative comparison with Figure \ref{fig:BA_opponents} cannot be made.

\begin{figure}[h!t]
    \centering
    \begin{subfigure}[b]{0.49\linewidth}
        \centering
        \includegraphics[width=\linewidth]{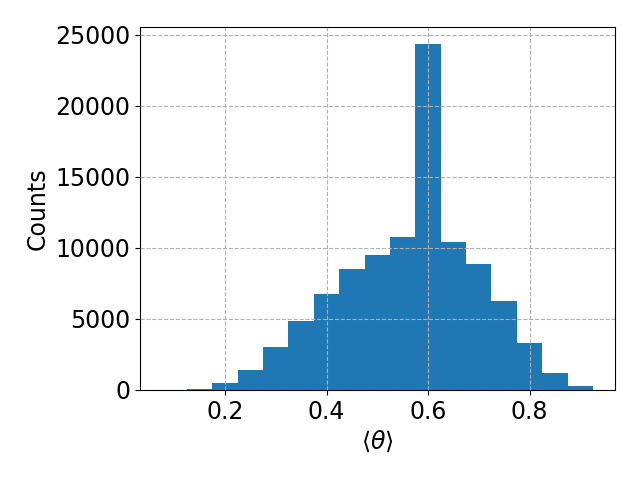}
        \caption{}
        \label{subfig:BA_opponents_hist_averaged}
    \end{subfigure}
    \begin{subfigure}[b]{0.49\linewidth}
        \centering
        \includegraphics[width=\linewidth]{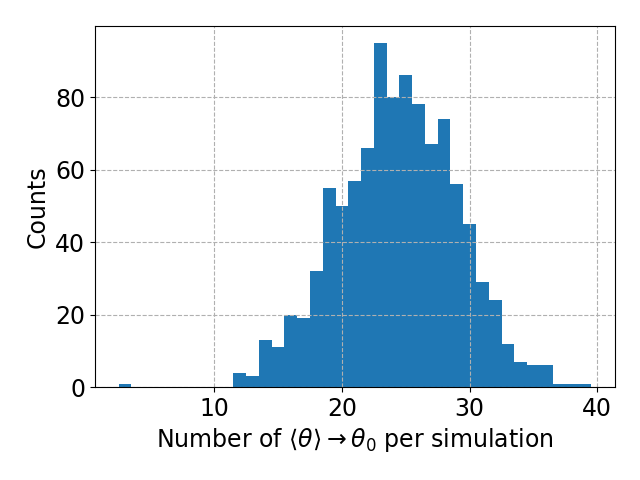}
        \caption{}
        \label{subfig:BA_opponents_num_correct}
    \end{subfigure}
    \caption{Distribution of beliefs once asymptotic learning is achieved by every agent in the \barabasi network with opponents only shown in Figure \ref{fig:BA_graph}. We accumulate $\mean$ of each of the $n=100$ agents over $10^3$ independent realizations with randomized $A_{ij}$, priors and coin tosses. (a) Histogram of $\mean$. (b) Number of agents out of $n=100$ attaining $\mean \rightarrow \theta_0$ per simulation.}
    \label{fig:BA_opponents_average}
\end{figure}

We observe that every agent in Figure \ref{fig:BA_opponents} achieves asymptotic learning, although they do not share the same final beliefs. We now ask: how are the final beliefs distributed? Figure \ref{subfig:BA_opponents_hist_averaged} shows the histogram of $\langle \theta \rangle$ of each of the $n=100$ agents at $t=t_{\rm a}$ accumulated over $10^3$ independent realizations using randomized $A_{ij}$, priors and coin tosses. The histogram has asymmetric tails with mean and median of $0.56$ and $0.60$ respectively. The histogram is similar to the $n=2$ network in Section \ref{sec:pair_opponents}, apart from the spike at $0.60$. Furthermore, we track the number of agents that settle at $\theta=\theta_0$ per simulation and the relevant histogram is shown in Figure \ref{subfig:BA_opponents_num_correct}. The histogram has a mean and median of 24, and a standard deviation of 5. The tails of the histogram are roughly symmetric around the mean.

\begin{figure}[h!t]
    \centering
    \begin{subfigure}[b]{0.49\linewidth}
        \centering
        \includegraphics[width=\linewidth]{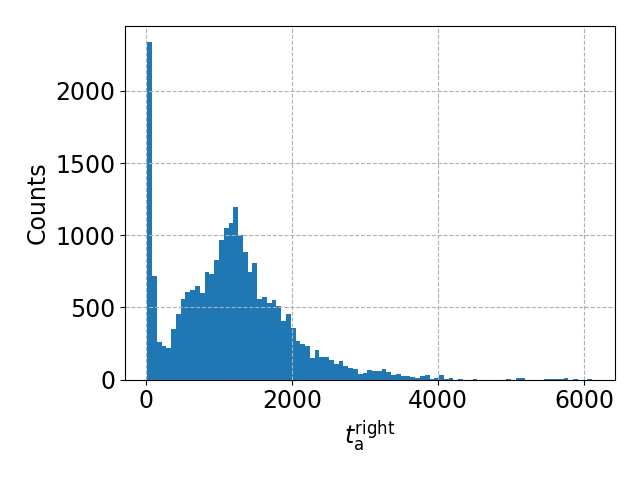}
        \caption{}
        \label{subfig:BA_opponents_average_correct_time}
    \end{subfigure}
    \begin{subfigure}[b]{0.49\linewidth}
        \centering
        \includegraphics[width=\linewidth]{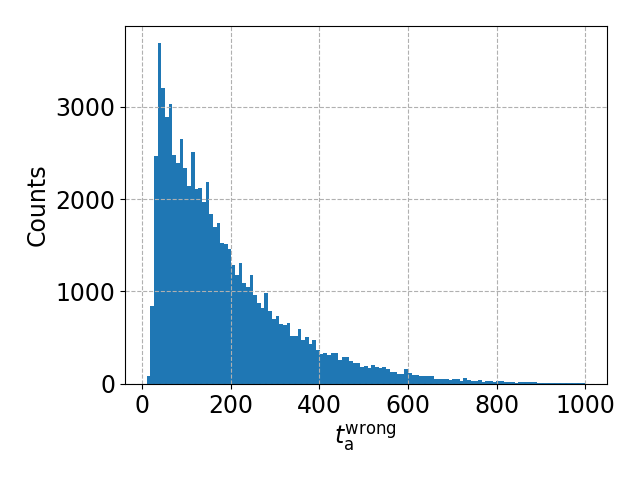}
        \caption{}
    \end{subfigure}
    \caption{Reaching the wrong conclusion first: histogram of (a) $t_{\rm a}^{\rm right}$ and (b) $t_{\rm a}^{\rm wrong}$ for $10^3$ independent simulations for a $n=100$, $m=3$ \barabasi network with opponents only. The same $10^3$ simulations that generate the histogram in Figure \ref{fig:BA_opponents_average} are also used to generate the above histograms. For visual clarity, we show only $t_{\rm a}^{\rm wrong} \leq 1000$ in (b), as there are relatively few $t_{\rm a}^{\rm wrong} > 1000$ data points.}
    \label{fig:BA_opponents_times_average}
\end{figure}

Confidence in the wrong conclusion is reached faster than confidence in the correct conclusion by some agents but not all of them. For the specific realization shown in Figure \ref{fig:BA_opponents}, the mean, median and standard deviation of $t_{\rm a}^{\rm right}$ are 571, 700 and 450 time steps respectively, while for $t_{\rm a}^{\rm wrong}$, they are 194, 156 and 134 time steps respectively. Hence, the wrong conclusion, on average, is reached quicker than the right conclusion. We also track $t_{\rm a}^{\rm right}$ and $t_{\rm a}^{\rm wrong}$ for the $10^3$ simulations that yield Figure \ref{fig:BA_opponents_average} and plot their histograms in Figure \ref{fig:BA_opponents_times_average}. We observe that $t_{\rm a}^{\rm right}$ has a mean, median and standard deviation of 1185, 1150 and 787 time steps respectively, cf. 209, 146 and 260 respectively for $t_{\rm a}^{\rm wrong}$. On average, the wrong conclusion is still reached quicker. The histogram of $t_{\rm a}^{\rm right}$ shows an interesting shape --- the modal value is at $t_{\rm a}^{\rm right}\approx 50$, but there are also two additional ``bumps'' at $t_{\rm a}^{\rm right} \approx 700$ and $t_{\rm a}^{\rm right} \approx 1200$.

\begin{table}[h!t]
\centering
\begin{tabular}{@{}lrr@{}}
\toprule
Property of $t_{\rm a}$ & $t_{\rm a}^{\rm right}$ & $t_{\rm a}^{\rm wrong}$   \\ \midrule
Mean (time steps) & 1185  & 209  \\
Standard deviation (time steps) & 787   & 260   \\
First quartile (time steps) & 648   & 77   \\
Median  (time steps) & 1150  & 146    \\
Third quartile (time steps) & 1599  & 258  \\ \midrule
Total & 24369 & 75631 \\ \bottomrule
\end{tabular}
\caption{Summary statistics of $t_{\rm a}^{\rm right}$ and $t_{\rm a}^{\rm wrong}$  for $10^5$ agents accumulated over $10^3$ simulations as in Figure \ref{fig:BA_opponents_times_average}. The corresponding histograms are visualized in Figure \ref{fig:BA_opponents_times_average}. The results are to be compared with Table \ref{tab:pair_wrong_conclusion} for $n=2$.}
\label{tab:ba_right_wrong_times}
\end{table}

On average, asymptotic learning is reached slightly quicker than for $n=2$ in Section \ref{sec:pair_opponents}. We observe that the means of $t_{\rm a}^{\rm right}$ and $t_{\rm a}^{\rm wrong}$ for opponents only with $n=100$ are 115 and 13 time steps respectively smaller than the means of $t_{\rm a}^{\rm right}$ and $t_{\rm a}^{\rm wrong}$ for a pair of opponents. Similarly, the quartiles of $t_{\rm a}^{\rm right}$ and $t_{\rm a}^{\rm wrong}$ for the $n=100$ network are smaller. Intuitively, more opponents give more opportunities for an agent's beliefs to get ``locked out'' of a wider range of $\theta$. As discussed in Section \ref{sec:pair_opponents}, an agent considering fewer $\theta$ values can achieve quicker asymptotic learning than an agent that has to consider all possible $\theta$ values. Additionally, $t_{\rm a}^{\rm wrong}$ in the $n=100$ network is more skewed, as its mode (39 time steps) lies further to the left of its median (146 time steps) compared to the pair of opponents (with mode and median of 150 and 190 respectively), even though the $n=100$ network has a smaller average $t_{\rm a}^{\rm wrong}$. Moreover, the histogram of $t_{\rm a}^{\rm right}$ for $n=2$ lacks the sharp spike at $t_{\rm a}^{\rm right} \approx 50$ for $n=100$.

Unlike the pair of opponents studied in Section \ref{sec:pair_opponents}, we do not attempt to analyze how the choice of priors influences who infers the correct bias in the same manner as Section \ref{sec:pair_opponents} due to the exponential computational cost of permuting 100 different priors. A systematic way of studying the choice of priors is an interesting focus for future research.

\subsection{Equal mix of allies and opponents: turbulent nonconvergence}
\label{sec:ba_mixed}

Having considered the special cases involving allies or opponents exclusively in \ref{sec:ba_allies} and Section \ref{sec:ba_opponents}, the natural next step is to investigate mixed networks where both types of interactions coexist. Many real world scenarios, such as political parties with mixed ideological stances \cite{ware2009dynamics}, can be modeled naturally in this way. To compare against a baseline, we analyze a network with the same structure as in Figure \ref{fig:BA_graph}. This time, each connection is randomly given a weight of $+1$ or $-1$ with equal probability. We again use the same priors and coin tosses as in \ref{sec:ba_allies} and Section \ref{sec:ba_opponents}.

\begin{figure}[h!t]
    \centering
    \begin{subfigure}[b]{0.45\linewidth}
        \centering
        \includegraphics[width=\linewidth]{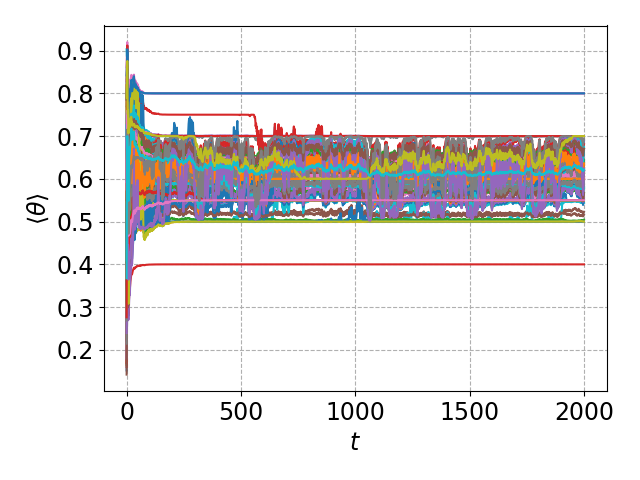}
        \caption{}
        \label{subfig:ba_mixed_mean}
    \end{subfigure}
    \begin{subfigure}[b]{0.45\linewidth}
        \centering
        \includegraphics[width=\linewidth]{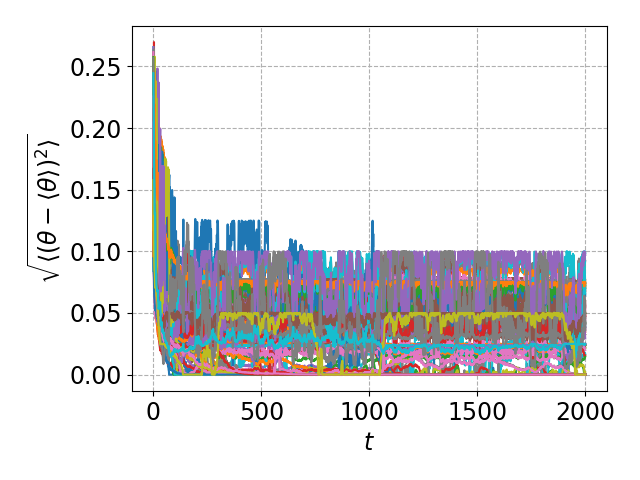}
        \caption{}
        \label{subfig:ba_mixed_stddev}
    \end{subfigure}
    \begin{subfigure}[b]{0.45\linewidth}
        \centering
        \includegraphics[width=\linewidth]{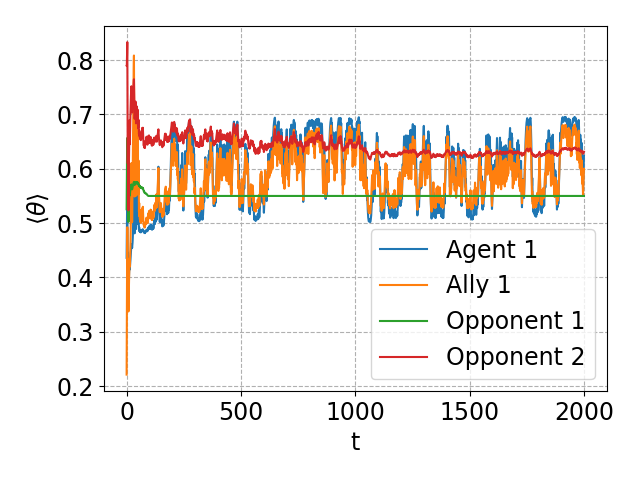}
        \caption{}
        \label{subfig:ba_mixed_mean_zoom}
    \end{subfigure}
    \begin{subfigure}[b]{0.45\linewidth}
        \centering
        \includegraphics[width=\linewidth]{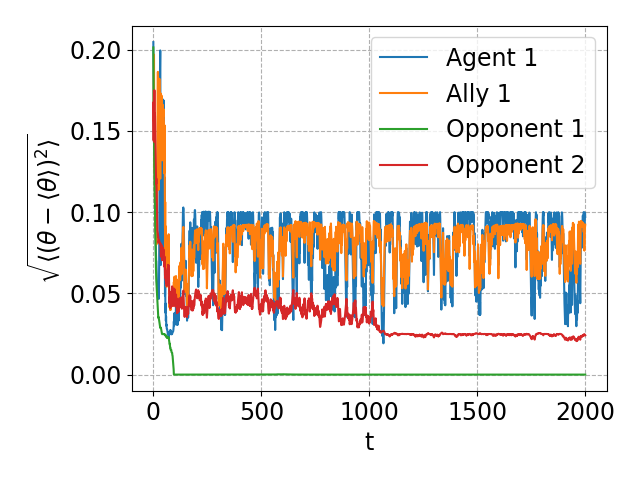}
        \caption{}
        \label{subfig:ba_mixed_stdded_zoom}
    \end{subfigure}
    \caption{Opinion formation in the \barabasi network in Figure \ref{fig:BA_graph} with a mix of allies and opponents, to be compared with the triad with internal tensions considered in Section \ref{sec:triad_unbalanced}. Evolution of the (a) mean $\mean$ and (b) standard deviation $\stddev$ of the beliefs of each agent, plotted as differently colored curves for each agent. As the curves are cluttered, we also plot the (c) mean and (d) standard deviation of the beliefs of a particular agent, as well as its one ally and two opponents. Opponent 1 achieves asymptotic learning, but agent 1, ally 1 and opponent 2 fail to do so. Only $0\leq t \leq 2000$ time steps are shown for clarity. Parameters: $n=100$, $m=3$, $\theta_0=0.6$, $\mu=0.25$, $A_{ij} \in \{-1,0, +1\}$ for all $1 \leq i,j \leq n$.}
    \label{fig:BA_mixed}
\end{figure}

Figures \ref{subfig:ba_mixed_mean} and \ref{subfig:ba_mixed_stddev} shows the evolution of $\mean$ and $\stddev$ of every agent in a particular realization. Turbulent nonconvergence is observed for 68 agents, while the other 32 agents achieve asymptotic learning. Of the 32 agents that achieves asymptotic learning, only eight correctly infer the coin bias, with $\mean \rightarrow \theta_0$. Every agent has zero confidence that $\theta_0$ lies outside the range $0.40 \leq \theta \leq 0.80$. This range is narrower than the opponents only network with $n=100$ in Section \ref{sec:ba_opponents}, which has $0.30 \leq \theta \leq 0.85$. Internal interactions between allies in the mixed network drive their beliefs closer together. The opponents only network lacks any mutually reinforcing behavior, which causes the wider spread of $\mean$. Figure \ref{subfig:ba_mixed_stddev} shows that the standard deviations of the beliefs of all agents at $t > 1000$ are constrained to $\stddev \leq 0.10$, because the beliefs of every agent who fails to achieve asymptotic learning are nonzero in the range $0.50 \leq \theta \leq 0.70$.

Figures \ref{subfig:ba_mixed_mean} and \ref{subfig:ba_mixed_stddev} are too cluttered to see the evolution of $\mean$ and $\stddev$ of individual agents, so we also plot $\mean$ and $\stddev$ for a particular agent, as well as their ally and opponents in Figures \ref{subfig:ba_mixed_mean_zoom} and \ref{subfig:ba_mixed_stdded_zoom}. We label this particular agent as agent 1. Agent 1's ally and opponents are not directly connected. Agent 1, its ally and one opponent (opponent 2 in Figure \ref{subfig:ba_mixed_mean_zoom}) fail to achieve asymptotic learning. Antagonistic interactions drive $x_1(t, \theta)$ towards zero for $0.55 \leq \theta \leq 0.65$. At $t>1000$, $x_1(t, \theta)$ is only nonzero at $\theta=0.50,0.70$. Opponent 2 achieves a steady state at $t>1000$, with nonzero beliefs at $\theta=0.60,0.65$ and $\stddev \approx 0.025$. For the triad with internal tensions in Section \ref{sec:triad_unbalanced}, the agent that does not achieve asymptotic learning always has exactly two allies, and those allies achieve asymptotic learning. In contrast, for mixed $n=100$ networks, it is possible for agents as well as their allies and opponents to fail to achieve asymptotic learning. More precisely, we observe that all agents who fail to achieve asymptotic learning have at least one ally.

Comparing to Martins et al.'s media-specific model \cite{martins2010mass} again, when there is a mix of allies and opponents, some of the agents' opinions fluctuate around the opinion promoted by the media. A similar phenomenon is observed in Figure \ref{fig:BA_mixed}, where some of the agents have constantly varying confidence in $0.50 \leq \theta \leq 0.70$, which contains $\theta_0$. While the agents in Martins et al.'s model are free to continuously vary their opinion around the opinion promoted by the media, agents in the model described in Sections \ref{sec:external_signal}--\ref{sec:internal_signals} may vary their confidence only in specific $\theta$ increments. The agent shown in Figures \ref{subfig:ba_mixed_mean_zoom} and \ref{subfig:ba_mixed_stdded_zoom} is confident in $\theta=0.50$ and $0.70$, but no other $\theta$ value. This complex belief cannot be described using a deterministic model.

\begin{figure}[h!t]
    \centering
    \includegraphics[width=0.5\linewidth]{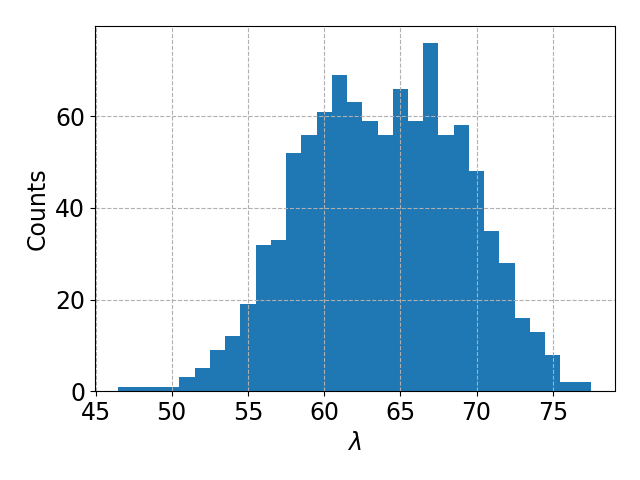}
    \caption{Turbulent nonconvergence in a mixed network of allies and opponents: histogram of $\lambda$, the number of agents (out of $n=100$ in total) that fail to achieve asymptotic learning per simulation. The data are collected over $10^3$ independent realizations with randomized priors and coin tosses, but identical $A_{ij}$. Each histogram bin has a width of unity.}
    \label{fig:ba_mixed_lambda}
\end{figure}

To study the statistics of how many agents fail to achieve asymptotic learning for the mixed $n=100$ network studied in this section, we run $10^3$ independent simulations with randomized priors and coin tosses, but retain the same $A_{ij}$. We track $\lambda$, the number of agents that fail to achieve asymptotic learning in each realization. Figure \ref{fig:ba_mixed_lambda} and Table \ref{tab:ba_mixed_lambda} displays the histogram and summary statistics respectively of $\lambda$. The mean and median of $\lambda$ both equal 64. In each of these $10^3$ simulations, at least 47 agents fail to achieve asymptotic learning.

\begin{table}[h!t]
\centering
\begin{tabular}{@{}lr@{}}
\toprule
Property of $\lambda$ & Value  \\ \midrule
Mean & 64    \\
Standard deviation & 5 \\
First quartile & 60   \\
Median & 64  \\
Third quartile  & 68  \\  \bottomrule
\end{tabular}
\caption{Turbulent nonconvergence: summary statistics of $\lambda$, the number of agents that fail to achieve asymptotic learning per simulation, based on the histogram shown in Figure \ref{fig:ba_mixed_lambda}.}
\label{tab:ba_mixed_lambda}
\end{table}

In the context of structural balance theory \cite{davis1967clustering}, the network studied in this section is unbalanced, similar to the triad with internal tensions in Section \ref{sec:triad_unbalanced}, while networks composed of allies only (Sections \ref{sec:pair_ally} and \ref{sec:ba_allies}) and opponents only (Sections \ref{sec:pair_opponents} and \ref{sec:ba_opponents}) are not. Likewise, turbulent nonconvergence is only observed in the mixed network. As mentioned in Section \ref{sec:triad_unbalanced}, Shi et al.'s model \cite{shi2016evolution} also predicts that turbulent nonconvergence occurs in unbalanced networks. In Section \ref{sec:balance} we study turbulent nonconvergence of networks with $n=100$ in the context of structural balance theory.

\subsection{Larger scale-free networks with $n>100$}

Although we analyze $n=100$ networks in this section as a representative case for networks of realistic sizes, some real-world networks are significantly larger. For example, a recent study, which aims to extract user opinions from Twitter during the 2016 US presidential election, utilizes tweets from more than a million unique users \cite{yaqub2017twitter}. It is already computationally expensive to produce the statistics behind Tables \ref{tab:ba_right_wrong_times} and \ref{tab:ba_mixed_lambda}, so it is unfeasible to analyze $n>100$ networks the same way as $n=100$ networks in Sections \ref{sec:ba_opponents} and \ref{sec:ba_mixed}. However, it is feasible to simulate one or two $n=1000$ networks as a check. Here, we run three simulations on a \barabasi network with $n=1000$ and attachment parameter $m=3$ under three different scenarios: all-allies, all-opponents and a mixture of half allies and half opponents. We keep the same priors, coin tosses and $|A_{ij}|$ between the three scenarios, and only vary the sign of $A_{ij}$. For consistency, we use the same sequence of coin tosses as the $n=2$ network studied in Section \ref{sec:pair_ally}. We present the results of the three simulations in \ref{section:1000_network}. Overall, we do not find any significant differences between the behaviors of the tested $n=1000$ and $n=100$ networks.

\subsection{Long-term intermittency}
\label{sec:intermittency}

\begin{figure}[h!t]
    \centering
    \begin{subfigure}[b]{0.45\linewidth}
        \centering
        \includegraphics[width=\linewidth]{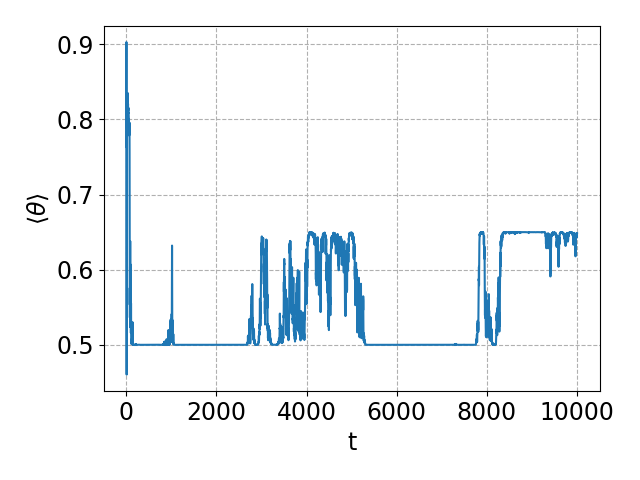}
        \caption{}
    \end{subfigure}
    \begin{subfigure}[b]{0.45\linewidth}
        \centering
        \includegraphics[width=\linewidth]{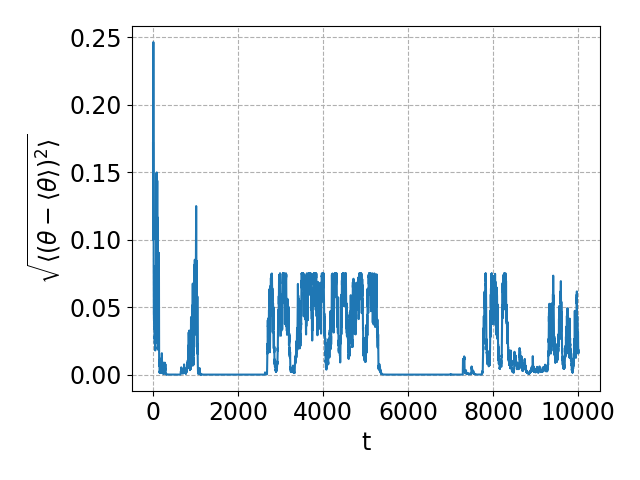}
        \caption{}
    \end{subfigure}
    \caption{Intermittent evolution of the beliefs of an agent in a mixed ally-opponent network with the topology in Figure \ref{fig:BA_graph}. Evolution for $t \leq 10^4$ of the (a) mean $\mean$ and standard deviation $\stddev$ of the belief of the intermittent agent. This agent is taken from the same simulation as shown in Figure \ref{fig:BA_mixed}.}
    \label{fig:ba_intermittency}
\end{figure}

In Section \ref{sec:ba_mixed}, one of the agents displays long-term intermittency: their beliefs remain constant for multiple iterations, then fluctuate for multiple iterations, then the cycle repeats. The behavior persists throughout the simulation, although it is not periodic. Figure \ref{fig:ba_intermittency} shows $\mean$ and $\stddev$ versus time for the intermittent agent. For $1200 < t < 2600$, the agent's belief is unchanging, with $\mean=0.5$ and $\stddev = 0$. In this time range, the agent's belief is singly peaked, with $x(t, \theta=0.50)=1$ and zero at all other $\theta$. For $2600 < t < 5500$, the agent's belief fluctuates, with $0.50 \leq \mean \leq 0.65$ and $0.000 \leq \stddev \leq 0.075$. For $5500 < t < 7000$, we once again observe that the belief does not change. This cycle of going back and forth between unchanging and fluctuating beliefs persists even when the simulation runs to $t=10^5$.

The long-term intermittency in Figure \ref{fig:ba_intermittency} is not caused by numerical inaccuracies. It can be understood in terms of the internal signals. The agent shown in Figure \ref{fig:ba_intermittency} has one ally and two opponents. We label the agent in Figure \ref{fig:ba_intermittency} as agent 1 and its ally as agent 2. At $1200 < t < 2600$, we have $x_1(t,\theta=0.50)=1$ and $x_1(t,\theta=0.65)=x_2(t,\theta=0.65)=0$. At $t\approx 2600$, due to agent 2's interactions with its allies and opponents, we have $x_2(t,\theta=0.65) > 0$. Agent 1's opponents have zero belief in $\theta=0.65$, so agent 2 ``drags'' agent 1 along to have $x_1(t,\theta=0.65) > 0$. The sudden change of $x_2(t,\theta=0.65) = 0$ to $x_2(t,\theta=0.65) > 0$ is due to agent 2 sharing an ally and opponents with nonzero beliefs at $\theta=0.65$. At $1200 < t < 2600$, agent 2's opponents' confidence in $\theta=0.65$ exceeds the ally's, so agent 2's confidence in $\theta=0.65$ remains unchanged with $x_2(t,\theta=0.65) = 0$. However, at $t\approx 2600$, the ally happens to become more confident in $\theta=0.65$ compared to the opponents, leading to $x_2(t,\theta=0.65) >0$. In the range $2600 < t < 5500$, agents 1 and 2 constantly change their confidence in $\theta=0.50$ and $0.65$. At $t\approx 5500$, agent 2's ally loses confidence in $\theta=0.65$ and gain confidence in $\theta=0.50$. At the same time, agent 2's opponents gain confidence in $\theta=0.65$. These changes in the allies' and opponents' beliefs drive $x_2(\theta=0.65)$ to zero, which in turn drags $x_1(\theta=0.65)$ to zero. We have another period of stability until $t=7000$, where agent 2's ally gains confidence in $\theta=0.65$ and the process repeats.

Although it is not explicitly discussed, this intermittency is also observed in Shi et al.'s model \cite{shi2016evolution}. As mentioned in Section \ref{sec:triad_unbalanced}, the beliefs of some agents in an unbalanced network can oscillate between the boundaries of the opinion space, $[-1,+1]$. There are periods of stability where the opinion stays at $\pm 1$. The model described in Sections \ref{sec:external_signal}--\ref{sec:internal_signals} provides additional information by evolving the probability distribution function of $\theta$. It tells us that agent 1 in Figure \ref{fig:ba_intermittency} during the periods of stability is completely confident in their belief, with $\stddevSub{1}=0$. It also quantifies the uncertainty in the agent 1's belief during periods of instability, with $0.0 \leq \stddevSub{1} \leq 0.075$.

\section{Structural balance and asymptotic learning}
\label{sec:balance}

\begin{figure}[h!t]
    \centering
    \begin{subfigure}[b]{0.32\linewidth}
        \centering
        \includegraphics[width=\linewidth]{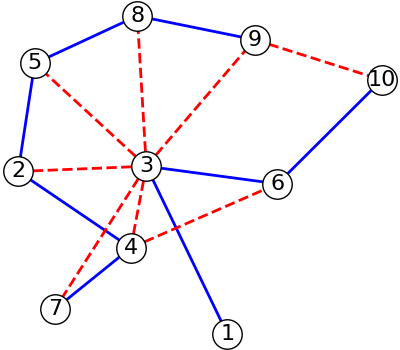}
        \caption{}
    \end{subfigure}
    \begin{subfigure}[b]{0.32\linewidth}
        \centering
        \includegraphics[width=\linewidth]{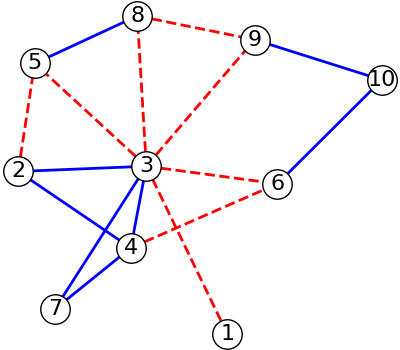}
        \caption{}
    \end{subfigure}
    \begin{subfigure}[b]{0.32\linewidth}
        \centering
        \includegraphics[width=\linewidth]{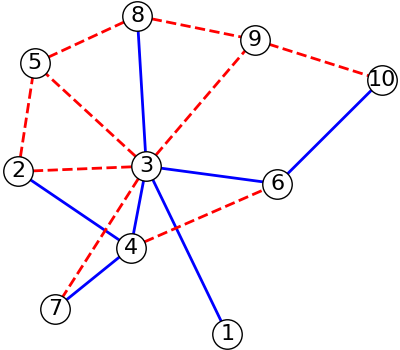}
        \caption{}
    \end{subfigure}
    \caption{Structural balance: examples of (a) strongly balanced, (b) weakly balanced and (c) unbalanced networks. Solid, blue and red, dashed lines denote alliances and antagonisms respectively. (a) The network can be split into two clusters, ${\sc C}_1=\{1, 3, 6, 10\}$ and ${\sc C}_2=\{2, 4, 5, 7, 8, 9\}$. (b) The network can be split into three clusters, ${\sc C}_1=\{2, 3, 4, 7\}$, ${\sc C}_2=\{1, 5, 8\}$ and ${\sc C}_3=\{6, 9, 10\}$. Note that agent 1 can belong to either ${\sc C}_2$ or ${\sc C}_3$, but either way, we still have three clusters. (c) It is impossible to split the agents into clusters. For instance, agents 2 and 3 should belong to the same cluster through their alliance with agent 4, but agents 2 and 3 should also belong to different clusters because they are opponents, which leads to a contradiction.}
    \label{fig:balance_examples}
\end{figure}

The original structural balance theory of Heider \cite{heider1946attitudes} concerns the social stability of relationships in a triad, and by extension, networks where every agent is an ally or opponent of every other agent. Under Heider's theory, triads are classified as either ``balanced'' or ``unbalanced''. It is predicted that unbalanced triads prefer to alter their relationships to become a balanced triad. One real-world example is the network of alliances between six particular countries prior to World War 1 which began as an unbalanced network, but settled as a balanced network \cite{antal2006balance}. The theory was later extended by Harary and Cartwright \cite{cartwright1956structural} to arbitrary networks. Davis \cite{davis1967clustering} extended the theory further to embrace the concept of weak balance. Namely, a network can be categorized based on how many clusters it has. A cluster is defined to be a nonempty set of agents such that any two agents in the same cluster are not opponents and any two agents from different clusters are not allies. Mathematically, the $i$-th cluster ${\sc C}_i$ is defined to be a set of agents such that one has $A_{ij} \geq 0$ for all $i,j \in {\sc C}_k$ and $A_{ij} \leq 0$ for $i \in {\sc C}_k,\; j \notin {\sc C}_k$. Then, a network is said to be (i) \textit{strongly balanced} if every agent can be grouped into one or two distinct clusters, (ii) \textit{weakly balanced} if every agent can be grouped into more than two distinct clusters and (iii) \textit{unbalanced} if it is impossible to group every agent into a cluster. Figure \ref{fig:balance_examples} depicts examples of the three discrete categories of structural balance.

Under Davis' definition, the allies-only networks studied in Section \ref{sec:pair_ally} and \ref{sec:ba_allies} are strongly balanced, while the opponents-only networks in Sections \ref{sec:pair_opponents} and \ref{sec:ba_opponents} are weakly balanced. The triad with internal tensions in Section \ref{sec:triad_unbalanced} and the mixed $n=100$ network in Section \ref{sec:ba_mixed} are unbalanced. It is an open question how to robustly quantify how ``balanced'' a network is on a continuous scale \cite{facchetti2011computing, kirkley2019balance}. A network with three clusters and another with four clusters are both weakly balanced, but we cannot say which of the two networks is more balanced than the other. We focus on Davis' three discrete categories and do not attempt to resolve the question of which networks in the same category are more or less balanced.

In Section \ref{sec:triad_unbalanced}, only the unbalanced triad fails to achieve asymptotic learning. Similarly, the mixed $n=100$ network, which is the only unbalanced network considered in Section \ref{sec:larger_networks}, experiences turbulent nonconvergence. Additionally, despite starting with identical priors, coin tosses and zero entries of $A_{ij}$, the $n=100$ network with allies only and the $n=100$ network with opponents only achieve different asymptotic learning times, with $t_{\rm a}=1433$ and $t_{\rm a}=1595$ respectively. 

Although the foregoing difference in $t_{\rm a}$ does not appear to be significant, we ask if turbulent nonconvergence and $t_{\rm a}$ are related to the structural balance of the network. To study this question, we examine the three categories in structural balance theory separately. We perform $10^3$ independent simulations with randomized priors, coin tosses and $A_{ij}$. The $A_{ij}$ are all randomly generated with the \texttt{barabasi\_albert\_graph(n=100, m=3)} function from the \texttt{networkx} package \cite{networkx}. We then randomly modify the unweighted network to fall into one of the categories. To obtain a strongly or weakly balanced network, we generate a clustered network based on a specified number of clusters. For a strongly balanced network, we always choose two clusters, while for a weakly balanced network, we randomly pick a number from the range $[3,n]$ inclusive. Other choices are equally valid, of course. Each agent is randomly assigned to one of the clusters and we ensure that each cluster contains at least one agent. The weight of the connections is then altered to satisfy the definition of a cluster. To obtain an unbalanced network, we randomly assign a weight of $+1$ or $-1$ to each nonzero $A_{ij}$ and check if the resulting network is indeed unbalanced.

\begin{figure}[h!t]
    \centering
    \begin{subfigure}[b]{0.32\linewidth}
        \centering
        \includegraphics[width=\linewidth]{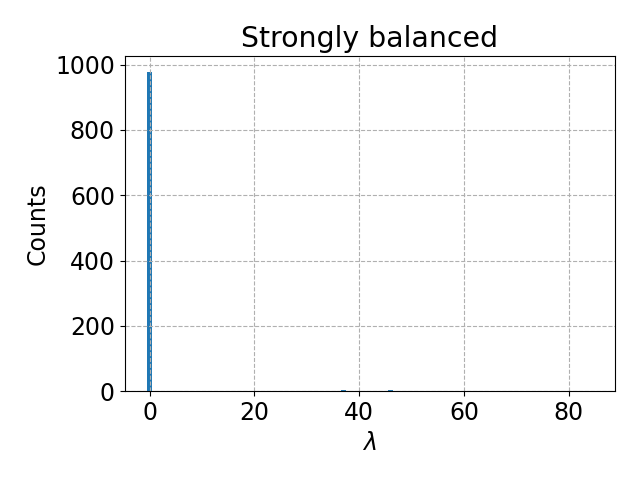}
        \caption{}
    \end{subfigure}
    \begin{subfigure}[b]{0.32\linewidth}
        \centering
        \includegraphics[width=\linewidth]{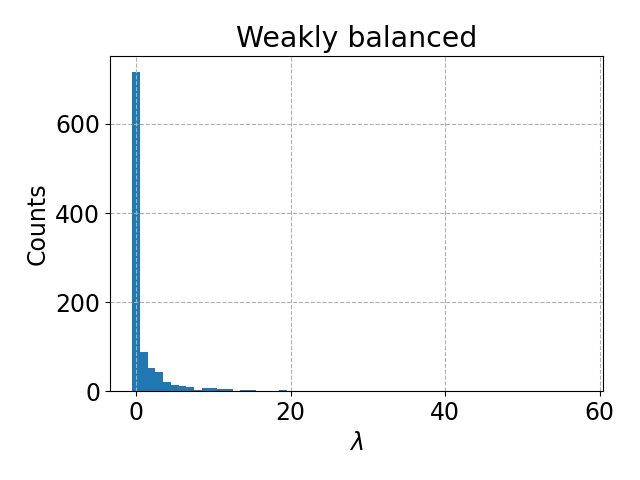}
        \caption{}
    \end{subfigure}
    \begin{subfigure}[b]{0.32\linewidth}
        \centering
        \includegraphics[width=\linewidth]{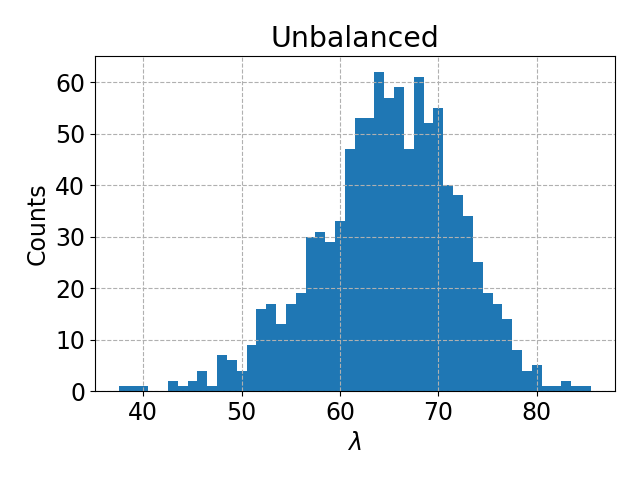}
        \caption{}
        \label{subfig:balance_lambda_unbalanced}
    \end{subfigure}
    \caption{Structural balance: histograms of $\lambda$, the number of agents (out of $n=100$ in total) experiencing turbulent nonconvergence per simulation, for (a) strongly balanced, (b) weakly balanced and (c) unbalanced networks. The data are collected from $10^3$ independent simulations with randomized priors, coin tosses and $A_{ij}$ for each structural balance. Each histogram bin has a width of unity.}
    \label{fig:balance_lambda}
\end{figure}

\begin{table}[h!t]
\centering
\begin{tabular}{@{}lrrr@{}}
\toprule
Property of $\lambda$ & Strongly balanced & Weakly balanced & Unbalanced \\ \midrule
Zero & 978 & 717 & 0 \\
Mean & 1 & 1 & 65 \\
Standard deviation & 6 & 3 & 7 \\
First quartile & 0 & 0 & 61 \\
Median & 0 & 0 & 65 \\
Third quartile & 0 & 1 & 70 \\ \midrule
Count & 1000 & 1000 & 1000 \\ \bottomrule
\end{tabular}
\caption{Structural balance: summary statistics of $\lambda$, the number of agents (out of $n=100$ in total) experiencing turbulent nonconvergence per simulation, according to the structural balance of the network, based on the histograms shown in Figure \ref{fig:balance_lambda}. Note that the 22 instances of nonzero lambda in the second column have a mean of 40, leading to a mean of unity overall.}
\label{tab:balance_lambda}
\end{table}

Figure \ref{fig:balance_lambda} and Table \ref{tab:balance_lambda} shows the histogram and summary statistics of $\lambda$, the number of agents who fail to achieve to asymptotic learning, grouped by the structural balance of the network. Unbalanced networks are the most likely of the three types of structural balance to experience turbulent nonconvergence, with all $10^3$ networks doing so, followed by weakly and strongly balanced networks, with $28\%$ and $2\%$ respectively, consistent with the predictions of structural balance theory, as discussed above. For unbalanced networks, the $\lambda$ histogram has a similar shape to the histogram shown in Figure \ref{fig:ba_mixed_lambda}. Although strongly and weakly balanced networks have similar means, their histograms are different. For strongly balanced networks, the 22 nonzero $\lambda$ values are spread out in the range $11 \leq \lambda \leq 84$, whereas for weakly balanced networks, $90\%$ of the 283 nonzero data points fall in the range $\lambda \leq 10$.

\begin{figure}[h!t]
    \centering
    \begin{subfigure}[b]{0.45\linewidth}
        \centering
        \includegraphics[width=\linewidth]{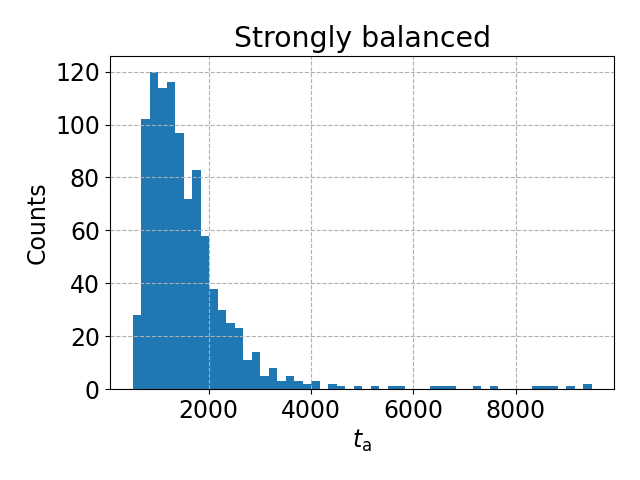}
        \caption{}
    \end{subfigure}
    \begin{subfigure}[b]{0.45\linewidth}
        \centering
        \includegraphics[width=\linewidth]{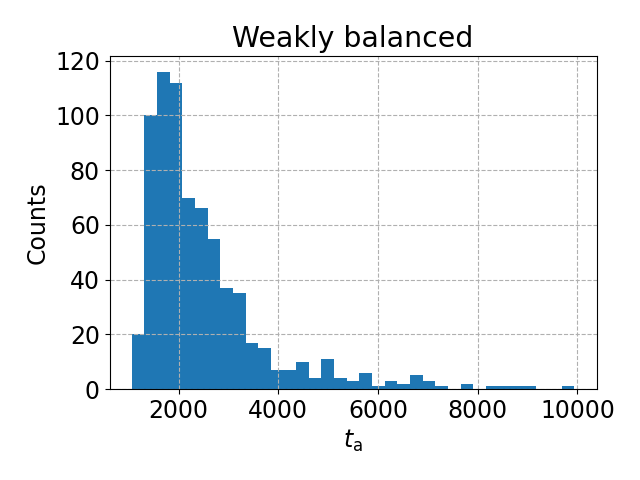}
        \caption{}
    \end{subfigure}
    \caption{Structural balance: histograms of $t_{\rm a}$ when the network does achieve asymptotic learning, for (a) strongly balanced and (b) weakly balanced networks. Unbalanced networks are excluded as they always experience turbulent nonconvergence. The data are collected in the same $10^3$ simulations performed for Figure \ref{fig:balance_lambda}.}
    \label{fig:balance_asymp_time}
\end{figure}

\begin{table}[h!t]
\centering
\begin{tabular}{@{}lrr@{}}
\toprule
Property of $t_{\rm a}$ (time steps) & Strongly balanced & Weakly balanced\\ \midrule
Mean & 1588 & 2500 \\
Standard deviation & 991 & 1260 \\
First quartile & 1009 & 1691  \\
Median & 1359 & 2126  \\
Third quartile & 1846 & 2830 \\ \midrule
Count & 978 & 717 \\ \bottomrule
\end{tabular}
\caption{Structural balance: summary statistics of $t_{\rm a}$ when the network achieves asymptotic learning, according to the structural balance of the network, based on the histograms shown in Figure \ref{fig:balance_asymp_time}.}
\label{tab:balance_asymp_time}
\end{table}

We also track $t_{\rm a}$, the time when the network achieves asymptotic learning. Figure \ref{fig:balance_asymp_time} and Table \ref{tab:balance_asymp_time} show the histogram and summary statistics respectively for $t_{\rm a}$, grouped by the structural balance of the network. The unbalanced networks are not shown because none of them achieves asymptotic learning. We observe that the histograms of $t_{\rm a}$ for both strongly and weakly balanced networks have similar shapes but they peak at different locations. The strongly balanced networks peak at a lower $t_{\rm a}$ with a median of 1359 time steps, compared to the weakly balanced networks with a median of 2126 time steps. There is a trend in $t_{\rm a}$: strongly balanced networks tend to achieve asymptotic learning the quickest, followed by weakly balanced networks. The unbalanced networks experience turbulent nonconvergence in each of the $10^3$ simulations. This trend again is consistent with the predictions of structural balance theory.

There are several proposed deterministic models that also note trends in the opinion dynamics based on the network's structural balance. One example mentioned above is Shi et al.'s model \cite{shi2016evolution}, where turbulent nonconvergence only occurs in unbalanced networks. In Altafini's model \cite{altafini2013consensus}, strongly balanced networks always result in agents settling on some real-valued $\theta_{\rm final}$ or its negative, $-\theta_{\rm final}$. However, for weakly balanced and unbalanced networks, agents always asymptotically settle on an opinion of zero in Altafini's model, which is interpreted as not ``picking a side''. In Aghbolagh et al.'s model \cite{aghbolagh2020balance}, where the network connections vary endogenously, weakly balanced and unbalanced networks always turn into  strongly balanced networks when asymptotic learning is achieved. The analysis in this paper cannot test this result, because we hold $A_{ij}$ fixed throughout the simulation. Relaxing this assumption is an interesting avenue for future work.

\section{Beyond scale-free networks: \erdos and square lattice networks}
\label{sec:other_networks}

The results of Sections \ref{sec:larger_networks} and \ref{sec:balance} are derived from simulations that are run exclusively on $n=100$ \barabasi networks. A question arises: are these results of Sections \ref{sec:larger_networks} and \ref{sec:balance} unique to \barabasi networks? The aim of this section is to investigate if the structure of $n=100$ networks influences the behavior of the automaton in Section \ref{sec:dynamics_algorithm}. We make no attempt to exhaustively investigate every type of network structure for $n=100$ networks. Instead, we focus on two types of networks: the popular random \erdos \cite{erdosrenyi1959, gilbert1959} model and square lattice networks.

\begin{figure}[!ht]
    \centering
    \begin{subfigure}[b]{0.45\linewidth}
        \centering
        \includegraphics[width=\linewidth]{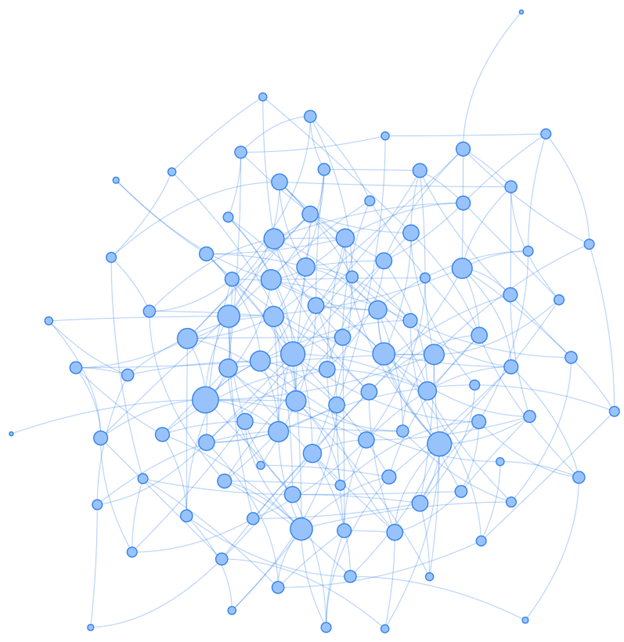}
        \caption{}
        \label{subfig:ER_graph}
    \end{subfigure}
    \begin{subfigure}[b]{0.45\linewidth}
        \centering
        \includegraphics[width=\linewidth]{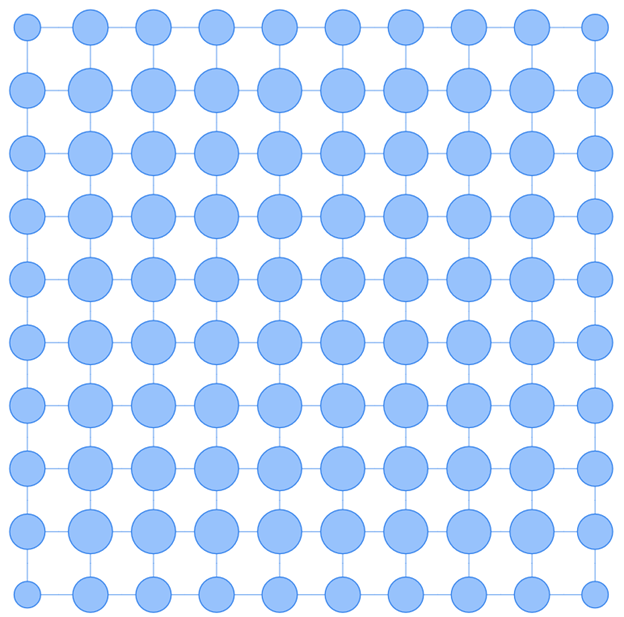}
        \caption{}
        \label{subfig:lattice_graph}
    \end{subfigure}
    \caption{The two types of networks investigated in this section. (a) A particular \erdos network generated with $n=100$ and 291 links, which is the same number of links as the \barabasi network shown in Figure \ref{fig:BA_graph}. (b) A non-periodic square lattice network with ten rows and ten columns. A periodic square lattice can be formed by connecting the agents in (b) at one end of the grid to the opposite end. In both (a) and (b), the circles represent the agents and the lines between them represent their connections. The size of a circle is proportional to the agent's degree.}
    \label{fig:other_graphs}
\end{figure}

Figure \ref{fig:other_graphs} visualizes a particular \erdos network and a non-periodic square lattice, both with $n=100$. The \erdos model operates on two assumptions: the probability that any given link exists is (i) equal to and (ii) independent of every other link. We can generate \erdos networks by either randomly selecting a specified number of links out of all the possible $n(n-1)/2$ links, or specifying the probability that each of the possible $n(n-1)/2$ links exists. Since we are concerned only with connected networks, we ignore any disconnected \erdos networks. Similar to the \barabasi model, the \erdos model generates $A_{ij}$ with entries 0 or $+1$, so further modifications are required to include $-1$ entries. By contrast, \barabasi networks (Figure \ref{fig:BA_graph}) have many agents with few connections and few agents with many connections, while the agents in \erdos networks (Figure \ref{subfig:ER_graph}) have relatively similar number of connections. In square lattice networks, all agents have exactly four links (except at the non-periodic edges), or three links if the network has two rows and 50 columns.

Using $n=100$ \erdos and square lattice networks, we investigate the behavior of opponents-only networks by studying the properties of $t_{\rm a}$, $\mean$ and the number of agents out of $n=100$ that settle at $\mean = \theta_0$ per simulation, as done in Figures \ref{fig:BA_opponents_average}, \ref{fig:BA_opponents_times_average} and Table \ref{tab:ba_right_wrong_times} from Section \ref{sec:ba_opponents}. We also investigate the properties of $\lambda$ and $t_{\rm a}$ in the context of structural balance, as done in Figures \ref{fig:balance_lambda}, \ref{fig:balance_asymp_time} and Tables \ref{tab:balance_lambda}, \ref{tab:balance_asymp_time} from Section \ref{sec:balance}. When randomly generating \erdos networks, we pick those with exactly 291 links because the \barabasi networks generated in Sections \ref{sec:larger_networks} and \ref{sec:balance} have 291 links. For the square lattice networks, we randomly pick between those with 2/4/5/10 rows and 50/25/20/10 columns. Due to the large volume of data produced, we present the results in \ref{sec:ER_lattice_stats} and summarize the main findings here. In \ref{sec:ER_lattice_stats}, we also briefly present an analysis of the dependence of the agent's behaviors on the network's diameter and clustering coefficient, in which we find no significant trends.

In general, we find no noticeable differences between the behaviors of \barabasi and \erdos networks. We also find no significant differences between the behaviors of periodic and non-periodic square lattice networks. In the opponents-only scenario, we find that, on average, six more agents in lattice networks infer the correct bias compared to \barabasi and \erdos networks. Furthermore, the agents in the lattice networks take a slightly longer time on average to achieve asymptotic learning, with $t_{\rm a}^{\rm right}$ and $t_{\rm a}^{\rm wrong}$ being $\sim 200$ and $\sim 30$ time steps slower respectively compared to both \barabasi and \erdos networks. A possible explanation may be related to the square lattice networks having fewer links compared to the \barabasi and \erdos networks. The tested periodic square lattice networks have either 200 or 150 links depending on the number of rows and columns, while the tested \barabasi and \erdos networks have 291 links. The fewer number of antagonistic links in the lattice networks mean that each agent has fewer opponents to ``lock out'' from the correct coin bias. Additionally, as argued in Sections \ref{sec:pair_opponents} and \ref{sec:ba_opponents}, having fewer opponents can lead to slower asymptotic learning times, which may explain the longer average $t_{\rm a}^{\rm right}$ and $t_{\rm a}^{\rm wrong}$ in lattice networks. A possible pathway for future work would be to systematically test if a higher number of links does correlate with fewer agents inferring the correct bias and quicker asymptotic learning for opponents-only networks.

In the context of structural balance, we find that $\sim 50\%$ and $\sim 90\%$ of strongly and weakly balanced square lattice networks respectively achieve asymptotic learning. This contrasts the behavior of \barabasi and \erdos networks, in which $\sim 90\%$ and $\sim 70\%$ of strongly and weakly balanced networks respectively achieve asymptotic learning. Moreover, strongly balanced lattice networks reach asymptotic learning somewhat slower than weakly balanced networks and much slower than \barabasi and \erdos networks. The strongly and weakly balanced lattice network has an average $t_{\rm a}$ of 2961 and 2466 time steps respectively, while both \barabasi and \erdos networks have $t_{\rm a} \approx 1600$ and 2500 time steps on average for strongly and weakly balanced networks respectively.

The behavior of lattice networks contrasts the behavior of \barabasi networks described in Section \ref{sec:balance}, in which strongly balanced networks are the most likely to achieve asymptotic learning, and they do so the quickest. For lattice networks, we instead find that weakly balanced networks are the most likely to achieve asymptotic learning, and they do so the quickest. However, we note that the main purpose of the model described in Sections \ref{sec:external_signal}--\ref{sec:internal_signals} is to emulate the behavior of human agents in a network, and square lattice networks may not be the best representative choice of a realistic network of human agents. Nonetheless, further scrutiny of lattice networks may provide useful insights in future work.

\section{Discussion: social implications}
\label{sec:discussion}

In Sections \ref{sec:smaller_networks}--\ref{sec:other_networks}, we investigate the behavior of a network of agents according to the model described in Sections \ref{sec:external_signal}--\ref{sec:internal_signals} under a variety of scenarios. We find that the networked agents display numerous interesting behaviors, some of which are not predicted by deterministic models. It must be emphasized again that the model described in Sections \ref{sec:external_signal}--\ref{sec:internal_signals} is highly idealized and cannot be expected to capture accurately the many psychological vagaries of human opinion formation. Nevertheless, with due reserve, we advance a small number of possible social implications which may flow from the results in Sections \ref{sec:external_signal}--\ref{sec:internal_signals}, for consideration by readers with expertise in the social sciences.

In Section \ref{sec:pair_opponents} we study the behavior of a pair of political opponents in inferring the bias of a coin. We find that the opponents behave counterintuitively: the agent who infers the wrong coin bias tends to do so quicker than the agent who correctly infers the true coin bias. To the authors' knowledge, this tendency to reach the wrong conclusion first is not predicted in deterministic models. It implies that political antagonism can cause agents to quickly and confidently assert their stance. The tendency to reach the wrong conclusion first is reminiscent of the ``backfire effect'' in social science, where a person who is exposed to information that is contradictory to their beliefs may instead become more confident in their initial beliefs \cite{nyhan2010backfire}. The results of Section \ref{sec:pair_opponents} also raise the possibility of an interesting second-round effect: cynical agents, who understand the network dynamics, may proactively form specific relationships (in alliance or opposition) with specific other agents in order to manipulate opinion formation by blocking others strategically from correct conclusions, e.g. by intentionally applying the Bayesian ``lock out'' dynamics described in Sections \ref{sec:pair_opponents} and \ref{sec:triad_unbalanced}. The study of this second-round effect goes beyond the scope of this paper, as it requires $A_{ij}$ to evolve with time, but the results in Section \ref{sec:pair_opponents} confirm that it can exist in principle.

The priors of the agents are not the sole factor governing who infers the correct coin bias, as shown in Section \ref{sec:pair_opponents}. Given fixed priors in a pair of opponents, a change in the sequence of coin tosses can change who correctly infers the coin bias. This implies that the first broadcast messages of a media organization are more influential in shaping the opinions of political opponents than subsequent messages, consistent with some research in social sciences \cite{holbrook2001primacy, druckman_fein_leeper_2012}. It is unclear at this stage whether the power of first impressions is a result of individual human psychology or a network effect in realistic human systems.

In the opponents-only networks studied in Sections \ref{sec:pair_opponents}, \ref{sec:ba_opponents} and \ref{sec:ba1000_opponents}, the networked agents always achieve asymptotic learning, which differs from the predictions of some deterministic models \cite{martins2010mass, shi2016evolution} where some opponents may experience turbulent nonconvergence. Instead, when dealing with Bayesian learners, a mixture of allies and opponents in the network is required to produce turbulent nonconvergence. One insight we can glean is that, under certain circumstances, the cooperative instincts of individual agents can promote indecision and uncertainty. If an agent strives to agree with every ally, but the allies are themselves opposed, the beliefs of the cooperative agent may fluctuate persistently --- and they may consequently drive fluctuations in the beliefs of other agents through second-round, knock-on effects. A related social implication is that intermittency --- discussed in Section \ref{sec:intermittency} and Figure \ref{fig:ba_intermittency}, generalizing previous deterministic work \cite{shi2016evolution} --- is partly a network effect. One may be tempted to attribute changing one's mind ``randomly'' to individual human psychology, but Section \ref{sec:balance} demonstrates that the network plays an important role, if agents are susceptible to peer pressure in the manner modeled by the automaton in Section \ref{sec:dynamics_algorithm}. Social programs to promote stability (and perhaps conformity) in opinions about factual matters (like the coin bias) through political allegiances (e.g. us-and-them messaging on social media) are therefore unlikely to succeed in practice, when the network comprises a mixture of allies and opponents. This collective aspect of intermittency will be studied further in future work.

In the context of structural balance, as studied in Section \ref{sec:balance}, we find that strongly balanced networks are the least likely to experience turbulent nonconvergence. However, it is somewhat counterintuitive that turbulent nonconvergence can occur in strongly balanced networks, as every agent by definition belongs to either one of the two groups where there is no antagonism between agents of the same group. Regardless, some agents of the same group may occasionally stubbornly disagree with each other due to antagonistic interactions with agents from the other group. A social implication is that, even though members of the same organization attempt to cooperate towards the same goal, disagreements may still arise due to different initial beliefs and different experiences in interacting with the opposing group.

We note that there are shortcomings in the analysis performed in Sections \ref{sec:smaller_networks}--\ref{sec:other_networks}. For example, we assume that the priors $x_i(t=0, \theta)$ of each agent are independent of their connections $A_{ij}$. This may not be a realistic assumption. For example, people who are part of the same social club may have relatively similar priors. Additionally, we assume that the network structure $A_{ij}$ never changes, but in real life, it is not unusual for a person's relationships to change. Furthermore, we assume that the external signal is noiseless and unfiltered, but there is no reason to believe that a media product is interpreted by all of its consumers in exactly the same manner. Relaxing these assumptions is an interesting avenue for future work.

\section{Conclusion}

In this paper, we map the real-world problem of a network of agents inferring media bias onto the idealized problem of inferring the bias of a coin. Many opinion dynamics models, such as the classic DeGroot \cite{degroot1974reaching} and Deffuant-Weisbuch \cite{deffuant2000mixing} models, consider opinions to be deterministic. We propose a probabilistic opinion dynamics model with a new Bayesian update rule that encapsulates the multi-faceted nature of opinions and the frustrated dynamics of networks that contain a mixture of allies and opponents. In this model, agents observe coin tosses and communicate with their allies and opponents to progressively refine their estimates of the coin bias, which obeys a probability density function. In other words, agents allow for uncertainty in their beliefs in this paper, complementing previous deterministic analysis.

The main findings of the paper are as follows. (i) In networks whose agents are all mutually opposed, we find that confidence in the wrong conclusion is gained quicker than in the right conclusion $96\%$ and $76\%$ of the time for networks with $n=2$ and $n=100$ (Barabási–Albert) respectively. This peculiar tendency to reach the wrong conclusion first is not reported in deterministic models. It occurs because agents attempt to minimize the overlap of their probability distribution functions with their opponents' and hence ``lock out'' certain ranges of beliefs.  (ii) In networks with a mixture of allies and opponents, we find that some of the agents experience turbulent nonconvergence. Specifically we find that in unbalanced triads, as defined in Davis' extension of structural balance theory \cite{davis1967clustering}, an agent who is allied to two mutually opposed agents always experiences turbulent nonconvergence. For randomized unbalanced $n=100$ \barabasi networks, $65\%$ of agents experience turbulent nonconvergence per simulation on average, whereas for strongly and weakly balanced networks, $1\%$ of agents on average experience turbulent nonconvergence. Additionally, $2\%$, $28\%$ and all of strongly balanced, weakly balanced and unbalanced $n=100$ \barabasi networks experience turbulent nonconvergence, consistent with structural balance theory \cite{davis1967clustering}. When asymptotic learning is reached, strongly balanced networks tend to do so quicker than weakly balanced networks. However, the behaviors of agents in square lattice networks are inconsistent with the predictions of structural balance theory: strongly balanced networks are less likely to achieve asymptotic learning compared to weakly balanced networks, and when asymptotic learning is achieved, they do so more slowly. (iii) We observe long-term intermittency in an unbalanced $n=100$ \barabasi network, where an agent cycles between ``eras'' of constant and fluctuating beliefs, and where each era lasts for many iterations. (iv) Turbulent nonconvergence only occurs in networks with a mix of allies and opponents. It does not occur in opponents-only networks, which contrasts some deterministic models \cite{martins2010mass, shi2016evolution}. (v) Complementing deterministic models, the model in this paper contains additional information about how the uncertainty in beliefs evolves. The standard deviation of an agent's beliefs approximately follows a power law in time and tends to zero at asymptotic learning in allies-only networks. In opponents-only networks, the standard deviations do not follow a power-law, but they also tend to zero. Our model also allows for complex, multi-modal belief structures that cannot be easily described by deterministic models, such as a person who beliefs that a media organization is either left or right wing, but not politically unbiased.

A key goal for future work is to formulate falsifiable tests of the idealized model in this paper using actual data from real-world networks. Ingredients may include: (i) quantitative measurements of the intrinsic bias in the published outputs of broadcast media; and (ii) real-time tracking of the evolution of beliefs as revealed by social media accounts. Automated analysis can be used to extract an estimate of the bias of a media organization (see \cite{Hamborg2019} for a review). Substantial efforts have been made to track the beliefs of social media users in real-time \cite{wang2012system, guerra2011sentiment, vilares2015sentiment, goel2016sentiment, DRAGONI20191103}, although they usually provide point estimates of the users' beliefs. Additionally, it would be interesting to extend the model to directed networks. There are many realistic examples of directed networks, such as the interactions between an opinion leader and their followers.

\section*{Acknowledgements}

Parts of this research are supported by the University of Melbourne Science Graduate Scholarship - 2020, the Australian Research Council (ARC) Centre of Excellence for Gravitational Wave Discovery (OzGrav) (project number CE170100004) and ARC Discovery Project DP170103625. 

\bibliographystyle{elsarticle-num}
\bibliography{ref.bib}

\appendix

\section{Relation to Fang et al.'s model}
\label{apendix:fang}

The model in this paper resembles the model advanced by Fang et al. \cite{fang2019social} in some respects and differs from it in others. We summarize the similarities and differences in this appendix for completeness. In Fang et al.'s model, the network topology is dynamic, unlike in the present paper. The adjacency matrix follows an update rule that depends on the difference of beliefs between the agent and its neighbor. Specifically, given $\lambda_{ij} = \text{max}_\theta|x_i(t,\theta) - x_j(t,\theta)|$ and two threshold values, $p<q$, agent $i$ categorizes other agents into three possible sets, the ``closer neighbor'' set ${\sc S}^{\rm C}_i(t)$, the ``farther neighbor'' set ${\sc S}^{\rm F}_i(t)$ and the ``not neighbor'' set ${\sc S}^{\rm N}_i(t)$:
    
\begin{align}
    {\sc S}^{\rm C}_i(t) &= \{ j | \lambda_{ij} \leq p \} \\
    {\sc S}^{\rm F}_i(t) &= \{ j | p < \lambda_{ij} < q \} \\
    {\sc S}^{\rm N}_i(t) &= \{ j | j \text{ not  in ${\sc S}^{\rm C}_i(t)$ or ${\sc S}^{\rm F}_i(t)$} \}
\end{align}
    
Additionally, an agent only considers another agent as its neighbor if, in the previous time step, the other agent is a neighbor or a neighbor's neighbor. The initial $A_{ij}(t=0)$ is specified. $A_{ij}$ is then updated as follows:
    
\begin{equation}
\label{eq:fang_update}
    A_{ij}(t+1) = 
    \begin{dcases}
        A_{ij}(t) + \Delta A_i(t)&, \text{if }\; j \in {\sc S}^{\rm C}_i(t) \; \text{ or }\; i=j\\
        rA_{ij}(t) &, \text{if }\; j \in {\sc S}^{\rm F}_i(t) \\
        0 &, \text{otherwise.}
    \end{dcases}
\end{equation}
    
\noindent In (\ref{eq:fang_update}), 

\begin{equation}
    \Delta A_i(t)=\frac{\sum_{j \in {\sc S}_i^C(t)} A_{ij}(t) + \sum_{j \in {\sc S}^{\rm F}_i(t)} (1 - r) A_{ij}(t) }{c_i(t) + 1}
\end{equation}

\noindent increases the weight of the ``closer neighbors'', $0<r<1$ reduces the weight of the ``farther neighbors'' and $c_i(t)$ is the number of agents in the set ${\sc S}^{\rm C}_i(t)$. Note that the dynamics of $A_{ij}(t)$ in Fang et al.'s model are endogenous.

The external signals in \cite{fang2019social} are private, unlike in the present paper, where the outcomes of the coin tosses are broadcast publicly. Applying the private protocol in \cite{fang2019social} to the model described in Sections \ref{sec:external_signal}--\ref{sec:internal_signals}, we would have a situation where each agent would possess their own personal coin with a true bias that may differ from the coins of other agents. In \cite{fang2019social}, each agent only observes the coin tosses of their own coin and cannot observe the coin tosses of other agents, but the agents have full access to the opinions of their allies, like the model described in Sections \ref{sec:external_signal}--\ref{sec:internal_signals}. Furthermore, \cite{fang2019social} does not account for antagonistic interactions.

\section{Validation exercise: consensus in allies-only networks with $n=100$}
\label{sec:ba_allies}

\begin{figure}[!ht]
    \centering
    \begin{subfigure}[b]{0.45\linewidth}
        \centering
        \includegraphics[width=\linewidth]{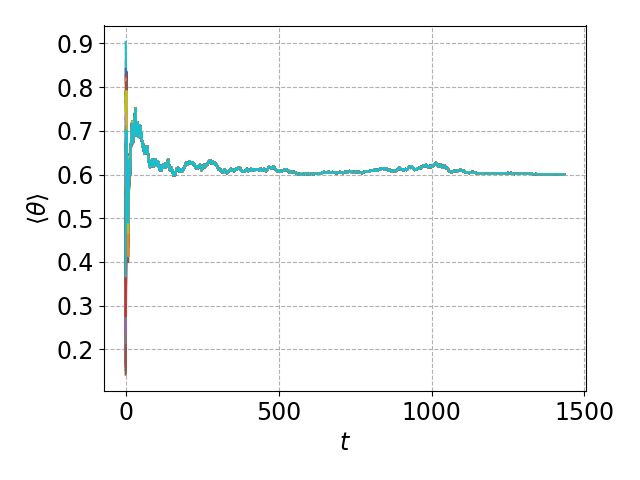}
        \caption{}
    \end{subfigure}
    \begin{subfigure}[b]{0.45\linewidth}
        \centering
        \includegraphics[width=\linewidth]{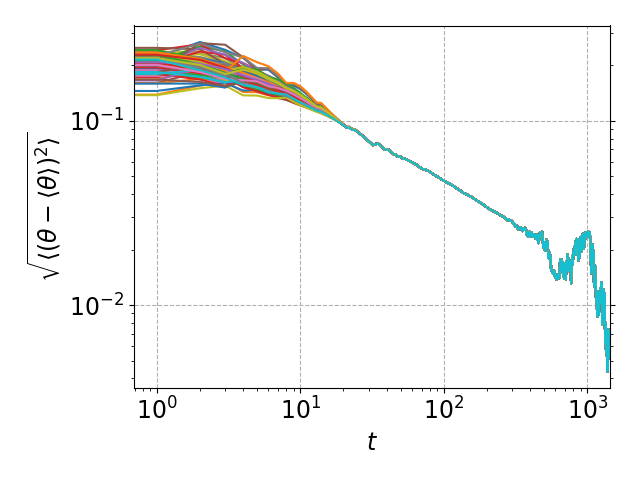}
        \caption{}
        \label{subfig:BA_allies_stddev}
    \end{subfigure}
    \caption{Opinion formation in the \barabasi network in Figure \ref{fig:BA_graph} with allies only, i.e. $A_{ij} \geq 0$ for all $i$ and $j$, to be compared with the allies-only study in Section \ref{sec:pair_ally} and Figure \ref{fig:pair_ally}. Evolution for $t \leq t_{\rm a} = 1433$ of the (a) mean $\mean$ and (b) standard deviation $\stddev$ of the beliefs of each agent, plotted as differently colored curves for each agent. The curves are indistinguishable by eye. Parameters: $n=100$, $m=3$, $\theta_0=0.6$, $\mu=0.25$, $A_{ij} \in \{0, +1\}$ for all $1 \leq i,j \leq n$.}
    \label{fig:BA_allies}
\end{figure}

In this section, we generalize the $n=2$ study in Section \ref{sec:pair_ally} to $n=100$ agents, all of whom are allies. Specifically, we set $A_{ij}=1$ for all nonzero entries in the \barabasi network shown in Figure \ref{fig:BA_graph}. We randomize the priors of each agent (as described in Section \ref{sec:dynamics_algorithm}), but we use the same sequence of coin tosses as the $n=2$ network studied in Section \ref{sec:pair_ally}. Figure \ref{fig:BA_allies} shows the evolution of $\mean$ and $\stddev$ of each agent for a particular realization. Each panel in Figure \ref{fig:BA_allies} overplots 100 curves in different colors, but they lie on top of one another and are indistinguishable by eye. At $t=t_{\rm a}$, the beliefs of each agent peak at $\theta=\theta_0$, with $x_i(t=t_{\rm a}, \theta=\theta_0)=1$ and $x_i(t=t_{\rm a}, \theta \neq \theta_0) \ll 1$.

The beliefs of the agents converge, as defined by equation (\ref{eq:belief_convergence}) with $\epsilon=10^{-3}$, in 60 time steps, which is slower than the pair of allies in Section \ref{sec:pair_ally}, who converge within 9 time steps. An intuitive reason is that, for $n=100$ with randomized priors, an agent is initially exposed to many different beliefs. An agent averages the beliefs of its allies, so the overall strength of the internal signal is weaker compared to a $n=2$ network, which causes slower convergence. Additionally, because of the existence of $A_{ij}=0$, agents cannot directly access the beliefs of every other agent, which introduces a further delay. At $t>60$, the external signals dominate the evolution of $\mean$ and $\stddev$ for each agent, so $\mean$ and $\stddev$ follow trajectories resembling those in Figure \ref{fig:pair_ally}. A least square fit to $\stddev$ for $50<t<500$ shows that it scales roughly as $t^{-0.47}$ for the first 500 time steps, similar to $t^{-0.46}$ in Figure \ref{subfig:pair_ally_std}.

There are no other notable quantitative or qualitative differences between the $n=100$ and $n=2$ networks for allies only. The discussions about $\stddev$ and the comparisons with deterministic models in Section \ref{sec:pair_ally} also apply to the $n=100$ network studied here.

\section{Even larger scale-free networks with $n=1000$}
\label{section:1000_network}

The aim of this section is to compare the behavior of a $n=100$ \barabasi network studied in Section \ref{sec:larger_networks} with a $n=1000$ \barabasi network, just to check that the case $n=100$ is free of any small-$n$ peculiarities. Due to the computational costs of running a simulation on a $n=1000$ network, we run three simulations on a \barabasi network with $n=1000$ and attachment parameter $m=3$ under three different scenarios: all-allies, all-opponents and an equal mixture of allies and opponents. We keep the same priors, coin tosses and $|A_{ij}|$ between the three scenarios, and only vary the sign of $A_{ij}$. For consistency, we use the same sequence of coin tosses as the $n=2$ network studied in Section \ref{sec:pair_ally}.

\subsection{Allies only: consensus}

\begin{figure}[!ht]
    \centering
    \begin{subfigure}[b]{0.45\linewidth}
        \centering
        \includegraphics[width=\linewidth]{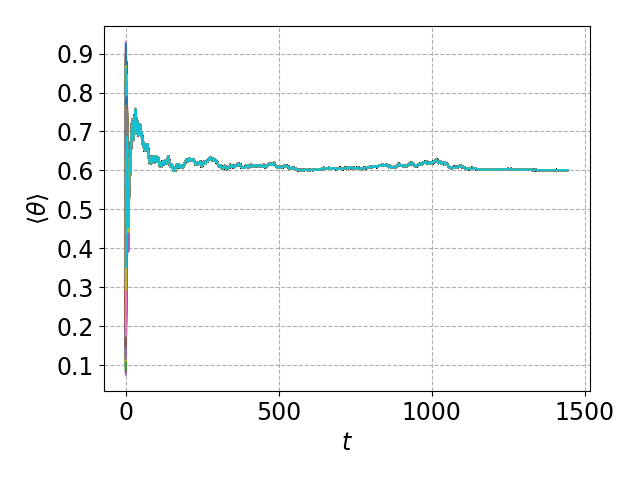}
        \caption{}
    \end{subfigure}
    \begin{subfigure}[b]{0.45\linewidth}
        \centering
        \includegraphics[width=\linewidth]{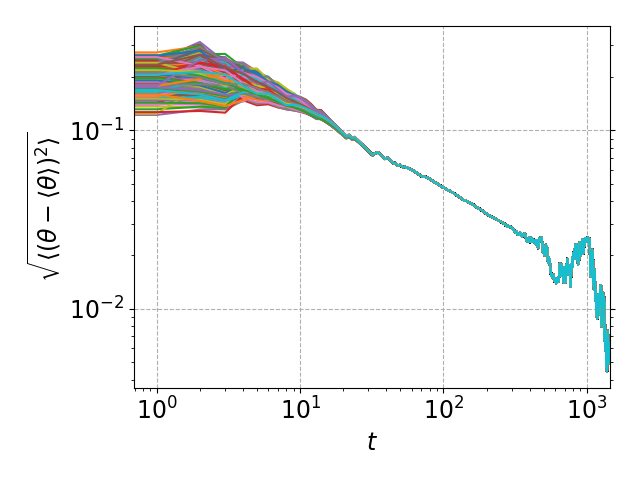}
        \caption{}
    \end{subfigure}
    \caption{Opinion formation in a $n=1000$ \barabasi network with allies only, i.e. $A_{ij} \geq 0$ for all $i$ and $j$, to be compared with the $n=100$ allies-only study in \ref{sec:ba_allies} and Figure \ref{fig:BA_allies}. Evolution for $t \leq t_{\rm a} = 1445$ of the (a) mean $\mean$ and (b) standard deviation $\stddev$ of the beliefs of each agent, plotted as differently colored curves for each agent. The curves are indistinguishable by eye. Parameters: $n=1000$, $m=3$, $\theta_0=0.6$, $\mu=0.25$, $A_{ij} \in \{0, +1\}$ for all $1 \leq i,j \leq n$.}
    \label{fig:BA1000_allies}
\end{figure}

To compare with the $n=2$ and $n=100$ allies-only networks shown in Figures \ref{fig:pair_ally} and \ref{fig:BA_allies} respectively, we start by looking at a $n=1000$ \barabasi network with $A_{ij}=1$ for all nonzero entries. Figure \ref{fig:BA1000_allies} shows the evolution of $\mean$ and $\stddev$ of each agent for a particular realization. Each panel in Figure \ref{fig:BA_allies} overplots 1000 curves in different colors, but they lie on top of one another and are indistinguishable by eye.

We find that the agents in the allies-only $n=1000$ network display identical behaviors to the agents in the allies-only $n=100$ and $n=2$ networks. At $t=t_{\rm a}$, the beliefs of each agent peak at $\theta=\theta_0$, with $x_i(t=t_{\rm a}, \theta=\theta_0)=1$ and $x_i(t=t_{\rm a}, \theta \neq \theta_0) \ll 1$. The beliefs of the agents converge, as defined by equation (\ref{eq:belief_convergence}) with $\epsilon=10^{-3}$, in 64 time steps, which is similar to the allies-only $n=100$ network in Section \ref{sec:pair_ally}, which converges within 60 time steps. A least square fit to $\stddev$ for $50<t<500$ shows that it scales roughly as $t^{-0.47}$ for the first 500 time steps, identical to the allies-only $n=2$ and $n=100$ networks in Figures \ref{subfig:pair_ally_std} and \ref{subfig:BA_allies_stddev} respectively. Hence, the conclusions of the $n=100$ allies-only network studied in \ref{sec:ba_allies} can also be applied here.

\subsection{Opponents only: reaching the wrong conclusion first}
\label{sec:ba1000_opponents}

\begin{figure}[!ht]
    \centering
    \begin{subfigure}[b]{0.49\linewidth}
        \centering
        \includegraphics[width=\linewidth]{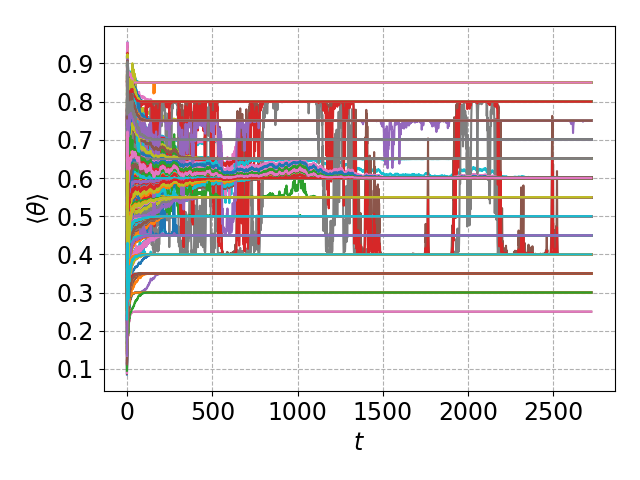}
        \caption{}
    \end{subfigure}
    \begin{subfigure}[b]{0.49\linewidth}
        \centering
        \includegraphics[width=\linewidth]{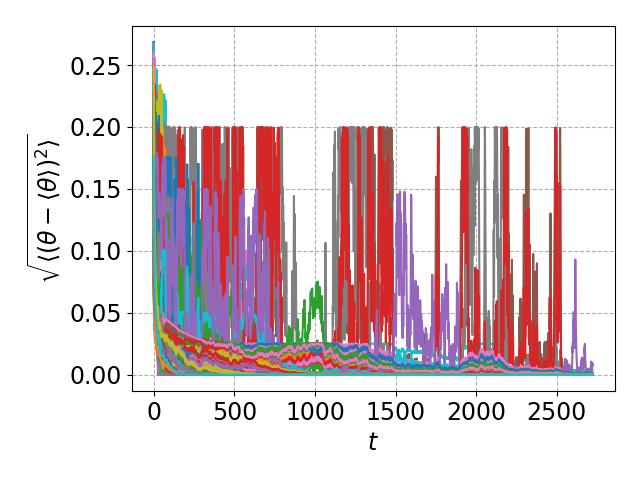}
        \caption{}
        \label{subfig:BA1000_opponents_stddev}
    \end{subfigure}
    \caption{Opinion formation in a $n=1000$ \barabasi network with opponents only, i.e. $A_{ij} \leq 0$ for all $i$ and $j$, to be compared with the $n=100$ opponents-only study in Section \ref{sec:ba_opponents} and Figure \ref{fig:BA_opponents}. Evolution for $t \leq t_{\rm a} = 2722$ of the (a) mean $\mean$ and (b) standard deviation $\stddev$ of the beliefs of each agent, plotted as differently colored curves for each agent. Most of the curves are indistinguishable by eye. Parameters: $n=1000$, $m=3$, $\theta_0=0.6$, $\mu=0.25$, $A_{ij} \in \{0, -1\}$ for all $1 \leq i,j \leq n$.}
    \label{fig:BA1000_opponents}
\end{figure}

Now, we look at the opposite extreme scenario, an opponents-only network with $A_{ij}=-1$ for all nonzero entries. Figure \ref{fig:BA1000_opponents} shows the evolution of $\mean$ and $\stddev$ of each agent for a particular realization. Each panel in Figure \ref{fig:BA1000_opponents} overplots 1000 curves in different colors, but many lie on top of one another and are indistinguishable by eye.

In the particular realization shown in Figure \ref{fig:BA1000_opponents}, the beliefs of the agents settle in the range $0.25 \leq \mean \leq 0.85$, which is similar to the $n=100$ case in Section \ref{sec:ba_opponents}. The shape of the histogram of $\mean$ at $t_{\rm a}$ for $n=1000$ is similar to the $n=100$ scenario shown in Figure \ref{subfig:BA_opponents_stddev}. The agents in the $n=1000$ network achieve asymptotic learning at $t_{\rm a}=2722$, which is significantly slower than the $n=100$ network with $t_{\rm a}=1595$. However, we note that an asymptotic learning time of $t_{\rm a}=2722$ is not unusual for $n=100$ networks, as seen in the histogram of $t_{\rm a}^{\rm right}$ in Figure \ref{subfig:BA_opponents_hist_averaged}, so we cannot say with certainty that $n=1000$ networks are generally slower to reach asymptotic learning. In Figure \ref{fig:BA1000_opponents}, 246 out of $n=1000$ agents infer the correct bias and there are 30 unique pairs of opponents that agree with each other, with $\meanSub{i}=\meanSub{j}$. Only one of these 30 pairs of opponents infer the correct coin bias. As seen in Figure \ref{subfig:BA_opponents_num_correct}, it is common for $\approx 25\%$ of agents in $n=100$ opponents-only \barabasi networks to infer the correct bias, so the observed number of agents who infer the correct bias in the $n=1000$ case is not significantly different. Overall, we find no significant difference between the behaviors of $n=1000$ and $n=100$ opponents-only \barabasi networks.

\subsection{Equal mix of allies and opponents: turbulent nonconvergence}

\begin{figure}[!ht]
    \centering
    \begin{subfigure}[b]{0.49\linewidth}
        \centering
        \includegraphics[width=\linewidth]{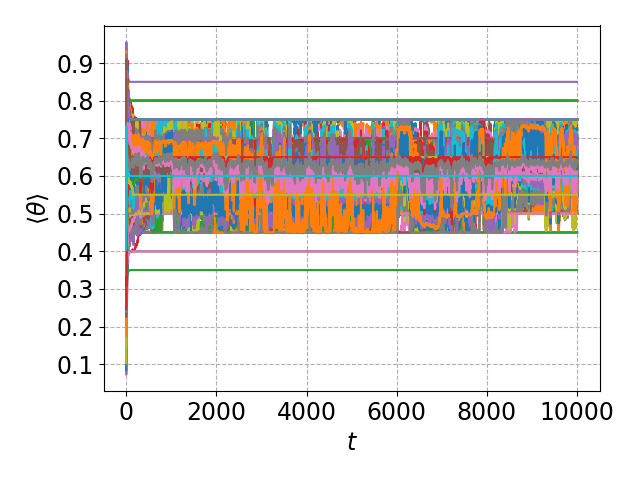}
        \caption{}
    \end{subfigure}
    \begin{subfigure}[b]{0.49\linewidth}
        \centering
        \includegraphics[width=\linewidth]{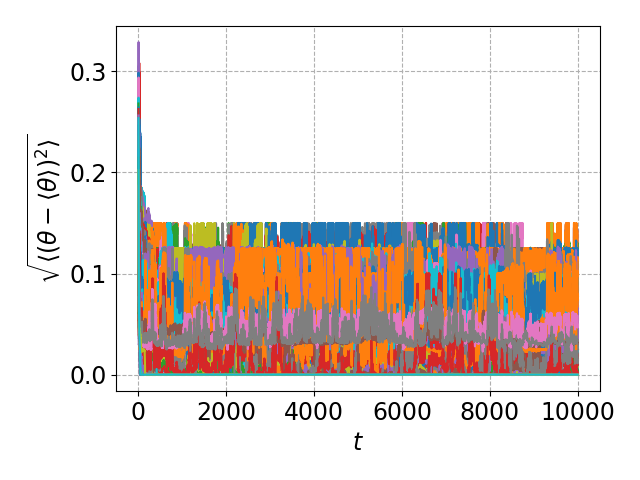}
        \caption{}
        \label{subfig:BA1000_mixed_stddev}
    \end{subfigure}
    \caption{Opinion formation in a $n=1000$ \barabasi network with an equal mix of allies and opponents, to be compared with the $n=100$ equal mix of allies and opponents study in Section \ref{sec:ba_mixed} and Figure \ref{fig:BA_mixed}. Evolution for $t \leq 10^3$ of the (a) mean $\mean$ and (b) standard deviation $\stddev$ of the beliefs of each agent, plotted as differently colored curves for each agent. Most of the curves are indistinguishable by eye. Parameters: $n=1000$, $m=3$, $\theta_0=0.6$, $\mu=0.25$, $A_{ij} \in \{0, -1\}$ for all $1 \leq i,j \leq n$.}
    \label{fig:BA1000_mixed}
\end{figure}

Here, we look at the last scenario considered in the $n=100$ \barabasi networks studied in Section \ref{sec:larger_networks}: an equal mix of allies and opponents. Figure \ref{fig:BA1000_mixed} shows the evolution of $\mean$ and $\stddev$ of each agent for a particular realization. Each panel in Figure \ref{fig:BA1000_mixed} overplots 1000 curves in different colors, but many lie on top of one another and are indistinguishable by eye.

In the particular realization shown in Figure \ref{fig:BA1000_mixed}, turbulent nonconvergence is observed for 697 agents, while 303 agents achieve asymptotic learning. Out of those 303 agents, only 69 agents infer the correct bias. In terms of the proportion of agents, the results for $n=1000$ are similar to $n=100$ (see Figure \ref{fig:BA_mixed}), where 32 agents achieve asymptotic learning and eight out of those 32 agents successfully infer the correct bias. For the $n=1000$ network, Figure \ref{subfig:BA1000_mixed_stddev} shows that the standard deviations of the beliefs of all agents at $t > 1000$ are constrained to $\stddev \leq 0.15$, a range which is slightly wider than $\stddev \leq 0.10$  for $n=100$ (see Section \ref{sec:ba_mixed}). However, with only one simulation for the $n=1000$ network, we cannot say if $n=1000$ generally has a wider spread of $\stddev$ compared to $n=100$ networks. Overall, we find no significant differences between the behaviors of $n=1000$ and $n=100$ \barabasi networks with an equal mix of allies and opponents.

\section{\erdos and square lattice networks}
\label{sec:ER_lattice_stats}

We aim to repeat the tests of Sections \ref{sec:larger_networks} and \ref{sec:balance} on \erdos and square lattice networks in order to verify that the topology of the network does not alter qualitatively the dynamics of opinion formation. Specifically, we investigate the behavior of opponents-only networks by studying the properties of $t_{\rm a}$, $\mean$ and the number of agents out of $n=100$ that settle at $\mean = \theta_0$ per simulation, as done in Figures \ref{fig:BA_opponents_average}, \ref{fig:BA_opponents_times_average} and Table \ref{tab:ba_right_wrong_times} from Section \ref{sec:ba_opponents}. We also investigate the properties of $\lambda$ and $t_{\rm a}$ in the context of structural balance, as done in Figures \ref{fig:balance_lambda}, \ref{fig:balance_asymp_time} and Tables \ref{tab:balance_lambda}, \ref{tab:balance_asymp_time} from Section \ref{sec:balance}. We find no significant qualitative differences between periodic and non-periodic square lattice networks, so we do not present the results of non-periodic square lattice networks.

\subsection{Opponents only: reaching the wrong conclusion first}
\label{sec:ER_lattice_stats_opponents}

\begin{figure}[h!tbp]
    \centering
    \begin{subfigure}[b]{0.49\linewidth}
        \centering
        \includegraphics[width=\linewidth]{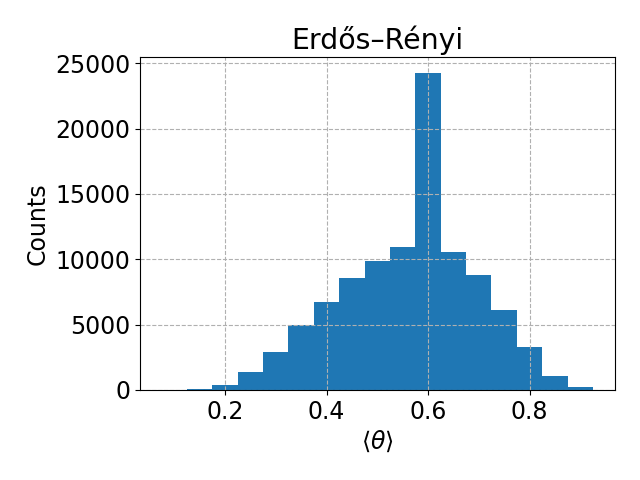}
        \caption{}
        \label{subfig:ER_opponents_means}
    \end{subfigure}
    \begin{subfigure}[b]{0.49\linewidth}
        \centering
        \includegraphics[width=\linewidth]{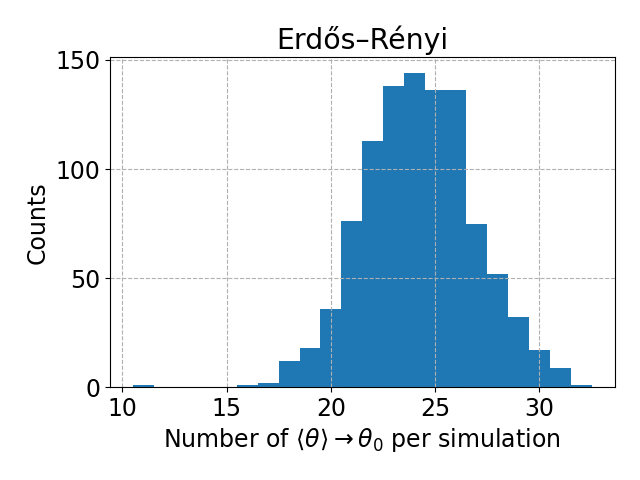}
        \caption{}
        \label{subfig:ER_opponents_num_correct}
    \end{subfigure}
    \begin{subfigure}[b]{0.49\linewidth}
        \centering
        \includegraphics[width=\linewidth]{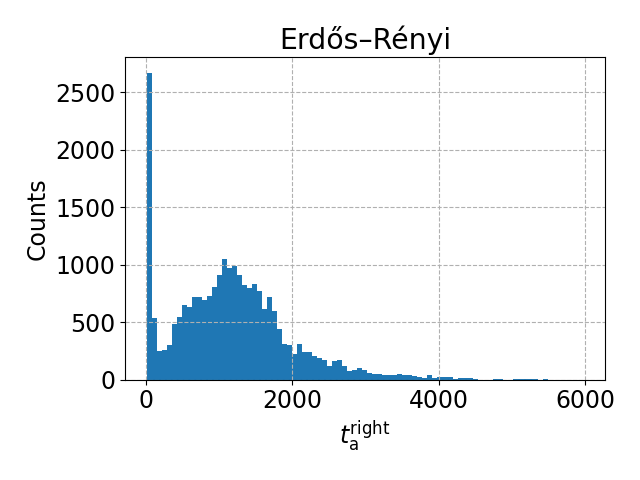}
        \caption{}
        \label{subfig:ER_opponents_correct_time}
    \end{subfigure}
    \begin{subfigure}[b]{0.49\linewidth}
        \centering
        \includegraphics[width=\linewidth]{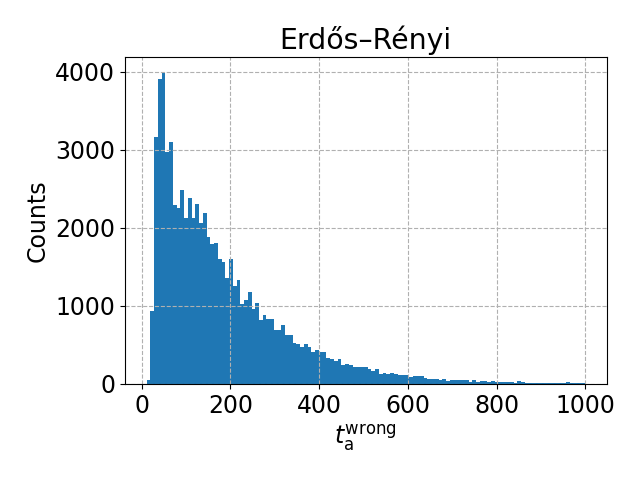}
        \caption{}
        \label{subfig:ER_opponents_incorrect_time}
    \end{subfigure}
    \caption{Opponents-only $n=100$ \erdos networks: distribution of beliefs once every agent achieves asymptotic learning, and the time taken to reach asymptotic learning, to be compared with Figures \ref{fig:BA_opponents_average} and \ref{fig:BA_opponents_times_average} (a) Histogram of $\mean$. (b) Histogram of the number of agents out of $n=100$ attaining $\mean \rightarrow \theta_0$. (c) Histogram of $t_{\rm a}^{\rm right}$. (d) Histogram of $t_{\rm a}^{\rm wrong}$.}
    \label{fig:ER_opponents}
\end{figure}

\begin{figure}[h!tbp]
    \centering
    \begin{subfigure}[b]{0.49\linewidth}
        \centering
        \includegraphics[width=\linewidth]{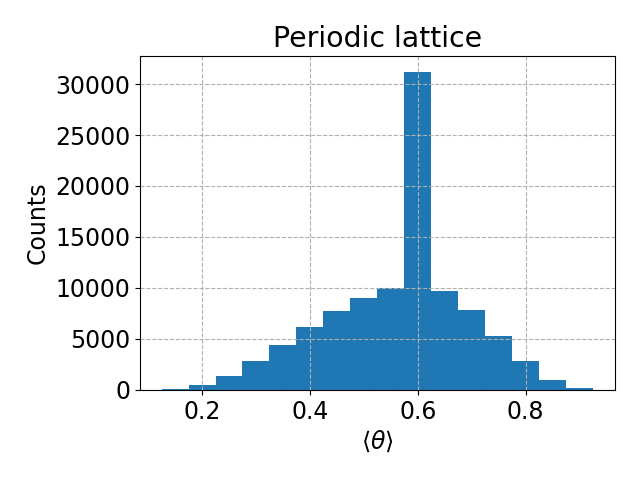}
        \caption{}
        \label{subfig:lattice_opponents_means}
    \end{subfigure}
    \begin{subfigure}[b]{0.49\linewidth}
        \centering
        \includegraphics[width=\linewidth]{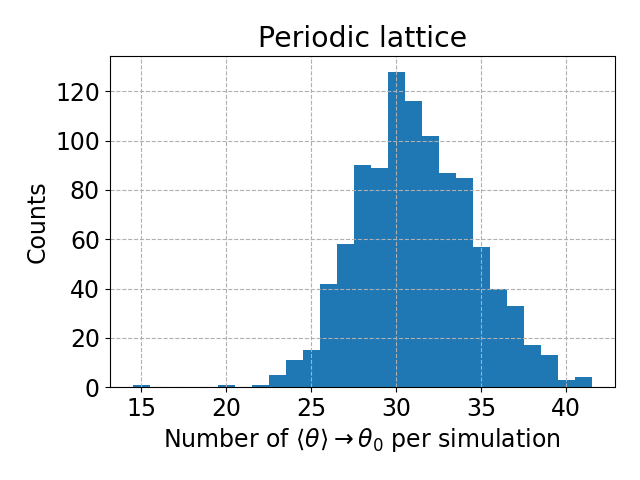}
        \caption{}
        \label{subfig:lattice_opponents_num_correct}
    \end{subfigure}
    \begin{subfigure}[b]{0.49\linewidth}
        \centering
        \includegraphics[width=\linewidth]{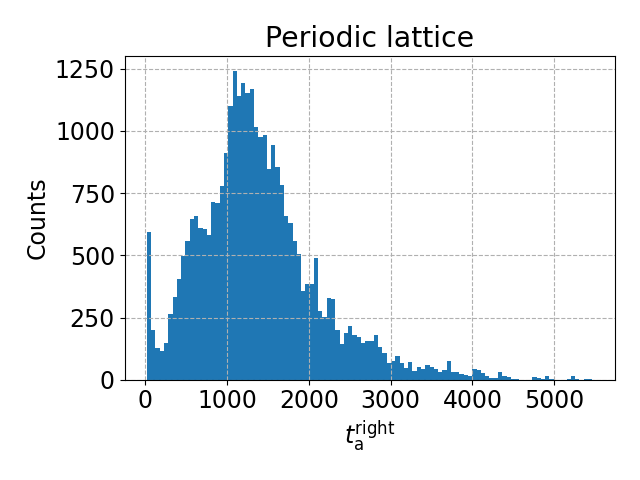}
        \caption{}
        \label{subfig:lattice_opponents_correct_time}
    \end{subfigure}
    \begin{subfigure}[b]{0.49\linewidth}
        \centering
        \includegraphics[width=\linewidth]{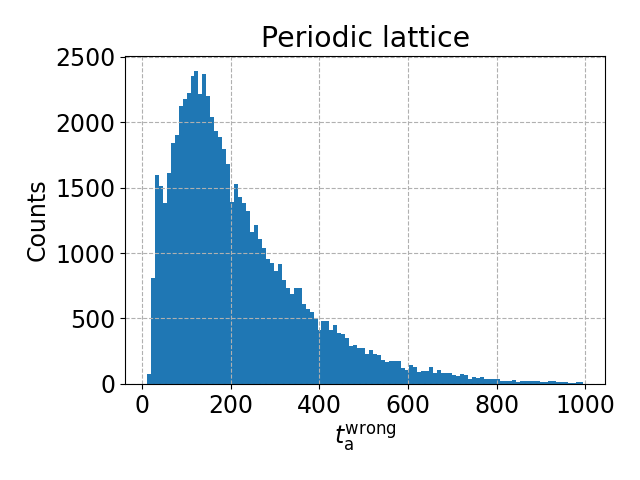}
        \caption{}
        \label{subfig:lattice_opponents_incorrect_time}
    \end{subfigure}
    \caption{Opponents-only $n=100$ periodic square lattice networks: distribution of beliefs once every agent achieves asymptotic learning, and the time taken to reach asymptotic learning, to be compared with Figures \ref{fig:BA_opponents_average} and \ref{fig:BA_opponents_times_average} (a) Histogram of $\mean$. (b) Histogram of the number of agents out of $n=100$ attaining $\mean \rightarrow \theta_0$. (c) Histogram of $t_{\rm a}^{\rm right}$. (d) Histogram of $t_{\rm a}^{\rm wrong}$.}
    \label{fig:lattice_opponents}
\end{figure}

First, we study the behavior of $n=100$ opponents-only \erdos and square lattice networks in order to draw comparisons with the \barabasi network studied in Section \ref{sec:ba_opponents}. Figures \ref{fig:ER_opponents} and \ref{fig:lattice_opponents} shows the histograms of $\mean$, the number of agents out of $n=100$ attaining $\mean \rightarrow \theta_0$ per simulation, $t_{\rm a}^{\rm right}$ and $t_{\rm a}^{\rm wrong}$ for \erdos and periodic square lattice networks respectively, to be compared with Figures \ref{fig:BA_opponents_average} and \ref{fig:BA_opponents_times_average}. Table \ref{tab:other_right_wrong_times} shows the summary statistics of $t_{\rm a}^{\rm right}$ and $t_{\rm a}^{\rm wrong}$ for both \erdos and periodic square lattice networks, to be compared with Table \ref{tab:ba_right_wrong_times}.

\begin{table}[h!tb]
\centering
\begin{tabular}{@{}lrrrr@{}}
\toprule
 &
  
  \multicolumn{2}{c}{\erdos} &
  \multicolumn{2}{c}{Square lattice} \\
Property of $t_{\rm a}$ &
  $t_{\rm a}^{\rm right}$ &
  $t_{\rm a}^{\rm wrong}$ &
  $t_{\rm a}^{\rm right}$ &
  $t_{\rm a}^{\rm wrong}$ \\ \midrule
Mean (time steps)             & 1182  & 206   & 1393  & 241   \\
Standard deviation (time steps) & 818   & 264   & 756   & 263   \\
First quartile (time steps)      & 608   & 73    & 913   & 109   \\
Median (time steps)              & 1125  & 144   & 1293  & 180   \\
Third quartile (time steps)      & 1609  & 256   & 1741  & 298   \\ \midrule
Total                           & 24250 & 75750 & 31208 & 68792 \\ \bottomrule
\end{tabular}
\caption{Summary statistics of $t_{\rm a}^{\rm right}$ and $t_{\rm a}^{\rm wrong}$ for $10^5$ agents accumulated over $10^3$ simulations each for \erdos and periodic square lattice networks. The corresponding histograms are visualized in Figures \ref{fig:ER_opponents} and \ref{fig:lattice_opponents}. The results are to be compared with Table \ref{tab:ba_right_wrong_times} for $n=100$ \barabasi networks.}
\label{tab:other_right_wrong_times}
\end{table}

We find \barabasi networks have a slightly wider spread in the number of agents attaining $\mean \rightarrow \theta_0$ compared to \erdos networks, with standard deviations of 5 and 3 respectively. Other than that, the properties of $\mean$, the number of agents out of $n=100$ attaining $\mean \rightarrow \theta_0$ per simulation, $t_{\rm a}^{\rm right}$ and $t_{\rm a}^{\rm wrong}$ of the \erdos networks shown in Figure \ref{fig:ER_opponents} are almost identical to the \barabasi networks shown in Figures \ref{fig:BA_opponents_average} and \ref{fig:BA_opponents_times_average}. By contrast, there are several slight qualitative differences in the behavior of square lattice networks compared to \barabasi networks. (i) On average, six more agents attain $\mean \rightarrow \theta_0$ in square lattice networks compared to \barabasi networks. (ii) The agents in the lattice networks take a slightly longer time on average to achieve asymptotic learning, with $t_{\rm a}^{\rm right}$ and $t_{\rm a}^{\rm wrong}$ being $\sim 200$ and $\sim 30$ time steps slower respectively compared to \barabasi networks. (iii) The characteristic spike in the histogram of $t_{\rm a}^{\rm right}$ near $t_{\rm a}^{\rm right} \approx 0$ for square lattice networks seen in Figure \ref{subfig:ER_opponents_correct_time} is roughly five time shorter compared to the \barabasi network counterpart in Figure \ref{subfig:BA_opponents_average_correct_time}.

\subsection{Structural balance and asymptotic learning}
\label{sec:ER_lattice_stats_balance}

\begin{figure}[h!tbp]
    \centering
    \begin{subfigure}[b]{0.32\linewidth}
        \centering
        \includegraphics[width=\linewidth]{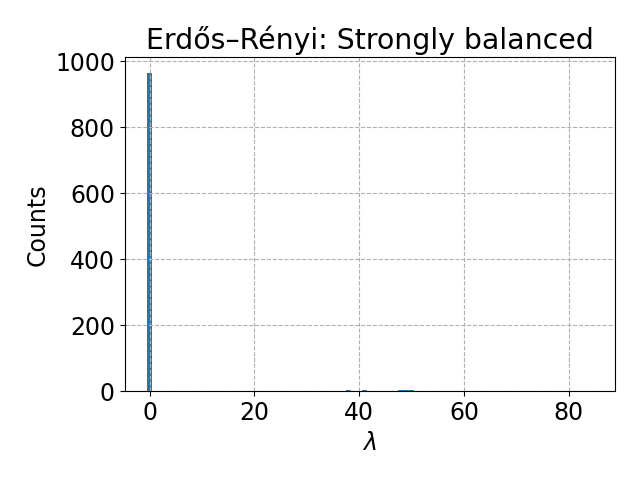}
        \caption{}
    \end{subfigure}
    \begin{subfigure}[b]{0.32\linewidth}
        \centering
        \includegraphics[width=\linewidth]{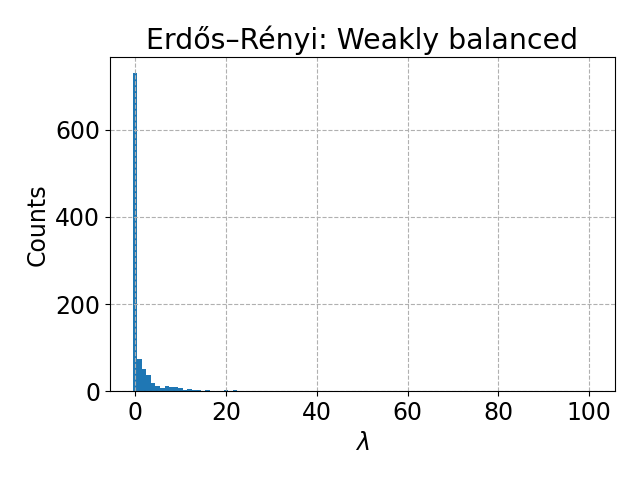}
        \caption{}
    \end{subfigure}
    \begin{subfigure}[b]{0.32\linewidth}
        \centering
        \includegraphics[width=\linewidth]{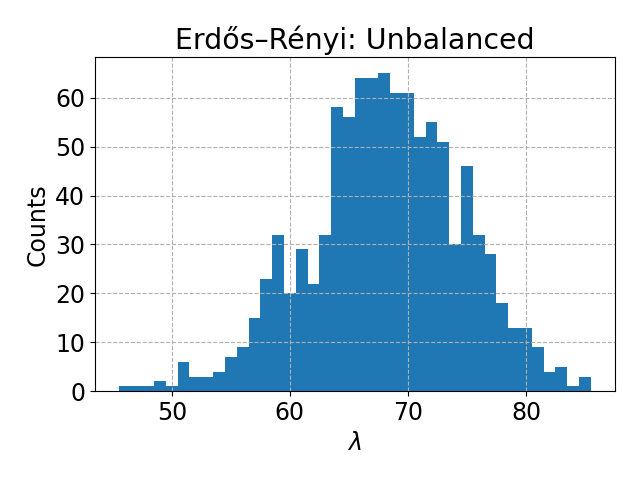}
        \caption{}
        \label{subfig:ER_balance_lambda_unbalanced}
    \end{subfigure}
    \caption{Structural balance: histograms of $\lambda$, the number of agents (out of $n=100$ in total) experiencing turbulent nonconvergence per simulation, for (a) strongly balanced, (b) weakly balanced and (c) unbalanced \erdos networks, to be compared with Figure \ref{fig:balance_lambda}. Each histogram bin has a width of unity.}
    \label{fig:ER_balance_lambda}
\end{figure}

\begin{figure}[h!tbp]
    \centering
    \begin{subfigure}[b]{0.32\linewidth}
        \centering
        \includegraphics[width=\linewidth]{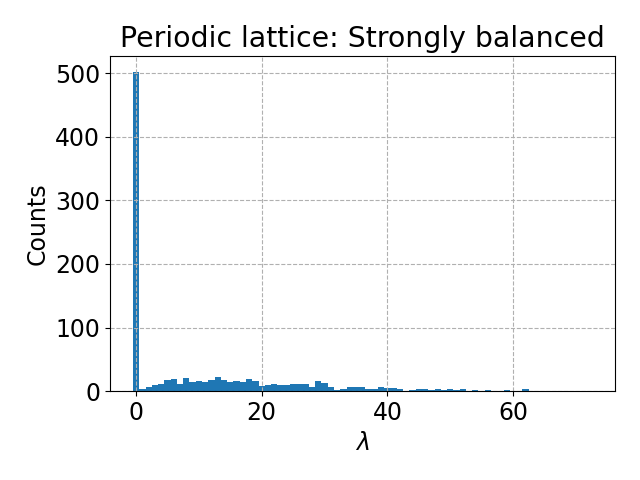}
        \caption{}
    \end{subfigure}
    \begin{subfigure}[b]{0.32\linewidth}
        \centering
        \includegraphics[width=\linewidth]{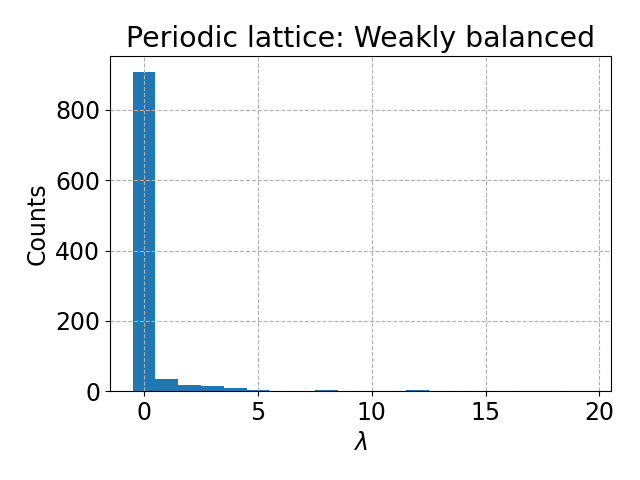}
        \caption{}
    \end{subfigure}
    \begin{subfigure}[b]{0.32\linewidth}
        \centering
        \includegraphics[width=\linewidth]{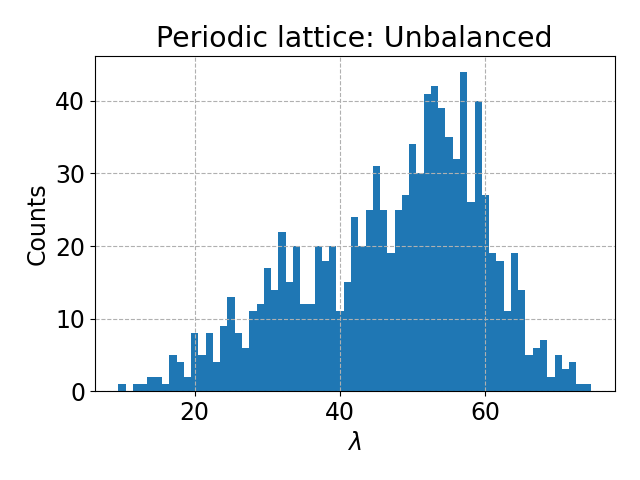}
        \caption{}
        \label{subfig:lattice_balance_lambda_unbalanced}
    \end{subfigure}
    \caption{Structural balance: histograms of $\lambda$, the number of agents (out of $n=100$ in total) experiencing turbulent nonconvergence per simulation, for (a) strongly balanced, (b) weakly balanced and (c) unbalanced periodic square lattice networks, to be compared with Figure \ref{fig:balance_lambda}. Each histogram bin has a width of unity.}
    \label{fig:lattice_balance_lambda}
\end{figure}

\begin{figure}[h!t]
    \centering
    \begin{subfigure}[b]{0.45\linewidth}
        \centering
        \includegraphics[width=\linewidth]{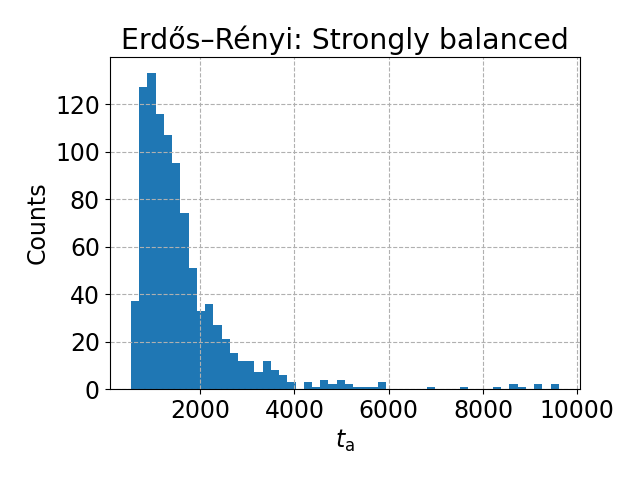}
        \caption{}
    \end{subfigure}
    \begin{subfigure}[b]{0.45\linewidth}
        \centering
        \includegraphics[width=\linewidth]{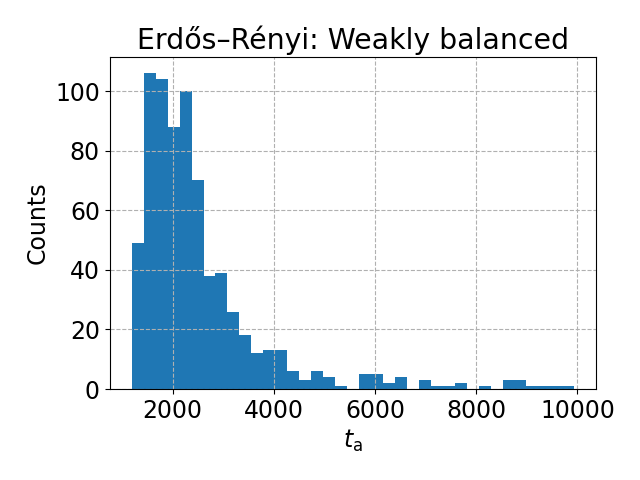}
        \caption{}
    \end{subfigure}
    \caption{Structural balance: histograms of $t_{\rm a}$ when the network does achieve asymptotic learning, for (a) strongly balanced and (b) weakly balanced \erdos networks, to be compared with Figure \ref{fig:balance_asymp_time}. Unbalanced networks are excluded as they always experience turbulent nonconvergence.}
    \label{fig:ER_balance_asymp_time}
\end{figure}

\begin{figure}[h!t]
    \centering
    \begin{subfigure}[b]{0.45\linewidth}
        \centering
        \includegraphics[width=\linewidth]{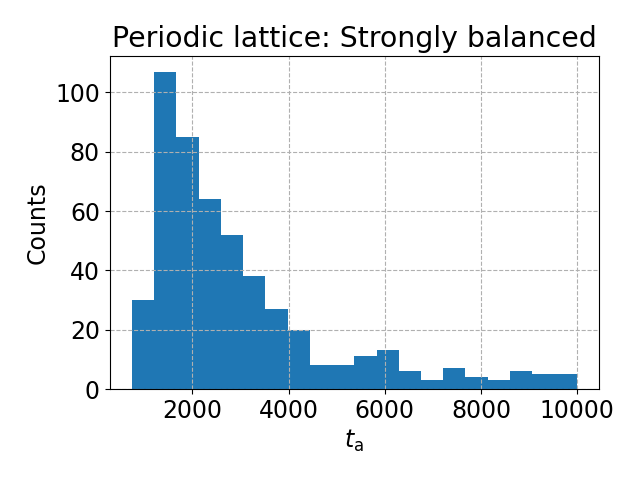}
        \caption{}
    \end{subfigure}
    \begin{subfigure}[b]{0.45\linewidth}
        \centering
        \includegraphics[width=\linewidth]{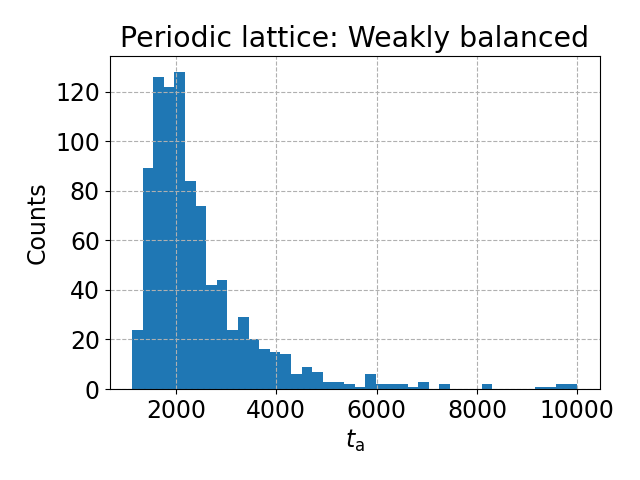}
        \caption{}
    \end{subfigure}
    \caption{Structural balance: histograms of $t_{\rm a}$ when the network does achieve asymptotic learning, for (a) strongly balanced and (b) weakly balanced periodic square lattice networks, to be compared with Figure \ref{fig:balance_asymp_time}. Unbalanced networks are excluded as they always experience turbulent nonconvergence.}
    \label{fig:lattice_balance_asymp_time}
\end{figure}

Next, we study the behavior of agents in $n=100$ \erdos and square lattice networks in the context of structural balance theory, to compare with \barabasi networks in Section \ref{sec:balance}. Here, we investigate the properties of $\lambda$ and $t_{\rm a}$ in \erdos and square lattice networks, with the same procedure used to generate Tables \ref{tab:balance_lambda} and \ref{tab:balance_asymp_time}. Figures \ref{fig:ER_balance_lambda} and \ref{fig:lattice_balance_lambda} shows the histograms of $\lambda$, the number of agents out of $n=100$ who fail to achieve asymptotic learning, for \erdos and periodic square lattice networks respectively according to their structural balance, while Table \ref{tab:other_balance_lambda} describes their summary statistics. For strongly and weakly balanced networks, Figures \ref{fig:ER_balance_asymp_time} and \ref{fig:lattice_balance_asymp_time} shows the histograms of $t_{\rm a}$ when asymptotic learning is achieved for \erdos and periodic square lattice networks respectively, while Table \ref{tab:other_balance_asymp_time} describes their summary statistics. 

\begin{table}[h!t]
\centering
\begin{tabular}{lrrrrrr}
\hline
                       & \multicolumn{3}{c}{\erdos} & \multicolumn{3}{c}{Periodic square lattice} \\
Property of  $\lambda$ & Strong   & Weak   & Unbalanced  & Strong    & Weak    & Unbalanced   \\ \hline
Zero                   & 964      & 730    & 0           & 502       & 908     & 0            \\
Mean                   & 2        & 1      & 68          & 10        & 0       & 47           \\
Standard deviation     & 8        & 5      & 6           & 14        & 1       & 12           \\
First quartile         & 0        & 0      & 64          & 0         & 0       & 38           \\
Median                 & 0        & 0      & 68          & 0         & 0       & 50           \\
Third quartile         & 0        & 1      & 73          & 18        & 0       & 57           \\ \hline
Count                  & 1000     & 1000   & 1000        & 1000      & 1000    & 1000         \\ \hline
\end{tabular}
\caption{Structural balance: summary statistics of $\lambda$, the number of agents (out of $n=100$ in total) experiencing turbulent nonconvergence per simulation, for strongly balanced (strong), weakly balanced (weak) and unbalanced \erdos and lattice networks, to be compared with Table \ref{tab:balance_lambda}. The summary statistics are based on the histograms shown in Figures \ref{fig:ER_balance_lambda} and \ref{fig:lattice_balance_lambda}.}
\label{tab:other_balance_lambda}
\end{table}

Similar to the \barabasi networks in Section \ref{sec:balance}, we observe that in every one of the $10^3$ unbalanced \erdos and square lattice networks, at least one agent fails to achieve asymptotic learning. The distribution of $\lambda$ for \erdos is almost identical to that of \barabasi networks. However, the periodic square lattice networks display an unexpected behavior that differs from \barabasi and \erdos networks: the strongly balanced networks are less likely to achieve asymptotic learning compared to weakly balanced networks, with 502 and 908 instances of $\lambda=0$ out of 1000 for strongly and weakly balanced networks respectively.

\begin{table}[h!tbp]
\centering
\begin{tabular}{lrrrr}
\hline
                        & \multicolumn{2}{c}{\erdos} & \multicolumn{2}{c}{Square lattice} \\
Property of  $t_{\rm a}$& Strong          & Weak          & Strong            & Weak           \\ \hline
Mean                    & 1641            & 2532          & 2961              & 2466           \\
Standard deviation      & 1123            & 1334          & 1951              & 1186           \\
First quartile          & 987             & 1736          & 1620              & 1736           \\
Median                  & 1331            & 2171          & 2331              & 2120           \\
Third quartile          & 1863            & 2820          & 3517              & 2777           \\ \hline
Count                   & 964             & 730           & 502               & 908            \\ \hline
\end{tabular}
\caption{Structural balance: summary statistics of $t_{\rm a}$ when the network achieves asymptotic learning, of strongly balanced (strong) and weakly balanced (weak) \erdos and lattice networks, to be compared with Table \ref{tab:balance_asymp_time}. The summary statistics are based on the histograms shown in Figures \ref{fig:ER_balance_asymp_time} and \ref{fig:lattice_balance_asymp_time}.}
\label{tab:other_balance_asymp_time}
\end{table}

When the agents in the strongly and weakly balanced networks do achieve asymptotic learning, the histogram of $t_{\rm a}$ of the \erdos networks is similar to that of the \barabasi networks in Section \ref{sec:balance}. The histogram of $t_{\rm a}$ for weakly balanced periodic square lattice networks is also similar to that of weakly balanced \barabasi and \erdos networks. However, the strongly balanced lattice networks tend to reach asymptotic learning significantly slower than their \barabasi and \erdos counterparts, with a mean $t_{\rm a}$ of 2961 time steps for lattice networks compared to a mean of $\approx 1600$ of both \barabasi and \erdos networks. The $t_{\rm a}$ has a wider distribution in strongly balanced lattice networks, with a standard deviation of $1951$ time steps, compared to 991 and 1123 time steps for \barabasi and \erdos networks respectively.

\subsection{Dependence on the clustering coefficient and diameter of the networks}

In \ref{sec:ER_lattice_stats_opponents} and \ref{sec:ER_lattice_stats_balance}, we investigate if the behaviors of the agents in \erdos and square lattice networks are different from those in \barabasi networks. In this section, we look at some common network properties to see if the behavior of the agents depends on these properties. We look at two specific network properties: diameter and clustering coefficient. Investigating other network properties lies outside the scope of this paper.

The diameter of a network is a measure of the distance of a network, which is related to the number of links that must be traversed to go from one agent to another. The distance between agents $i$ and $j$ is defined to be the fewest number of links that must be traversed to get from agent $i$ to $j$. In other words, if the opinion of agent $i$ is to influence agent $j$, then the distance is related to the minimum number of agents that the opinion of agent $i$ must diffuse through before reaching agent $j$. Then, the diameter of the network is defined to be the longest distance between all pairs of agents in the network \cite{diestel2000graph}. The diameter of a network can be calculated using \texttt{networkx}'s \texttt{diameter} function \cite{networkx}. The diameter of a network takes integer values, with a maximum possible value of $n-1$.

Loosely, the clustering coefficient of a network measures how tightly connected a network is. There are many definitions of the clustering coefficient. Here, we use the global clustering coefficient, which is also called transitivity. The transitivity aims to measure the following: given that agents 1 and 2 are linked to agent 3, what are the odds that agent 1 and 2 are also linked? We measure the transitivity of a network using \texttt{networkx}'s \texttt{transitivity} function \cite{networkx}, which takes a value in the range $[0,1]$. Note that square lattice networks have a transitivity of zero. Although there are generalized definitions of transitivity to account for signed networks, such as \cite{opsahl2009clustering}, we stick to \texttt{networkx}'s implementation for simplicity, which ignores whether the agents are allies or opponents.

\begin{figure}[h!t]
    \centering
    \begin{subfigure}[b]{0.45\linewidth}
        \centering
        \includegraphics[width=\linewidth]{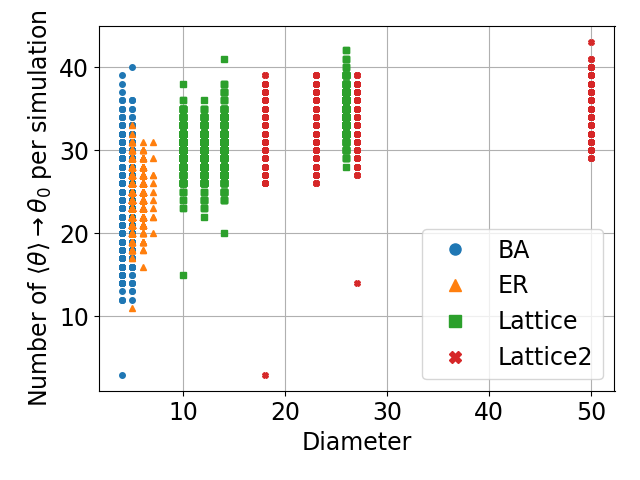}
        \caption{}
    \end{subfigure}
    \begin{subfigure}[b]{0.45\linewidth}
        \centering
        \includegraphics[width=\linewidth]{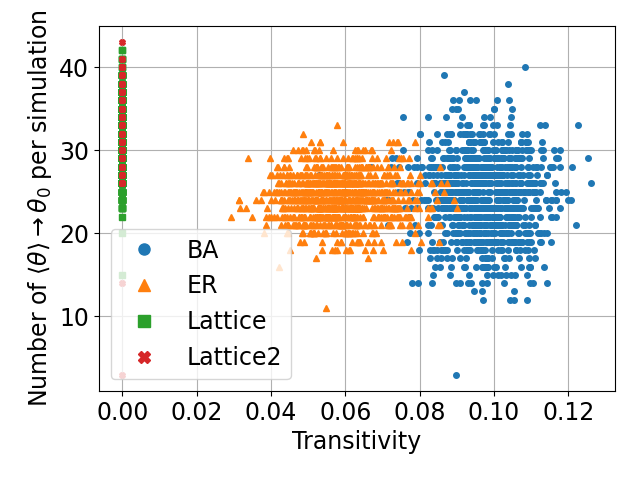}
        \caption{}
    \end{subfigure}
    \caption{Opponents-only $n=100$ networks: the number of agents attaining $\mean \rightarrow \theta_0$ out of $n=100$ per simulation plotted against the (a) diameter and (b) transitivity of the network. The blue circle data points (BA) represent \barabasi networks, the orange triangle points (ER) represent \erdos networks, while the green square (lattice) and red cross (lattice2) data points represent periodic and non-periodic square lattice networks respectively. Note that the periodic and non-periodic lattice network data points in (b) lie on top of each other and are indistinguishable by eye.}
    \label{fig:other_properties_opponents}
\end{figure}

Using the opponents-only networks studied in Figures \ref{fig:BA_opponents_average}, \ref{fig:ER_opponents} and \ref{fig:lattice_opponents}, we calculate the diameter and transitivity of each of the generated \barabasi, \erdos and square lattice network and compare these properties with the networks' respective $\mean$, the number of agents attaining $\mean \rightarrow \theta_0$ and $t_{\rm a}$. Figure \ref{fig:other_properties_opponents} shows the number of agents out of $n=100$ attaining $\mean \rightarrow \theta_0$ per simulation plotted against the network's diameter and transitivity. The lattice networks always have a greater diameter than the tested \barabasi and \erdos networks, but have zero transitivity. The tested \erdos networks tend to have a larger diameter but lower transitivity compared to the \barabasi networks. We note that a large portion of the parameter space is unexplored in this data set --- we are missing networks with diameters in the range $(30, 50)$ and networks with a transitivity greater than 0.13. Additionally, we do not investigate every single possible network of a given diameter or transitivity, which may cause sampling bias. For example, the non-periodic square lattice networks with 50 rows and two columns are the only networks in our data set with a diameter of 50. Nonetheless, Figure \ref{fig:other_properties_opponents} does not show any significant trend in the number of agents out of $n=100$ attaining $\mean \rightarrow \theta_0$ per simulation with the network's diameter or transitivity. A similar analysis of the influence of the diameter and transitivity of opponents-only networks on $\mean$ and $t_{\rm a}$ reveals no significant trend. Likewise, in the context of structural balance theory studied in Section \ref{sec:balance} and \ref{sec:ER_lattice_stats_balance}, there are no notable trends in $\lambda$ and $t_{\rm a}$ with the network's diameter and transitivity. A more thorough analysis of the dependence of the agents' behaviors on the various network properties is an interesting avenue for future research endeavors.

\end{document}